\pdfoutput=1
\documentclass[aoas,authoryear,preprint]{imsart}

\RequirePackage[OT1]{fontenc}
\RequirePackage{amsthm,amsmath}
\RequirePackage[numbers]{natbib}
\RequirePackage[colorlinks,citecolor=blue,urlcolor=blue]{hyperref}

\usepackage{amssymb}
\usepackage{booktabs}
\usepackage{bm}
\usepackage{color}
\usepackage{enumerate}
\usepackage{fontenc}
\usepackage{float} 
\usepackage[hmargin=2.8cm, vmargin=2.8cm]{geometry}
\usepackage{graphicx}      
\usepackage{lscape}
\usepackage{mathrsfs}
\usepackage{mdframed}
\usepackage{verbatim}
\usepackage[dvipsnames]{xcolor}
\newcommand{\T}{\ensuremath{\mathrm{\scriptscriptstyle T}}}

\newcommand\invisiblesection[1]{%
  \refstepcounter{section}%
  \addcontentsline{toc}{section}{\protect\numberline{\thesection}#1}%
  \sectionmark{#1}}

\def\dif{\mathrm{d}}
\def\e{\mathrm{e}}
\newcommand{\pr}{\textsf{P}}
\newcommand{\E}{\textsf{E}}
\newcommand{\sizeA}{0.2}
\newcommand{\sppart}[1]{\begin{center}{ \parskip = 2\baselineskip \small \MakeUppercase{#1}}\\[0.3cm]\end{center}}
\startlocaldefs
\numberwithin{equation}{section}
\theoremstyle{plain}

\newtheorem{example}{Example}

\endlocaldefs

\begin{document}

\begin{frontmatter}
\title{Time-Varying Extreme Value Dependence \\ 
  with Application to Leading European Stock Markets}
\runtitle{Time-Varying Extreme Value Dependence}

\begin{aug}
  \author{\fnms{Daniela}~\snm{Castro}~\snm{Camilo}\ead[label=e1]{daniela.castro@kaust.edu.sa}},
  \author{\fnms{Miguel}~\snm{de Carvalho}\ead[label=e2]{miguel.decarvalho@ed.ac.uk}} \and
  \author{\fnms{Jennifer}~\snm{Wadsworth}\ead[label=e3]{j.wadsworth@lancaster.ac.uk}}
  \runauthor{Castro Camilo, de Carvalho and Wadsworth}
  
  \affiliation{
    King Abdullah University of Science and Technology, \\
    University of Edinburgh and \\ 
    Lancaster University
  }
  
  \address{D.~Castro Camilo \\
    CEMSE Division \\
    King Abdullah University of Science and Technology \\
    Thuwal, Kingdom of Saudi Arabia \\
    \printead{e1}}
  
\address{M.~de Carvalho \\
School of Mathematics\\
University of Edinburgh \\
Edinburgh, UK\\
\printead{e2}}
  
  \address{J.~Wadsworth \\
    Department of Mathematics and Statistics \\
    Lancaster University \\
    Lancaster, UK \\
    \printead{e3}}
\end{aug}

\begin{abstract}
Extremal dependence between international stock markets is of particular interest in today's global financial landscape. However, previous studies have shown this dependence is not necessarily stationary over time. We concern ourselves with modeling extreme value dependence when that dependence is changing over time, or other suitable covariate. Working within a framework of asymptotic dependence, we introduce a regression model for the angular density of a bivariate extreme value distribution that allows us to assess how extremal dependence evolves over a covariate. We apply the proposed model to assess the dynamics governing extremal dependence of some leading European stock markets over the last three decades, and find evidence of an increase in extremal dependence over recent years.
\end{abstract}


\end{frontmatter}

\section{Introduction}
\label{Introduction}
In recent years, international stock markets have been registering unprecedented levels of turbulence. Episodes such as the subprime crisis and the Greek debt crisis may have boosted this turbulence a little further, and led many to fear a financial doomsday. The situation has been extraordinarily delicate in Europe, where evidence of increasing extremal dependence was found by \cite{PAL03, PAL04} before the most recent financial crisis. We look to update suitable parts of their analysis and in particular analyze the time-varying extremal dependence in a more complete manner than has been done before. To achieve this goal, we propose an approach for modeling nonstationarity in the extreme value dependence structure.

Statistical modeling of univariate extreme values has been in development since the 1970s \citep{NERC75}. Fundamental to practical application to complex problems has been the development of methodology to account for nonstationarity in the distributions of interest, which was first strongly advocated by \citet{DS90}. Typical approaches to this problem are based around the generalized linear modeling idea of allowing the parameters of a marginal distribution to depend on covariates; more flexible approaches involving generalized additive modeling were introduced by \citet{CD05}. \citet{ET09} present related ideas where data are preprocessed according to their dependence on covariates. 

Statistical methods for modeling multivariate extreme values were introduced by \citet{T88}, and developed in \citet{T90} and \citet{CT91}. Since this time, much work has been done on developing dependence modeling frameworks for extremes, yet surprisingly little has focused on how to incorporate nonstationarity into the (extremal) dependence structure. Exceptions include \citet{E09}, who introduces a conditionally independent hierarchical model, \citet{JAL14}, who develop methodology for including covariates in the model of \citet{HT04}, and \cite{CD14}, who develop a semiparametric model for settings where several multivariate extremal distributions are linked through the action of a covariate on an unspecified baseline distribution. In addition, \cite{HG14}  developed nonstationary models for spatial extremes where covariates can be included. In this work we add to the literature on modeling nonstationarity in the dependence
structure by proposing flexible methodology for a simple set-up. Working within a tail dependence framework known as \emph{asymptotic dependence}, we suppose that the relevant bivariate extreme value distribution evolves over a certain covariate of interest. The approach that we take is fully nonparametric, which is advantageous since neither the form of the bivariate distribution at a given covariate, nor the form of dependence on the covariate can be parametrically specified.

Our methodology is particularly tailored for assessing temporal changes in extremal dependence, which is the situation that we would like to investigate in our motivating example. \citet{PAL03, PAL04} studied the dependence between stock market returns in the US, UK, France, Germany, and Japan. The main focus of their works was to highlight that not all markets exhibit a sufficient strength of tail dependence to be asymptotically dependent, and to propose alternative dependence summaries. However, considering only the European markets, they noted that there was evidence for relatively strong left tail dependence, and we also find evidence for asymptotic dependence in the left tails of these major European markets. As noted by \citet{PAL03}, the dependence is not stationary in time, and a main focus of this work is to explore this nonstationarity using a full model for the time-varying dependence structure, rather than simply summary statistics.

In the next section we provide a background on dependence modeling for extreme values, and introduce our proposed framework for incorporating nonstationarity. In Section~\ref{Estimation and inference} we introduce our estimation and inference methods; numerical illustrations follow in Section~\ref{Simulation study}. The focus of Section~\ref{Application} is on applying the proposed methods to returns from three major European stock markets---using CAC, DAX, and FTSE---to assess the evolution of their extremal dependence structure over time. We conclude in Section~\ref{Discussion}.

\section{Conditional modeling for bivariate extremes}
\label{modeling}
\subsection{Bivariate statistics of extremes}
\label{bse}

Let $\{(Y_{i,1},Y_{i,2})\}_{i=1}^{N}$ be a collection of independent and 
identically distributed random vectors with continuous marginal 
distributions $F_{Y_1}$ and $F_{Y_2}$. We are
concerned with assessments of the extremal dependence between the
components of the vectors, and thus without loss of generality we
shall suppose that they have standard {Fr\'echet} margins, i.e.,
${\pr(Y_{j} > y)=\exp(-1/y)}$, for $y>0$ and $j=1,2$. 
Let 
\begin{equation*}
  (M_{N,1},M_{N,2}) = 
  \frac{1}{N}\biggl(\max_{1\leqslant i \leqslant N}\{Y_{i,1}\}, 
  \max_{1\leqslant i \leqslant N}\{Y_{i,2}\}\biggr)
\end{equation*}
be the standardized vector of componentwise maxima. Then if
\begin{align}
\label{eq:conv}
  \pr(M_{N,1}\leqslant y_1,M_{N,2}\leqslant y_2) \to G(y_1,y_2), \quad \text{as~} N\to\infty,
\end{align}
where $G$ is a non-degenerate distribution function, $G$ has the form
\begin{align}
 G(y_1,y_2) = \exp\left\{- 2 \int_{[0,1]}
 \max\left(\frac{w}{y_{1}},\frac{1-w}{y_{2}}\right)\,H(\dif w)\right\}, \quad 
  y_{1},y_{2} > 0. 
\label{eq:maxconv}
\end{align}
Here, $G(y_1,y_2)$ is the so-called \emph{bivariate extreme value distribution} and $H$ is a probability measure---known as the {\emph{angular measure}}. A consequence of Pickands' (\citeyear{P81}) representation theorem is that the angular measure needs to obey the following marginal moment constraint
\begin{align}
 \int_{[0,1]} w\,H(\dif w) = 1/2;
  \label{const_1}
\end{align}
see, for example, \citet[][Theorem~8.1]{C01}. Let $R = Y_1 + Y_2 $ and $W = Y_1/(Y_1+Y_2)$. \citet{HR77} have shown that the convergence {in~\eqref{eq:conv}} is equivalent to
\begin{align}
\pr\left(W\in\cdot \mid  R> u \right) \to H(\cdot), \quad u \to\infty. \label{eq:wconv}
\end{align}
In practice, convergence \eqref{eq:wconv} is more often useful than \eqref{eq:conv} and tells us that when the `radius' $R$ is large, the `pseudo-angles' $W$ are approximately distributed according to $H$, and approximately independent of $R$. The distribution of mass of $H$ on $[0,1]$ describes the extremal dependence structure of the random vector $(Y_1,Y_2)$. The extreme cases of this distribution are given by \emph{asymptotic independence}, whereby all mass is placed at the vertices of $[0,1]$, giving $G(y_1,y_2) = \exp\{-(y_1^{-1}+y_2^{-1})\}$,  and by complete dependence, whereby all mass is placed at the center of the interval, yielding $G(y_1,y_2) =  \exp\{-\max(y_1^{-1},y_{2}^{-1})\}$. We refer to situations where $H$ has mass away from the vertices as \emph{asymptotic dependence} and this will be the framework of our modeling. Nevertheless, asymptotic independence is a relatively common situation in practice, and can be detected when $R$ and $W$ are not found to be independent for any values of $R$, with the mass of $W$ moving closer to $0$ and $1$ as events become more extreme. In this situation, no models for $H$ will provide useful information on the extremal dependence structure. Finally, a standard assumption for statistical modeling is that $H$ is absolutely continuous with \emph{angular density} $h = \dif H/ \dif w$, and this will be our framework.

Functionals of interest of the angular measure include the bivariate extreme value distribution \eqref{eq:maxconv}, which also represents the extreme value \textit{copula}, $C_{\text{EV}}$, \citep[e.g.][]{GS10} in Fr\'echet margins, i.e., $G(y_1, y_2)  = C_{\text{EV}}(\e^{-1/y_1}, \e^{-1/y_2})$. Other functionals include the \citet{P81} dependence function $A(w) = 1 - w + 2 \int_0^w H(u) \, \dif u$, and the extremal coefficient $C = 2 A(1/2)$. Extreme value independence corresponds to $A(w) = 1$, whereas perfect dependence corresponds to $A(w) = \max(w, 1 - w)$.  

\subsection{Conditional modeling framework}\label{definitions}
We define the conditional bivariate extreme value (BEV) distribution as
 \begin{align}\label{biv_gev}
  G_x(y_1,y_2) \equiv  G(y_1,y_2 \mid X = x) =  \exp\left\{-2\int_{[0,1]} \max\bigg(\frac{w}{y_1},\frac{1-w}{y_2}\bigg) H(\dif w \mid X = x)\right\},
 \end{align}
for {$x \in \mathcal{X} \subseteq \mathbb{R}$}, and $y_1,y_2 > 0$. Here $H_x(\cdot) \equiv H(\cdot \mid X = x)$ are conditional  probability measures satisfying
\begin{equation}
  \int_{[0,1]} w H_x(\dif w) = 1/2, \quad x \in \mathcal{X}.
  \label{moment.constraint}
\end{equation}
If $H_x(w) \equiv H_x[0,w]$ is absolutely continuous, its conditional angular density is $h_x = \dif H_x/\dif w$. Further aspects of conditional angular measures are discussed in \cite{C16}. 

\begin{figure}
  \vspace{-1.2cm}
  \begin{minipage}[c]{6.8cm}
  \hspace{-3.5cm}
    \includegraphics[width=2\textwidth]{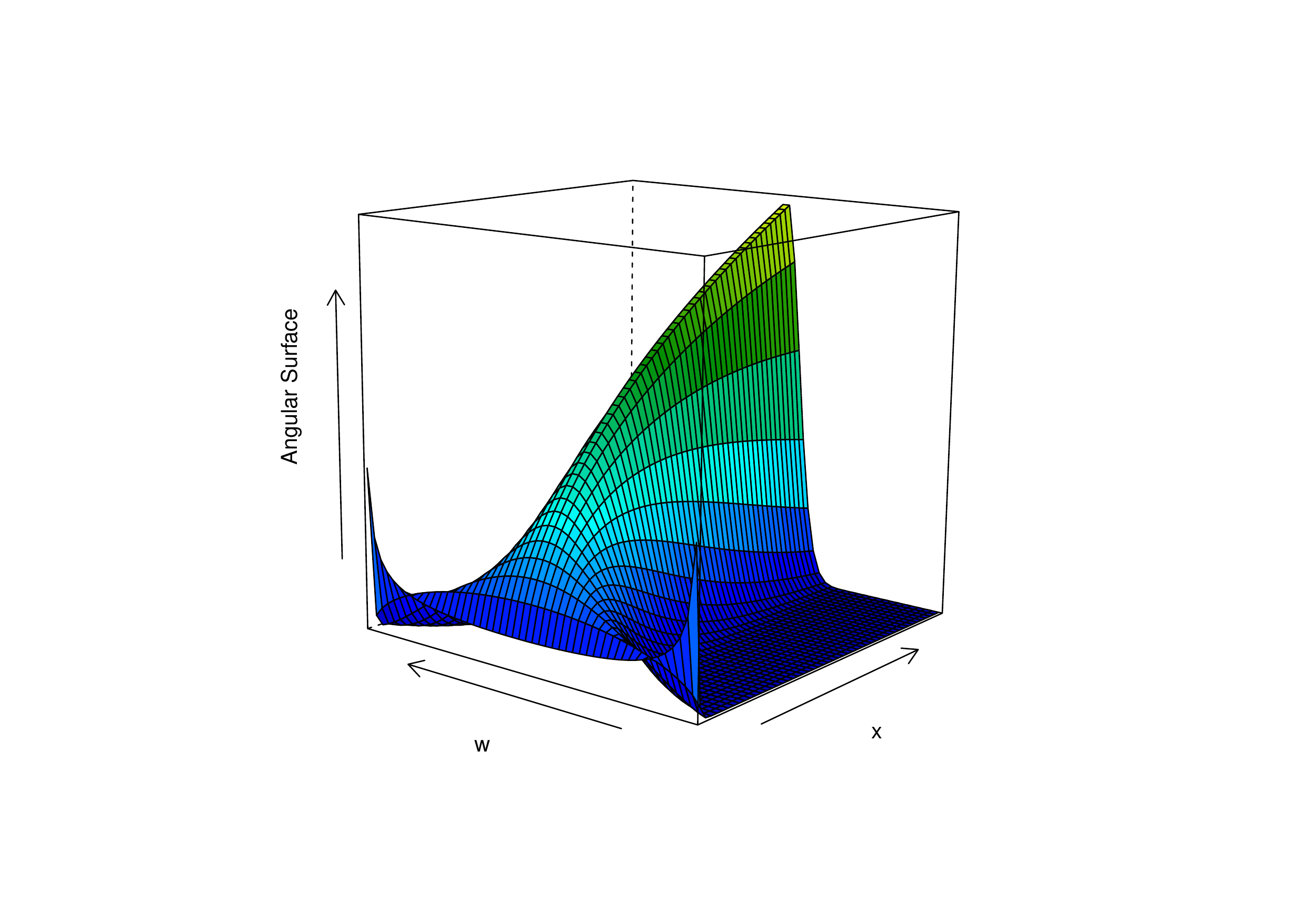}
    \text{\hspace{0cm}}
    \label{claims}
  \end{minipage} 
  \begin{minipage}[c]{6.8cm}
    \hspace{-2.5cm}
    \includegraphics[width=2\textwidth]{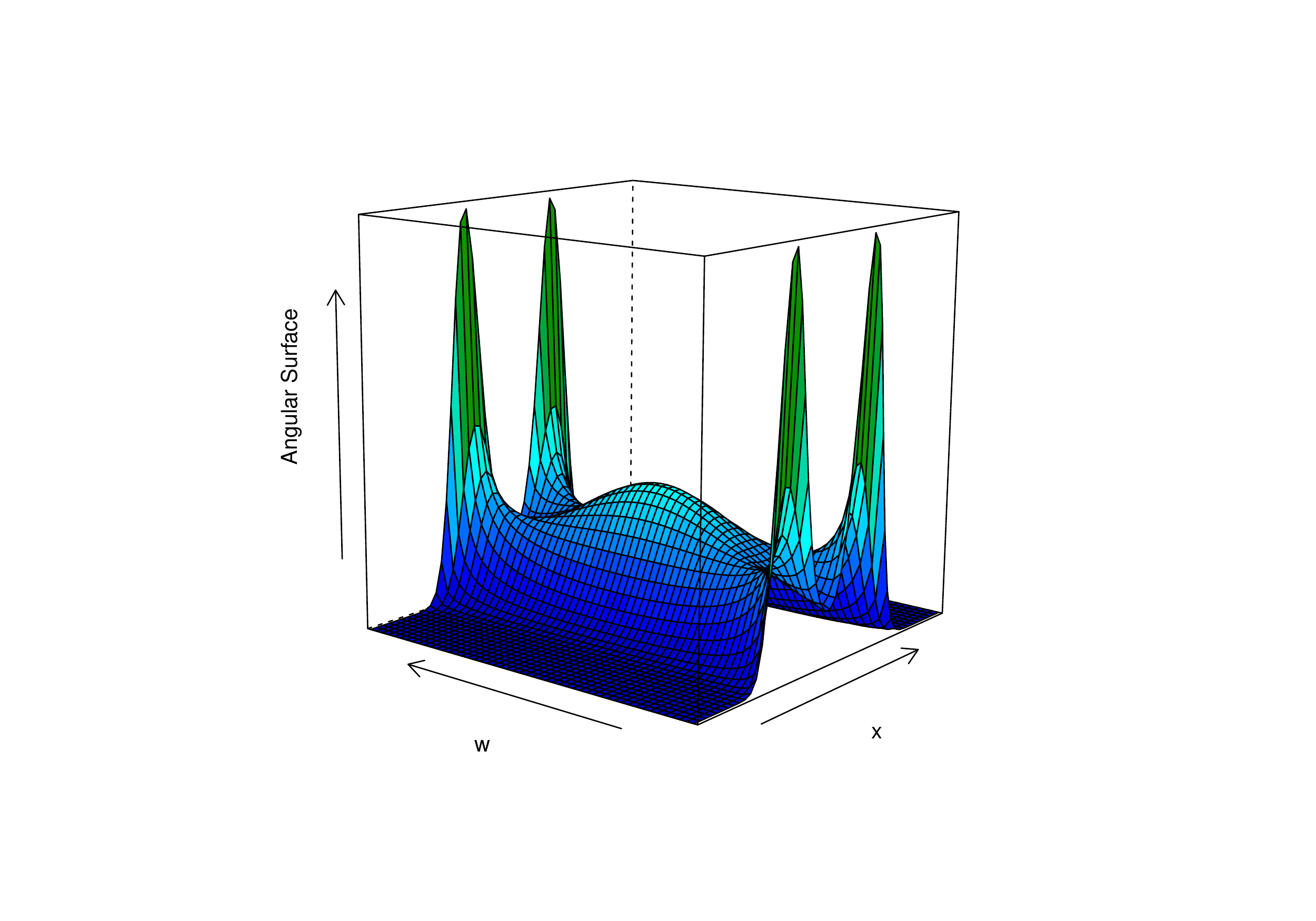}
    \label{claims}
  \end{minipage} 
  \vspace{-1.2cm}
  \caption{\footnotesize (a) Angular surface from a conditional beta family, with $\mu_x=x$ , for $x \in \mathcal{X} = [0.5, 50]$. (b) Angular surface from a conditional logistic family, with $\alpha_x=\Phi(x^2)$, for $x \in \mathcal{X} = [-3, 3]$.}
  \label{spec.examples}
\end{figure}
  
Our main modeling object of interest will be the set of conditional angular densities $\{h_x(w) :w\in[0,1], x\in \mathcal{X}\}$, which we will refer to as the \textit{angular surface}. A simple angular surface can be obtained with the conditional angular density $h_x(w) = \beta(w; \mu_x, \mu_x)$, where $\mu: \mathcal{X} \mapsto (0, \infty)$, and $\beta(\cdot; p, q)$ denotes the beta density with shape parameters $p, q > 0$. In Figure~\ref{spec.examples} (a), we represent an angular surface based on this model, with  $\mu_x=x$, for $x \in \mathcal{X} = [0.5, 50]$. As can be seen, larger values of the predictor $x$ lead to stronger levels of extremal dependence. Other angular surfaces can be readily constructed from parametric models for the angular density.

\begin{example}[conditional logistic model]\normalfont
The logistic angular surface is a covariate-adjusted extension of the logistic model \citep[][p.~146]{C01}, and it is based on the conditional angular density
\begin{equation}\label{logistic}
  h_x(w) = \frac{1}{2} \bigg(\frac{1}{\alpha_x} - 1\bigg)\{w(1-w)\}^{-1-1/\alpha_x} \{w^{-1/\alpha_x} + (1-w)^{-1/\alpha_x}\}^{\alpha_x - 2}, \quad w \in (0,1),
\end{equation}
where $\alpha: \mathcal{X} \mapsto (0,1]$. The closer $\alpha_x$ is to 0, the higher the level of extremal dependence, while the closer $\alpha_x$ is to 1, the closer we get to independence. Angular surfaces with simple `shapes' can be obtained by modeling $\alpha_x$ with either a distribution function, $F(x)$, or a survivor function, $1 - F(x)$. More sophisticated shapes can be obtained with $\alpha_x = (F \circ G)(x)$, for a certain continuous function $G: \mathcal{X} \mapsto \mathbb{R}$. In Figure~\ref{spec.examples} (b) we represent the logistic angular surface in \eqref{logistic} with $\alpha_x = \Phi(x^2)$, for $x \in \mathcal{X} = [-3, 3]$, where $\Phi$ denotes the standard normal distribution function.
\label{Logistic}
\end{example}

\begin{example}[conditional Dirichlet model]\normalfont The Dirichlet angular surface is a covariate-adjusted extension of the Dirichlet model \citep{CT91}, and it is based on the conditional angular density
  \begin{equation}
    h_x(w) = \frac{a_x b_x \Gamma(a_x + b_x + 1) (a_x w)^{a_x - 1} \{b_x (1-w)\}^{b_x - 1}}
    {2 \Gamma(a_x) \Gamma(b_x) \{a_x w + b_x (1-w)\}^{a_x + b_x + 1}}, \quad w \in (0,1),
  \end{equation}
  where $a: \mathcal{X} \mapsto (0, \infty)$ and $b: \mathcal{X} \mapsto (0, \infty)$. Angular surfaces with simple shapes can be obtained with $a_x = b_x = \exp(x)$, while if more complex dynamics are desirable, can be based on $a_x = \exp\{\mathcal{A}(x)\}$, $b_x = \exp\{\mathcal{B}(x)\}$, where $\mathcal{A}: \mathcal{X} \mapsto \mathbb{R}$ and $\mathcal{B}: \mathcal{X} \mapsto \mathbb{R}$ are continuous functions. 
\label{Dirichlet}
\end{example}

The basic idea of a conditional angular measure is not especially complicated, and inference for such would be simple if: (i) we knew our data conform to a particular parametric family, and (ii) we knew precisely how that family depended on $x$. However, since we do not have knowledge of either of these things, the natural approach to take is a nonparametric one. We assume that $h_x$ varies smoothly with $x$, and thus kernel smoothing becomes a natural option. We describe our estimation strategy in Section~\ref{Estimation and inference}.  

\subsection{Related conditional objects of interest}\label{Related conditional objects of interest}
Our estimation target $\{h_x(w) :w\in[0,1], x\in \mathcal{X}\}$ can be used for constructing other objects of interest when modeling bivariate extremes. For example, a conditional version of \citet{P81} dependence function can be defined as 
\begin{equation*}
  A_x(w) = 1 - w + 2 \int_0^w H_x(u) \, \dif u, \quad x \in \mathcal{X}, \quad w \in [0,1],
\end{equation*}
leading to the conditional extremal coefficient $C_x = 2 A_x(1/2)$. A covariate-adjusted extreme value copula can be readily constructed from~\eqref{biv_gev}. Although much theoretical and applied work has been devoted to time-dependent copulas \citep{P06, VAL07, AAL11, FM12}, the amount of work dedicated to time-varying extreme value copulas is by comparison fairly reduced, but of obvious relevance in a wealth of contexts of applied interest. The latter setup is the one of interest in the current manuscript. 

\begin{example}\normalfont
Using the conditional angular density from Example~\ref{Logistic}, we obtain $A_x(w) = \{(1-w)^{1/\alpha_x} + w^{1/\alpha_x}\}^{\alpha_x}$ and $C_x = 2^{\alpha_x}$, while the logistic angular surface is based on the conditional BEV distribution,
\begin{equation*}
  G_x(y_1, y_2) = \exp\{-(y_1^{-1/\alpha_x} + y_2^{-1/\alpha_x})^{\alpha_x}\}, \quad x \in \mathcal{X}, \quad y_1,y_2 > 0.
\end{equation*}
\end{example}

\section{Estimation and inference}\label{Estimation and inference}
\subsection{Derivation of pseudo-angles}\label{dpa}
Consider Eq.~\eqref{eq:wconv}. We are now supposing nonstationarity in the dependence structure such that
\begin{align}\label{eq:wconvns}
\pr\left(W \in\cdot \mid  R > u, X = x\right) \to H_x(\cdot), \quad u \to\infty. 
\end{align}
Note that we still assume that $R$ and $W$ are derived from $Y_1, Y_2$ with standard {Fr\'echet} margins. Typically, when stationarity in the extremal dependence structure is assumed, one searches for a high threshold in $R$, such that $W$ and $R$ are approximately independent above the threshold, and uses all $W$ associated to threshold exceedances of $R$ for inference. Supposing that $x$ does not impact upon the rate of convergence in the limit~\eqref{eq:wconvns}, a similar approach is justified here. However, for prudence, we assess the dependence of $R$ on $x$ using quantile regression \citep{K05}. {To be consistent with the nonparametric nature of our approach, we fit a nonparametric quantile regression using regression splines. This method flexibly fits a piecewise cubic polynomial to estimate the 95\% quantile of $R$. If any relationship between $R$ and $x$ is detected, then we take the $W$ associated to exceedances of the fitted threshold by $R$ for inference.}
Below we use $n = o(N)$ to denote the number of pseudo-angles that resulted from thresholding $R_i = Y_{i, 1} + Y_{i, 2}$, for $i = 1, \ldots, N$. Further details on the derivation of pseudo-angles for our data application can be found in Section~\ref{tmed}. 

We note that we are not allowing for the margins to change over the predictor. This is however a sensible modeling assumption for our data application, because (filtered) returns are known to be approximately stationary. Indeed, as posed by \citet[][p.~7]{R07}~``Returns have more attractive statistical properties than prices such as stationarity.'' See Section~\ref{exploratory} for details on the filtering methods used in our data application.

\subsection{Conditional angular density estimation}\label{pdsde}
Here we outline our estimator for the family of densities $\{h_x(w) :w\in[0,1], x\in \mathcal{X}\}$. Assume observations $\{(X_i, W_i)\}_{i=1}^n$, where the covariates $X_i$ {are continuous and  in $\mathcal{X} \subseteq \mathbb{R}$}. Let $K_b(x)=(1/b)K(x/b)$ be a kernel with bandwidth $b>0$. For any $x \in \mathcal{X}$, we define the estimator
\begin{equation}\label{eq:h*}
 \widehat{h}_x(w) = \sum_{i=1}^n \pi_{b,i}(x) \beta(w; \nu W_{i} \theta_b(x) + \tau, \nu\{1-W_{i} \theta_b(x)\} + \tau), \quad w \in (0,1), 
\end{equation}
where 
\begin{align*}
\theta_b(x) = \frac{1/2}{\sum_{i=1}^n \pi_{b,i}(x) W_{i}}, \quad  \pi_{b,i}(x) = \frac{K_b(x - X_i)}{\sum_{j=1}^n K_b(x-X_j)}, \quad i = 1,\ldots, n.
\end{align*}
The moment constraint \eqref{moment.constraint} is satisfied, since 
\begin{align*}
\int_0^1 w \widehat{h}_x(w) \, \dif w = \frac{\sum_{i=1}^n K_b(x-X_i)\{\nu W_{i} \theta_b(x) + \tau\}}{(\nu + 2\tau)\sum_{i=1}^n K_b(x-X_i)} = \frac{\nu/2+\tau}{\nu+2\tau}=1/2,
\end{align*}
for all valid $\tau \geqslant 0$, upon substitution of $\theta_b(x)$.

The two kernels ($K_b$ and $\beta$) and the three parameters involved in our estimator can be interpreted as follows. The bandwidth $b > 0$ is the scale parameter of the kernel $K_{b}$ and controls the amount of smoothing in the $x$-direction. The choice of the kernel $K_b$ is subject to the typical considerations. In principle, $K_b$ should be symmetric and unimodal, since there is a sense in which density estimators based on kernels that do not satisfy these requirements are inadmissible~\citep{C88}. While there are many kernel functions that do satisfy these basic requirements, it is well known that the choice of the kernel has little impact on the corresponding estimators; see~\citet[][Ch.~2]{WJ95} and references therein. The parameter $\nu > 0$ is asymptotically inversely proportional to the variance of the kernel $\beta$ and has the main role of controlling the amount of smoothing in the $w$-direction. The additional parameter $\tau \geqslant 0$ has the role of adjusting slightly the center of the kernel, allowing more flexible estimation, whilst not affecting the imposition of the moment constraint. Note that $\tau=0$ yields a kernel with mean equal to $W_i$, whilst $\tau=1$ yields a kernel with mode $W_i$. In addition, $\theta_b (x)$ assesses by how much we deviate from the moment constraint \eqref{moment.constraint}. To see this, note that $\theta_b(x) = (1/2)/\widehat{\E}(W \mid X = x)$, where $\widehat{\E}(W \mid X = x) = \sum_{i=1}^n \pi_{b,i}(x) W_{i}$ is the Nadaraya--Watson estimator \citep{N64, W64} of $\E(W \mid X = x) = \int_{[0,1]} w H_x(\dif w) = 1/2$, for all $x \in \mathcal{X}$. 

Plug-in estimators for the related conditional objects of interest discussed in Section~\ref{Related conditional objects of interest} can be readily obtained; particularly
\begin{equation*}
  \widehat{H}_x(w) = \sum_{i=1}^n \pi_{b,i}(x) B(w; \nu W_i \theta_b(x) + \tau, \nu\{1-W_i \theta_b(x)\} + \tau), \quad w \in (0,1),
\end{equation*}
where $B(w; p,q)$ is the regularized incomplete beta function, with $p, q>0$; in addition, the plug-in estimators for the conditional Pickands dependence function, extremal coefficient, and bivariate extreme value distribution can be written as

\begin{equation}\label{plugin}
  \begin{split}
    \widehat{A}_x(w) &= 1 - w + 2 \sum_{i=1}^n  \pi_{b,i}(x)\int_0^w B(u; \nu W_i \theta_b(x) + \tau, \nu\{1-W_i \theta_b(x)\} + \tau) \,\dif u, \\
    \widehat{C}_x &= 2 \widehat{A}_x(1/2) = 1 + 4 \sum_{i=1}^n  \pi_{b,i}(x) \int_0^{1/2} B(u; \nu W_i \theta_b(x) + \tau, \nu\{1-W_i \theta_b(x)\} + \tau) \,\dif u, \\ 
    \widehat{G}_x(y_1, y_2) &= 
    \exp\bigg\{-2\int_0^1 \max\bigg(\frac{u}{y_1},\frac{1-u}{y_2}\bigg) \\ & \hspace{3cm} \times \sum_{i=1}^n
      \pi_{b,i}(x)\beta(u; \nu W_i \theta_b(x) + \tau, \nu\{1-W_i \theta_b(x)\} + \tau) \, \dif u\bigg\},
  \end{split}
\end{equation}
for $x \in \mathcal{X}$, and $y_1,y_2 > 0$. 

\subsection{Connections to smoothing on the unit interval}
Kernel density estimation on the unit interval is a challenging problem; see \cite{C99}, \cite{JH07}, \cite{CAL13}, \cite{G14}, and the references therein. In this section we contrast a stationary version of our estimator \eqref{eq:h*} with that of \cite{C99}, and comment on the connections with the smooth Euclidean likelihood angular density of \cite{CAL13}. The latter can be regarded as a moment constrained kernel density estimator on the unit interval, in the sense that it obeys \eqref{const_1}. 

If all covariates $x$ take the same value, so that the estimation problem reduces to one of estimating the angular density for an identically-distributed set of pseudo-angles $\{W_i\}_{i=1}^n$, then~\eqref{eq:h*} becomes
\begin{align}
 \widehat{h}(w) = \frac{1}{n}\sum_{i=1}^n \beta\bigg(w; \nu \frac{W_i}{2\overline{W}} + \tau, \nu \bigg\{1-\frac{W_i}{2\overline{W}} \bigg\} + \tau\bigg),  \quad w \in (0,1). \label{eq:h*ss}
\end{align}
{The version of our estimator in Eq.~\eqref{eq:h*ss} differs from Chen's beta kernel \citep{C99}:}
\begin{equation}
{h}^\star(w) = \frac{1}{n} \sum_{i = 1}^n \beta\bigg(W_i; \frac{w}{s} + 1, \frac{1 - w}{s} + 1 \bigg), 
\label{eq:chen}  
\end{equation}
{where $s > 0$ is a bandwidth. Indeed, \eqref{eq:chen} puts the mode of the kernel at $W_i$ and so does our estimator in \eqref{eq:h*ss}, if we set $\tau = 1$. Yet, in \eqref{eq:h*ss} $w$ is the argument of $\beta(\cdot)$, whereas in \eqref{eq:chen}, $W_i$ is the argument of $\beta(\cdot)$. Estimator~\eqref{eq:h*ss} has closer} connections with the smooth Euclidean angular density estimator in \citet[][p.~1190]{CAL13}, and which is given by 
\begin{equation}\label{smooth.Euclidean}
  \begin{split}
    \widetilde{h}(w) &= \frac{1}{n} \sum_{i=1}^n \{ 1- (\overline{W} -
    1/2)S^{-2} (W_i - \overline{W}) \} \, \beta\{w; W_i \nu,
    (1-W_i)\nu\} \\
   &= \frac{1}{n} \sum_{i=1}^n 
    \beta\{w; W_i \nu, (1-W_i)\nu\} - \frac{1}{n} \sum_{i=1}^n (\overline{W} - 1/2)S^{-2} (W_i - \overline{W}) \} \beta\{w; W_i \nu, (1-W_i)\nu\},
  \end{split}
\end{equation}
for $w \in (0,1)$; here $\overline{W}$ and $S^2$ are the sample mean and sample variance of $W_1, \ldots, W_n$, that is,
\begin{equation*}
  \overline{W} = \frac{1}{n} \sum_{i=1}^n W_i, \quad S^2 = \frac{1}{n} \sum_{i=1}^n (W_i - \overline{W})^2.
\end{equation*}
A heuristic argument can be used to see this, by focusing on the case $\tau = 0$. The right-hand term in \eqref{smooth.Euclidean} enforces the moment constraint, and hence it is asymptotically negligible, so that for large $n$, we have $\widetilde{h}(w) \approx (1/n) \sum_{i=1}^n \beta\{w; W_i \nu, (1-W_i)\nu\}$; on the other hand, we also have that for large $n$, $\widehat{h}(w) \approx (1/n) \sum_{i=1}^n \beta\{w; W_i \nu, (1-W_i)\nu\}$, since $\overline{W} = 1/2 + o_{\text{p}}(1)$, as $n \to \infty$.  {While both \eqref{eq:h*ss} and \eqref{smooth.Euclidean}} obey the moment constraint \eqref{moment.constraint}, they impose it through different approaches: our estimator enforces \eqref{const_1} by rescaling the pseudo-angles with a factor of $(2 \overline{W})^{-1}$; the smooth Euclidean angular density enforces \eqref{const_1} additively, through the right-hand term in \eqref{smooth.Euclidean}. {To our knowledge, it is not straightforward to impose the moment constraint on Chen's kernel in \eqref{eq:chen}.}

\subsection{Tuning parameter selection and bootstrap}\label{tuning}
We select the tuning parameters {via maximum likelihood $K$-fold cross-validation (MLCV) \citep[][Section~7.10.1]{HAL01}}. Specifically, let $\{\mathbf{W}_1,\ldots,\mathbf{W}_K\}$ be the full sample of pseudo-angles split into $K$ blocks. In the analyses in Sections~\ref{Simulation study} and \ref{Application}, we split the blocks according to the values of the accompanying covariate $x$, so that each $\mathbf{W}_k = (W_{k,1}, \ldots, W_{k,n_k})$ is in a similar part of the covariate space. Letting $ \widehat{h}_{x(-k)}$ denote the estimator leaving out the $k$th sample, $\mathbf{W}_k$, of length $n_k$, we select
\begin{align}\label{mlcv}
  (\widehat{b}, \widehat{\nu}, \widehat{\tau}) = \arg \min_{(b,\nu,\tau)\in\mathcal{R}_{\mathcal{X},n}}  \sum_{k=1}^{K} \sum_{j=1}^{n_k} -\log \widehat{h}_{X_{k,j}(-k)}(W_{k,j}),
\end{align}
with
\begin{align}\label{Rset}
  \mathcal{R}_{\mathcal{X},n} &= \{(b,\nu,\tau)\in(0, \infty)^{3}: \nu W_i\theta_b(x)+\tau>0,\nu \{1-W_i\theta_b(x)\}+\tau>0, \text{for }i=1,\dots, n; x\in \mathcal{X}\}\nonumber\\
& =\{(b,\nu,\tau)\in(0, \infty)^{3}: \nu \{1-W_i\theta_b(x)\}+\tau>0, \text{for }i=1,\dots, n; x\in \mathcal{X}\}.
\end{align}
The constrained optimization yields well-defined estimates, since it guarantees the positivity of the beta parameters in our estimator. The latter equality in \eqref{Rset} follows from noticing that $\nu W_i\theta_b(x)+\tau>0$, for all $x\in \mathcal{X}$; further details on practical implementation of tuning parameter selection are given in Section~\ref{Finite sample performance}. 
  {It is known that for density estimation, MLCV can produce estimates with suboptimal performance leading to undersmoothed density estimates, especially when the true density has unbounded support~\citep[][Section~32.10.1]{dasgupta2008asymptotic}. Computational experiments in the supplementary materials show that the main findings in Section~\ref{Application} are very similar regardless of whether we use MLCV or least-squares cross-validation (LSCV) \citep[][Section~32.10.2]{dasgupta2008asymptotic}. Better results than the ones in Section~\ref{Simulation study} are to be expected if LSCV is used. However, LSCV would not be theoretically grounded for non-square integrable densities (e.g., $h_x(w) = \beta(w; x, x)$, for $x \in (0, 1/2)$).}


An uncertainty assessment can be performed by simulating from kernel density estimates themselves---in the spirit of the so-called smoothed bootstrap \citep{SY87}. The procedure detailed below, allows us to generate $B$ bootstrap angular surfaces. For $r\in\{1,\ldots, B\}$:
\begin{enumerate}
\item[1.~] Sample $j^\star$ from a discrete uniform distribution over $\{1, \hdots, n\}$.
\item[2.~] Sample $X_j^{r}\sim K_{\widehat{b}}(\cdot - X_{j^\star})$.
\item[3.~] Sample $W_j^{r}\sim \widehat{h}_{X_j^{r}}$ with 
  \begin{equation*}
    \widehat{h}_{X_j^{r}}(w) = \sum_{i=1}^n \pi_{\widehat{b},i}(X_j^{r}) \beta(w; \widehat{\nu} W_i \theta_{\widehat{b}}(X_j^{r}) + \widehat{\tau}, \widehat{\nu}\{1-W_i \theta_{\widehat{b}}(X_j^{r})\} + \widehat{\tau}), \quad w \in (0,1),
  \end{equation*}
where 
\begin{align*}
  \theta_{\widehat{b}}(X_j^{r}) = \frac{1/2}{\sum_{i=1}^n \pi_{\widehat{b},i}(X_j^{r}) W_i}, \quad  \pi_{\widehat{b},i}(X_j^{r}) = \frac{K_{\widehat{b}}(X_j^{r} - X_i)}{\sum_{k=1}^n K_{\widehat{b}}(X_j^{r}-X_k)}, \quad i = 1,\ldots, n.
\end{align*}
\item[4.~] Repeat Steps 1--3 $n$ times to obtain the $r$th bootstrap sample $({\bf{X}}^r, {\bf{W}}^r)$, with  ${\bf{X}}^r = (X_1^r,\ldots,X_n^r)^{\T}$ and ${\bf{W}}^r = (W_1^r,\ldots,W_n^r)^{\T}$.
\item[5.~] Use $({\bf{X}}^r, {\bf{W}}^r)$ and~\eqref{mlcv} to obtain bootstrap estimates $(\widehat{b}^r, \widehat{\nu}^r, \widehat{\tau}^r)$. 
\end{enumerate}
Using the bootstrap samples $\{({\bf{X}}^r, {\bf{W}}^r)\}_{r=1}^B$, the bootstrap estimates $\{(\widehat{b}^r, \widehat{\nu}^r, \widehat{\tau}^r)\}_{r=1}^B$, and~\eqref{eq:h*}, we can construct $B$ bootstrap angular surfaces $\widehat{h}^{1}_x, \ldots \widehat{h}^{B}_x$. {For computational convenience, Step~2 considers only a single bandwidth, $\widehat{b}$, but it is known \citep[see, e.g.,][]{P01} that smoothed bootstrap resamples need not be generated from the kernel density estimate with the same bandwidth. Indeed, the so-called calibration methods are known to perform well, but they require one to construct a resample over a sequence of bandwidths, and thus are computationally costlier.} Visualizing uncertainty of angular surfaces can be awkward, but cross sections of the angular
surface (i.e., conditional angular density estimates at fixed values of $x$) can be easily summarized using, for example, functional boxplots~\citep{sun2012functional}. Details on constructing functional boxplots for angular densities are given in Section~\ref{Finite sample performance}.
%


\subsection{A local-linear version of the estimator}\label{loclinear}
A local linear version of our estimator can be readily constructed by replacing the Nadaraya--Watson weights in \eqref{eq:h*} with 
\begin{equation}
  \label{eq:loclinwei}
   \pi_{b, i}(x) =  \frac{1}{n}
   \frac{\{\widehat{s}_2(x; b) - \widehat{s}_1(x; b)(X_i - x)\}K_b(X_i - x)}
   {\widehat{s}_2(x; b) \widehat{s}_0(x; b) - \widehat{s}_1^2(x; b)},
\end{equation}
where $\widehat{s}_m(x; b) = n^{-1} \sum_{i = 1}^n (X_i - x)^mK_b(X_i - x)$, for $m=0,1,2$. Local linear regression is often presented as a solution to mitigate boundary bias issues of the Nadaraya--Watson estimator \citep[][Section~5.5]{WJ95}. Throughout, we consider both Nadaraya--Watson and local linear weights to illustrate their relative performance.

\section{Simulation study}\label{Simulation study}
\subsection{Data-generating configurations and preliminary experiments}
\label{Data configuration}
\begin{figure}
\begin{center}
  \begin{footnotesize}
      {\textbf{Logistic angular surface}}
  \end{footnotesize}
  \end{center}
  \vspace{-1cm}
\begin{tabular}{ccc}
  \begin{minipage}[c]{7cm}
  \hspace{-2.5cm}
    \includegraphics[width=1.3\textwidth]{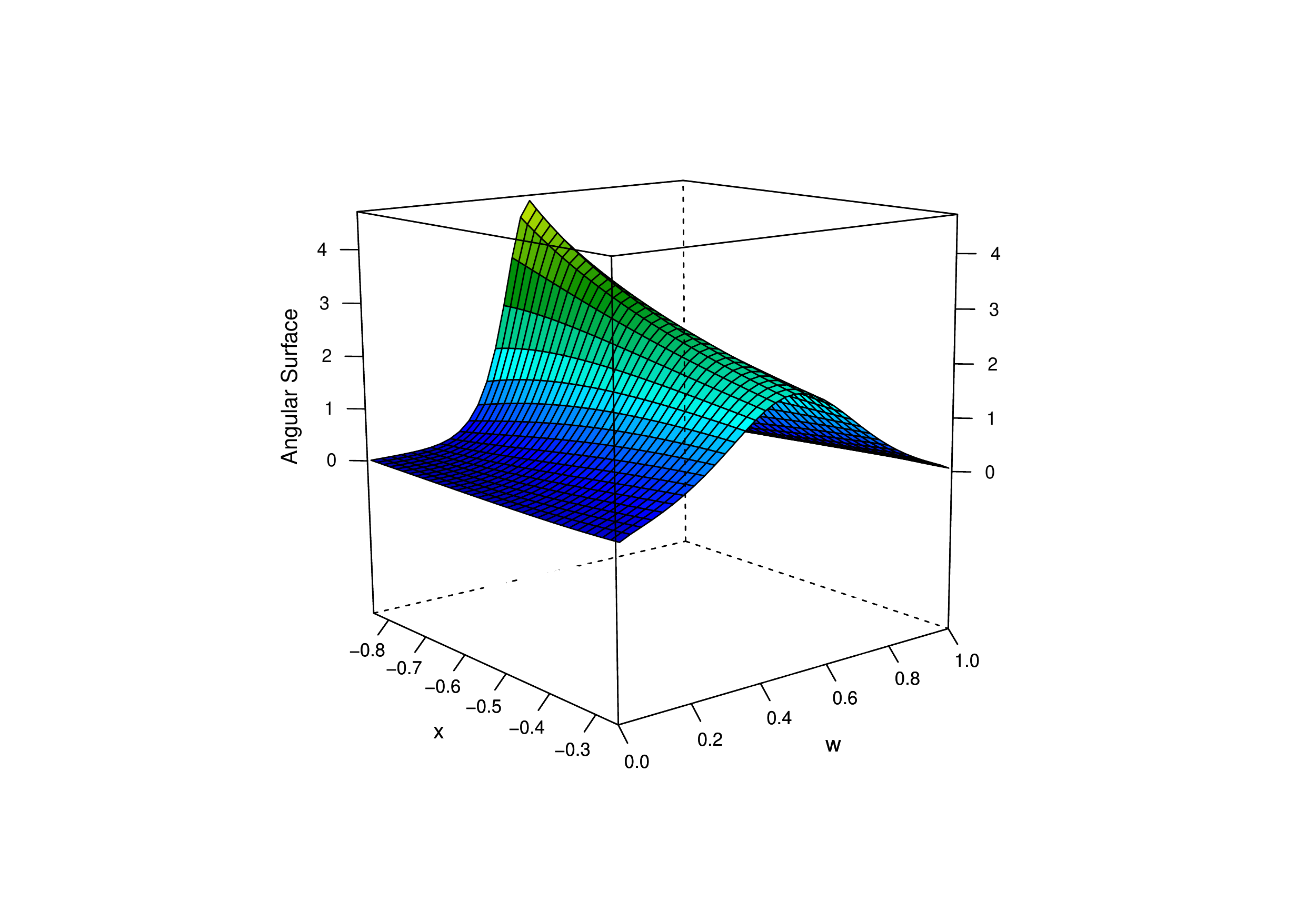}
  \end{minipage} &
  \begin{minipage}[c]{7cm}
    \hspace{-4.3cm}
    \includegraphics[width=1.3\textwidth]{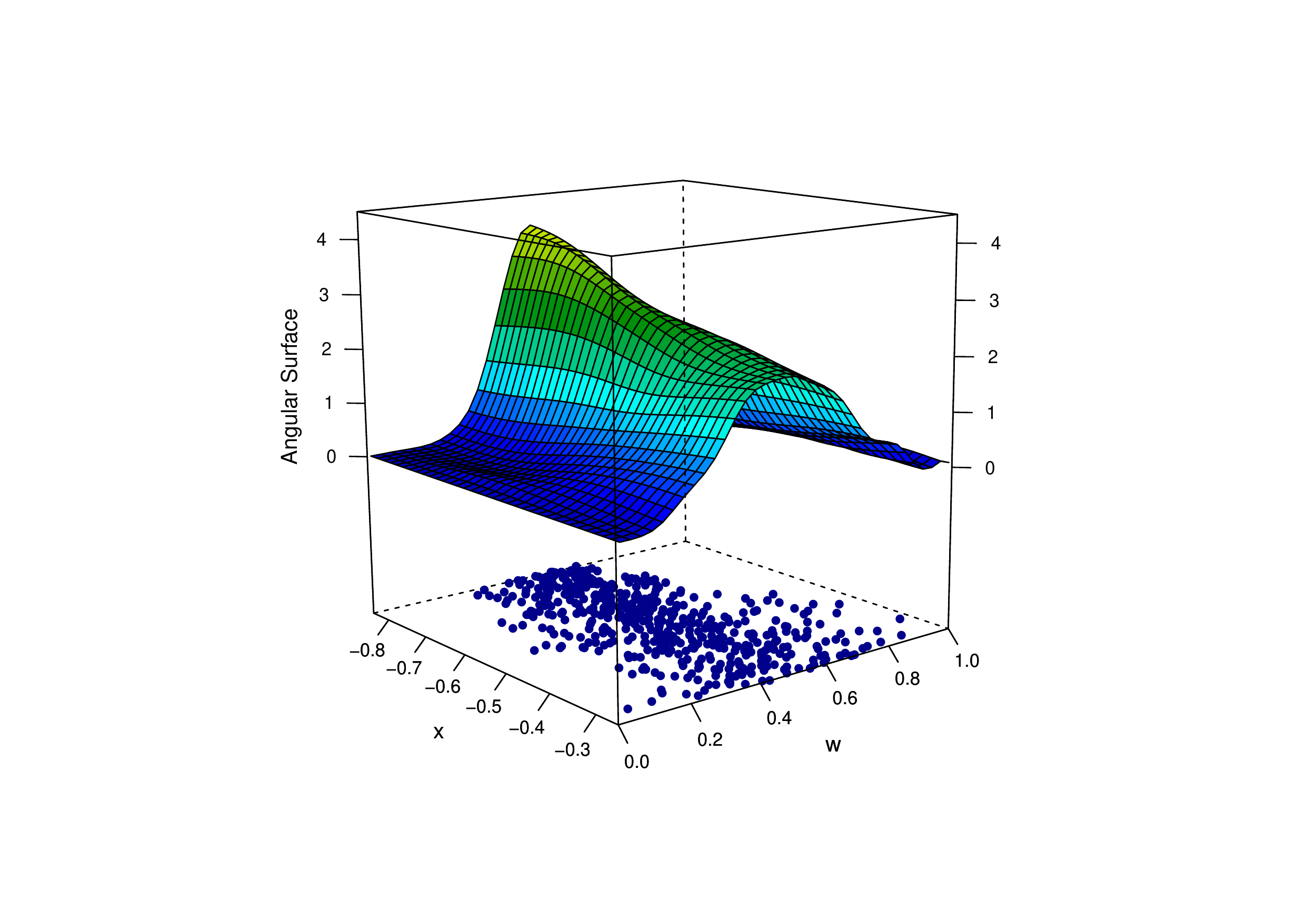}
  \end{minipage} &
  \begin{minipage}[c]{7cm}
  \hspace{-6cm}
    \includegraphics[width=1.3\textwidth]{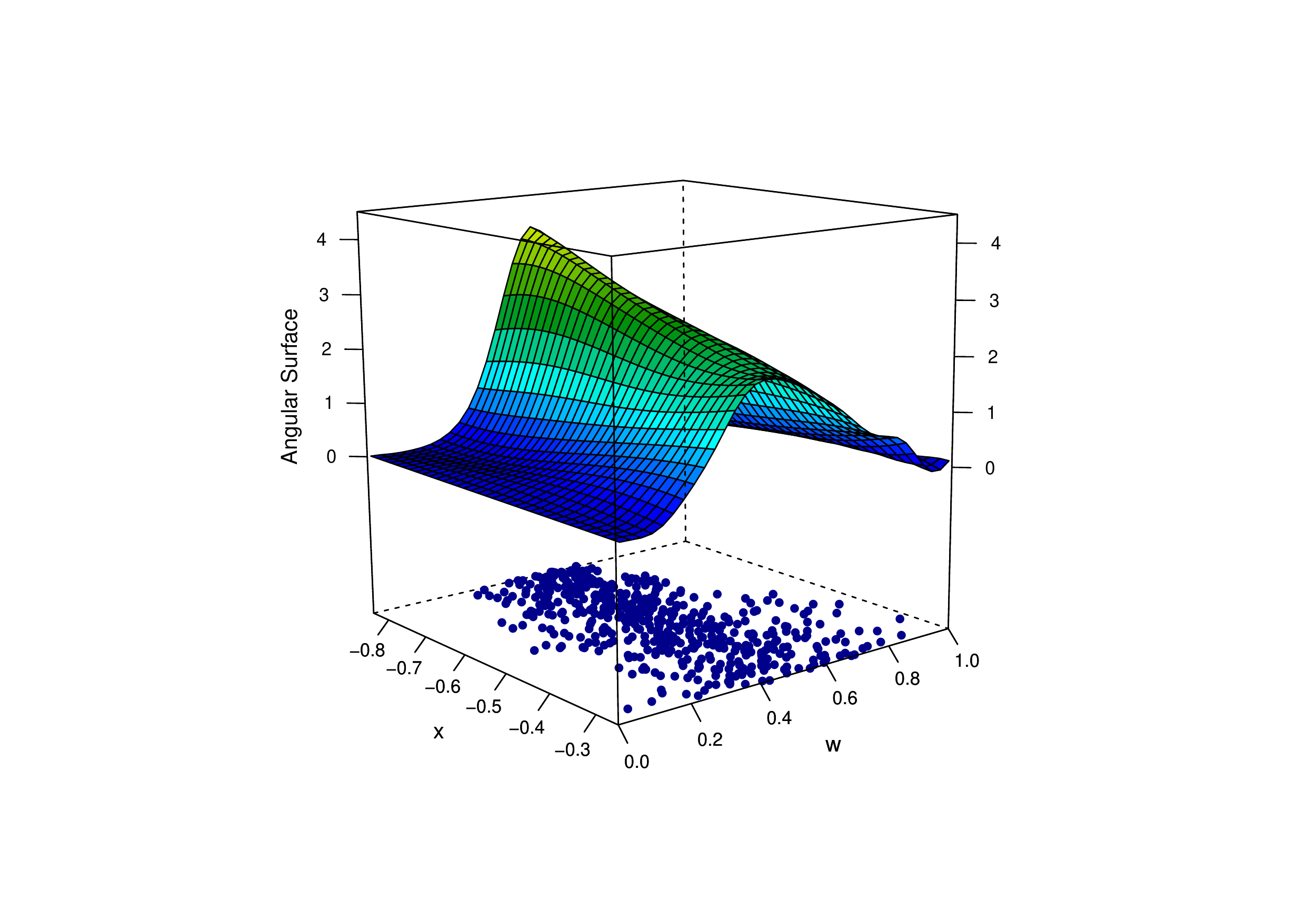}
  \end{minipage} \\
  \end{tabular}
    \vspace{-.2cm}
  \begin{center}
  \begin{footnotesize}
      {\textbf{Symmetric Dirichlet angular surface}}
  \end{footnotesize}
  \end{center}
    \vspace{-1cm}
\begin{tabular}{ccc}
    \begin{minipage}[c]{7cm}
      \hspace{-2.5cm}
    \includegraphics[width=1.3\textwidth]{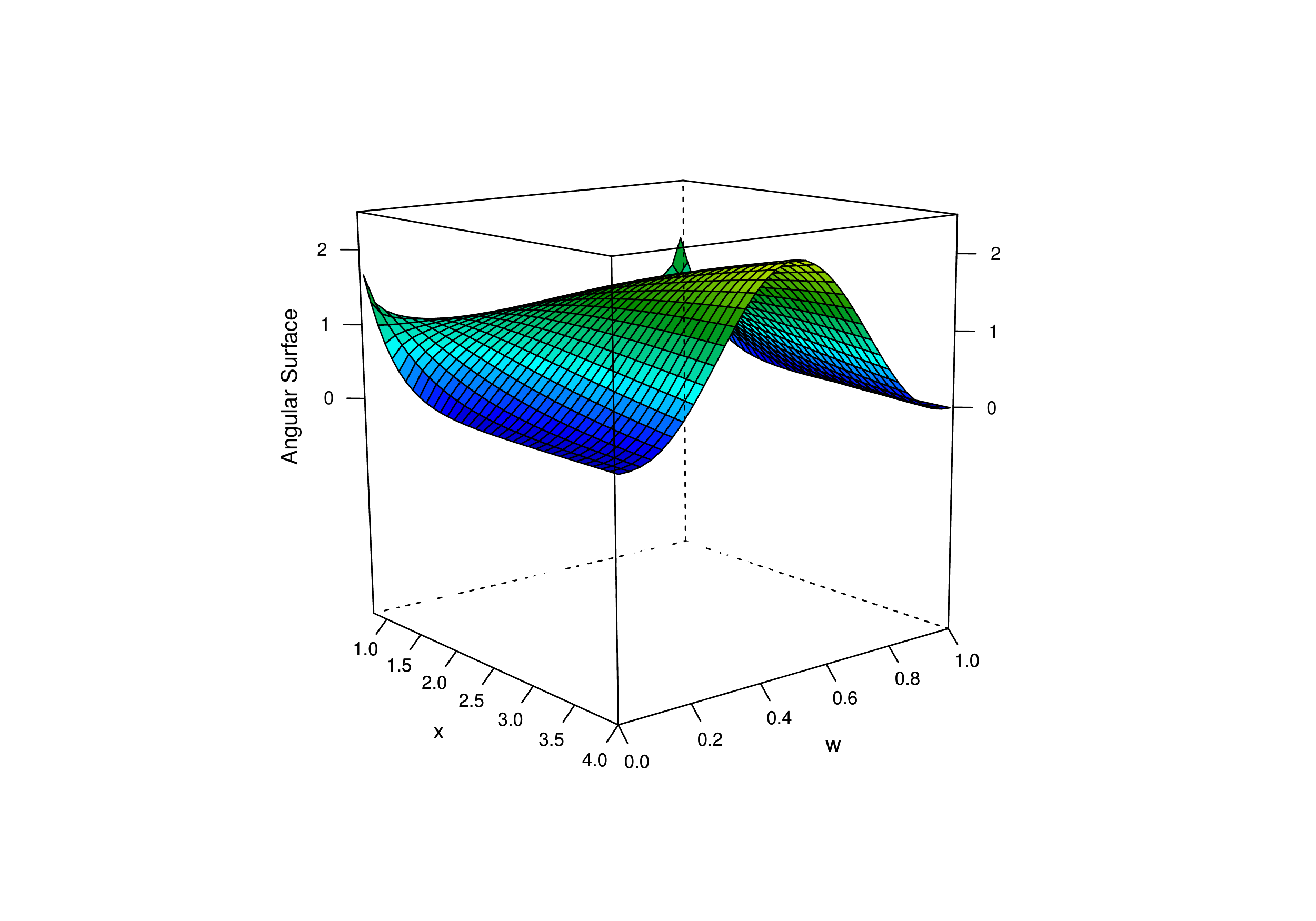}
  \end{minipage} &
  \begin{minipage}[c]{7cm}
    \hspace{-4.3cm}
    \includegraphics[width=1.3\textwidth]{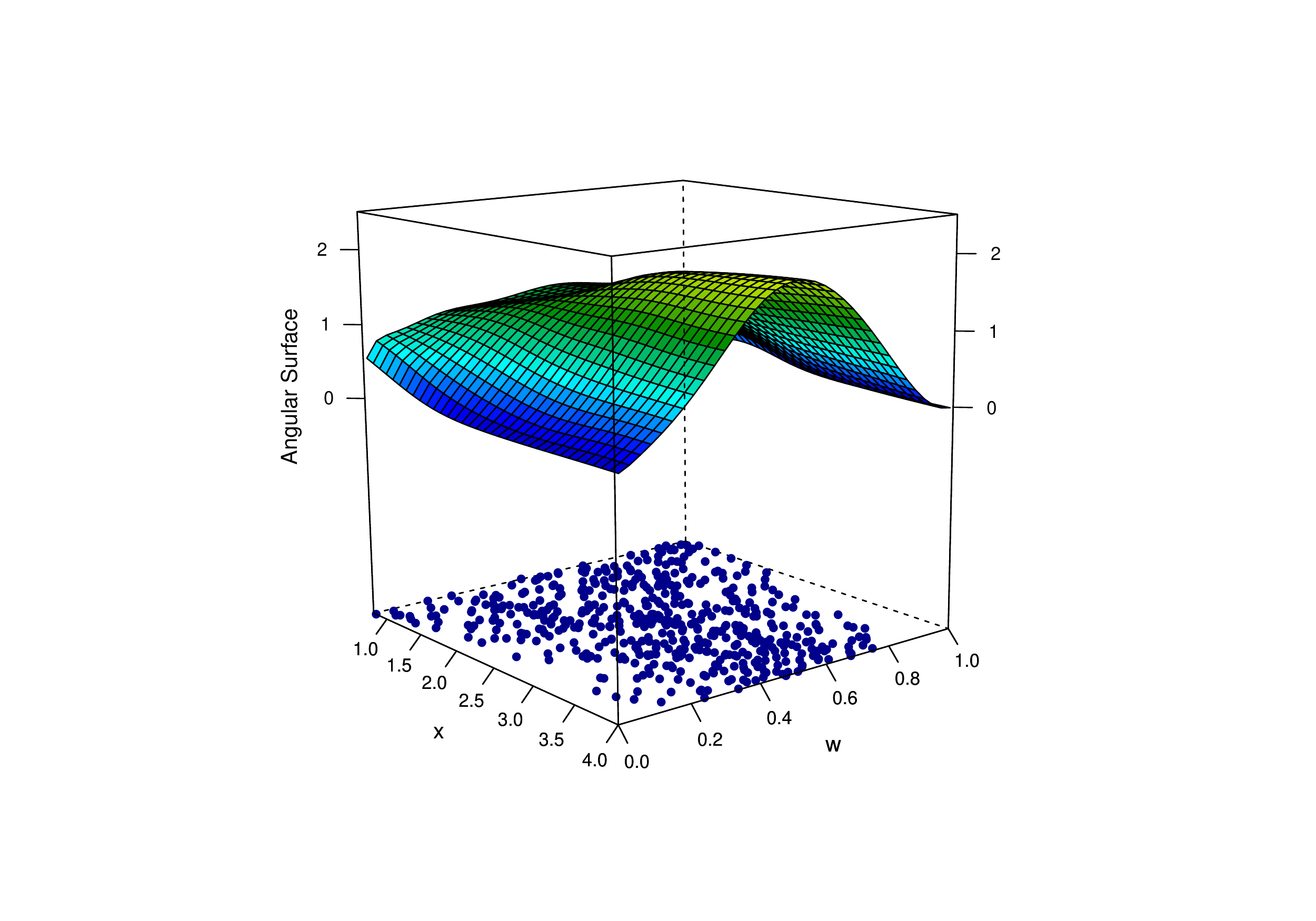}
  \end{minipage} &
  \begin{minipage}[c]{7cm}
      \hspace{-6cm}
    \includegraphics[width=1.3\textwidth]{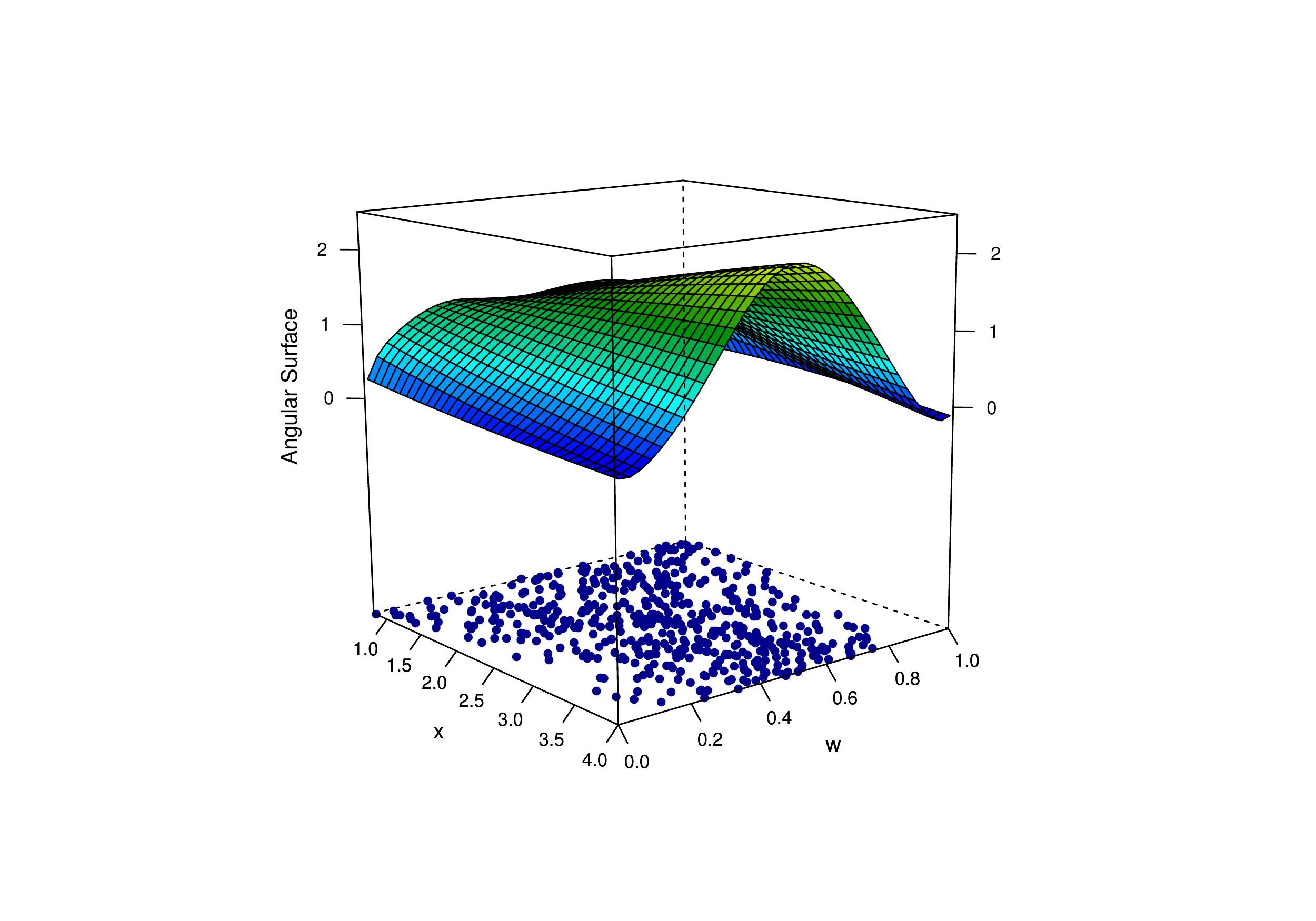}
  \end{minipage} \\
    \end{tabular}
    \vspace{-.2cm}
    \begin{center}
  \begin{footnotesize}
      {\textbf{Asymmetric Dirichlet angular surface}}
  \end{footnotesize}
  \end{center}
    \vspace{-1cm}
\begin{tabular}{ccc}
    \begin{minipage}[c]{7cm}
      \hspace{-2.5cm}
    \includegraphics[width=1.3\textwidth]{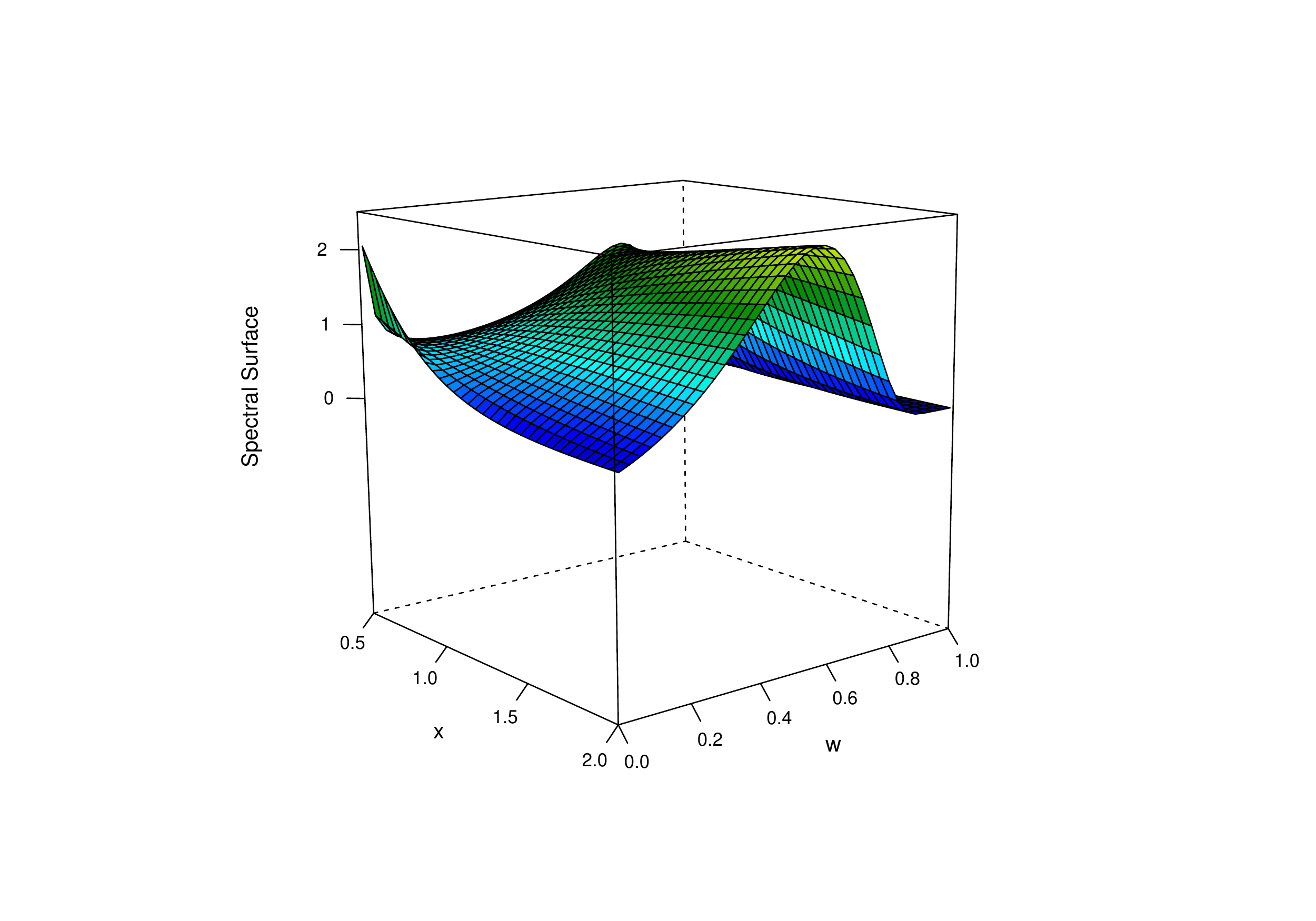}
  \end{minipage} &
  \begin{minipage}[c]{7cm}
    \hspace{-4.3cm}
    \includegraphics[width=1.3\textwidth]{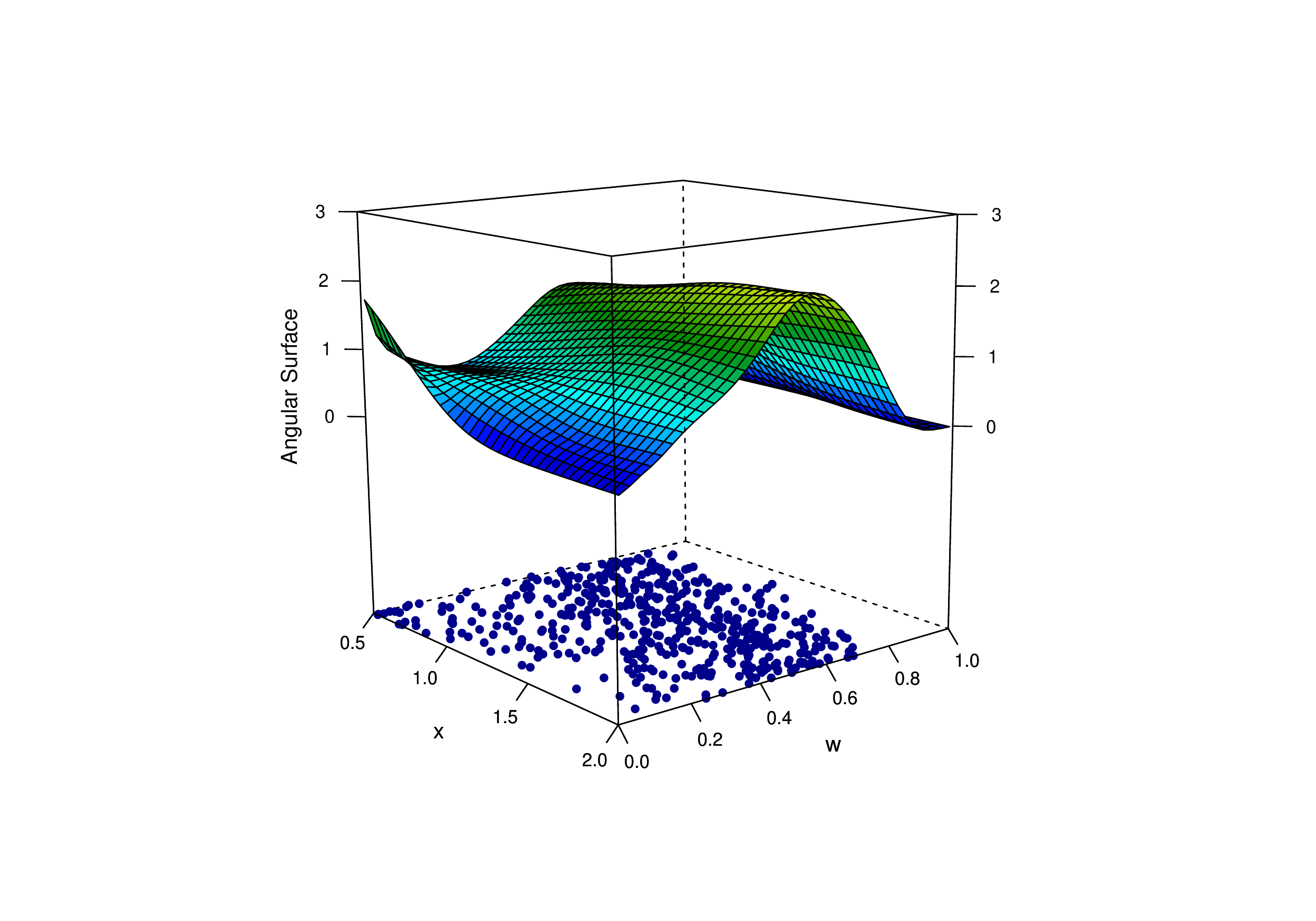}
  \end{minipage}&
  \begin{minipage}[c]{7cm}
      \hspace{-6cm}
    \includegraphics[width=1.3\textwidth]{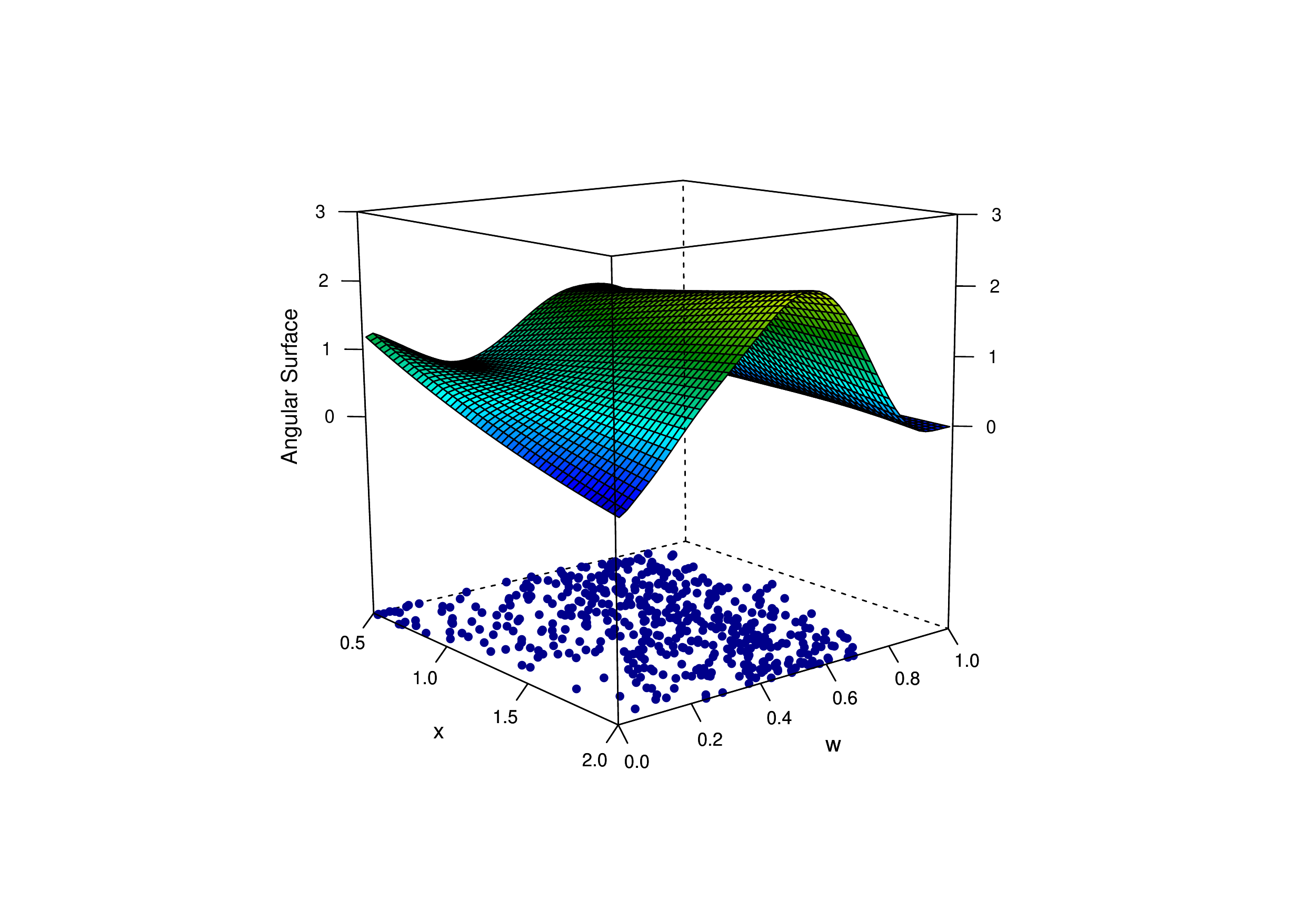}
  \end{minipage}\\
      \end{tabular}
      \vspace{-.5cm}
\caption{\footnotesize True angular surfaces (left) and corresponding estimates using Nadaraya--Watson weights (middle) and local linear weights (right). Top panel: conditional logistic model with $\alpha_x = \Phi(x)$, for $x \in \mathcal{X}_{\text{logistic}} = [\Phi^{-1}(0.2),\Phi^{-1}(0.4)]$. Center panel: conditional Symmetric Dirichlet model with $(a_{x},b_{x})=(x,x)$, for $x \in \mathcal{X}_{\text{sDir}} = [0.8,4]$. Bottom panel: conditional Asymmetric Dirichlet model with $(a_x,b_x)=(x,100)$, for $x \in \mathcal{X}_{\text{aDir}} \in [0.5,2]$. The simulated pseudo-angles based on which the estimates are produced are overlaid on the bottom of the boxes.}
\label{true_est.pdf}
\end{figure}

We study the performance of our methods under the logistic and Dirichlet conditional models introduced in Examples~\ref{Logistic} and \ref{Dirichlet}. Regarding the logistic conditional model, we take $\alpha_x = \Phi(x)$ and consider $x \in \mathcal{X}_{\text{logistic}} = [\Phi^{-1}(0.2),\Phi^{-1}(0.4)]$. For the Dirichlet conditional model we consider two scenarios: a symmetric Dirichlet angular surface with $(a_{x},b_{x})=(x,x)$, for $x \in \mathcal{X}_{\text{sDir}} = [0.8,4]$ and an asymmetric Dirichlet angular surface with $(a_x,b_x)=(x,100)$, for $x \in \mathcal{X}_{\text{aDir}} = [0.5,2]$. In Figure~\ref{true_est.pdf} we plot the true and estimated angular surfaces for the three cases described above on a single experiment with $n=500$. The top panel of Figure~\ref{true_est.pdf} corresponds to the logistic angular surface, where extremal dependence decreases as a function of the predictor. The center panel shows the symmetric Dirichlet angular surface, where we observe weaker dependence for lower values of the covariate, whereas stronger dependence prevails for higher values. Finally, an increasing asymmetric dependence dynamic is displayed in the bottom panel, where we have plotted the asymmetric Dirichlet angular surface. 

The single run experiment in Figure~\ref{true_est.pdf} allows us to illustrate strengths and limitations with the methods. Even though there is a good fit---which is discussed in further detail in Section~\ref{Finite sample performance}---we can anticipate from this figure that our estimator suffers from limitations inherent to kernel-based estimators. For example, pointwise estimation using the Nadaraya--Watson weights (middle column of Figure~\ref{true_est.pdf}) underperforms when the angular surface peaks, but this is mostly due to the boundary bias of $K_b$ which is a drawback of kernel-based estimators on bounded domains; see \citet[][Section 4.4]{H90} and references therein. To mitigate this issue we also compute our estimator using local linear weights, as described in Section~\ref{loclinear} (right column of Figure~\ref{true_est.pdf}). We see that the performance in the upper boundaries of the covariate space is slightly improved for both Dirichlet angular surfaces, but it remains almost the same for the logistic angular surface.
The estimator using local linear weights seems to produce smoother estimates for the asymmetric Dirichlet model. This relative improvement is corroborated in Table~\ref{tableMIAE}, where we assess the mean performance of both estimators. Estimates for the other two models tend to be better (in terms of mean performance) using the Nadaraya--Watson weights. In terms of computations, the runtime of the estimator using the Nadaraya--Watson weights outperforms its local linear counterpart by at least a factor of 10. In spite of these limitations, both estimators successfully recover the shape of the true angular surface, and thus are able to reproduce accurately the evolution of extremal dependence over the covariate.

\begin{figure}
  \begin{center}
    \begin{footnotesize}
      {\textbf{Logistic angular surface}}
    \end{footnotesize}
  \end{center}
  \vspace{-0.2 cm}
\footnotesize \rotatebox{90}{\textbf{\hspace{-1cm}N--W weights}}
  \begin{minipage}[c]{0.4\linewidth}
    \hspace{-.1cm}
    \includegraphics[scale = 0.2]{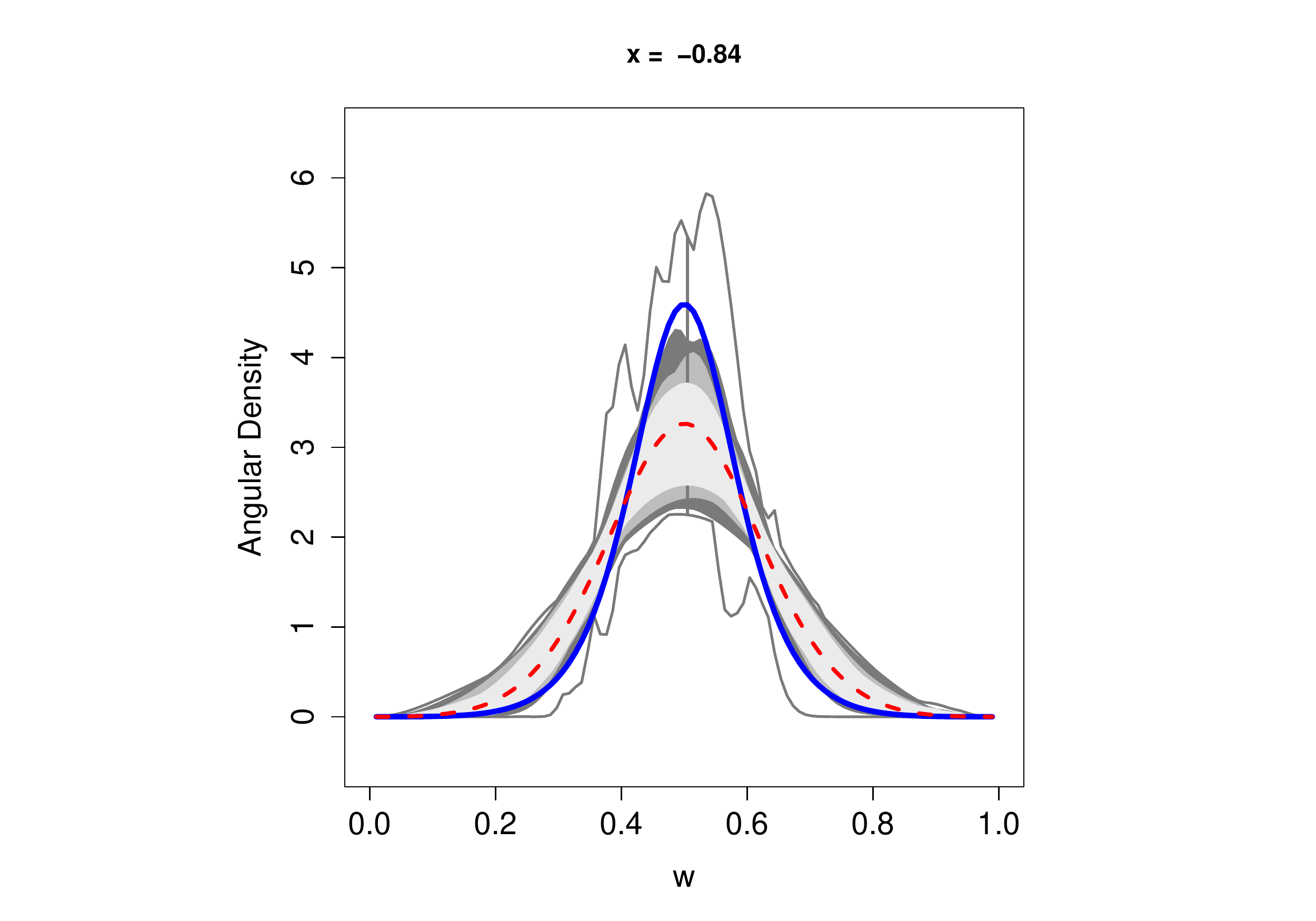}
  \end{minipage}%
  \hspace{-2cm}
  \begin{minipage}[c]{0.4\linewidth}
   \hspace{.2cm}
    \includegraphics[scale = 0.2]{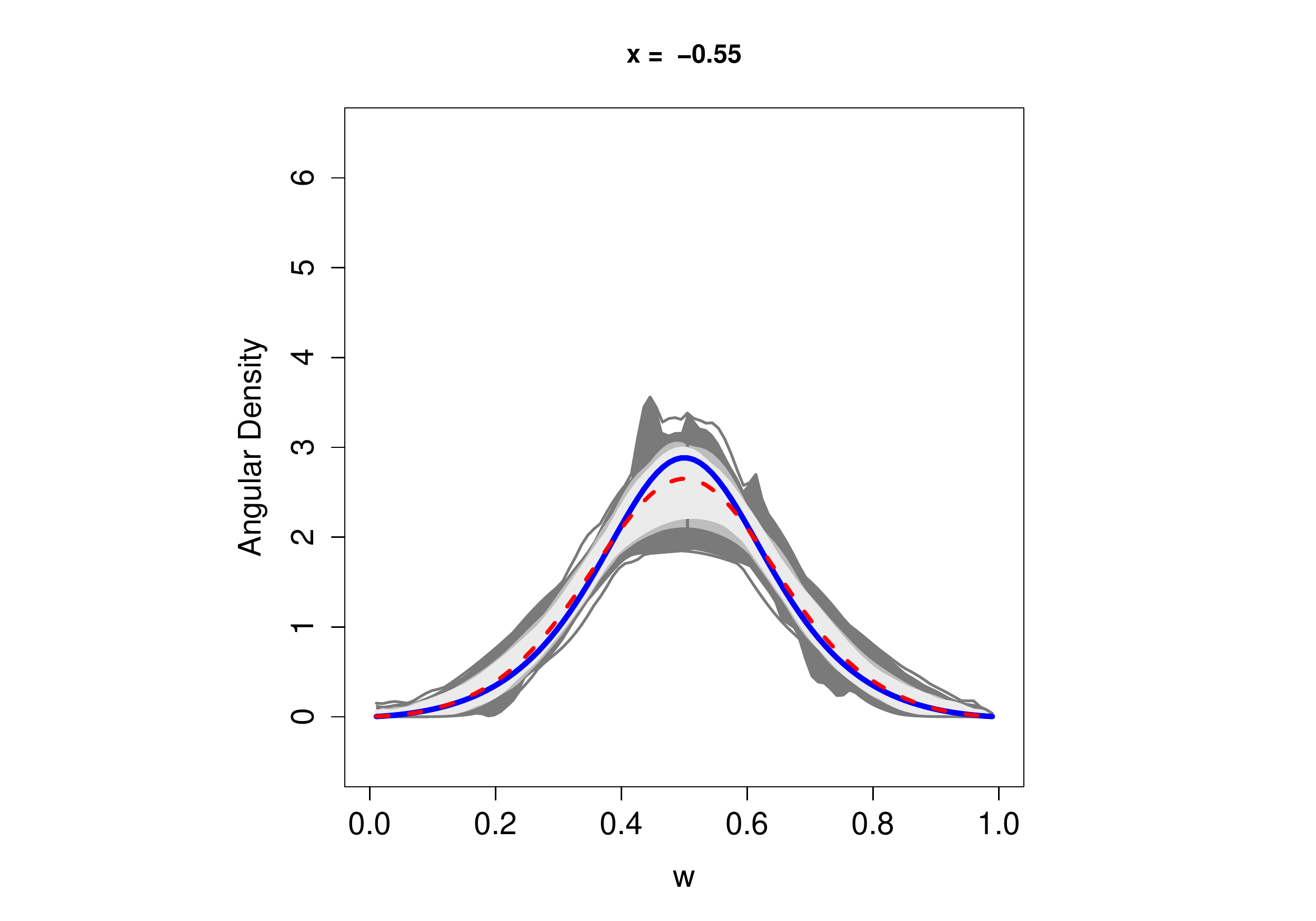}
  \end{minipage}%
  \hspace{-1.8cm}
  \begin{minipage}[c]{0.4\linewidth}
 \hspace{.2cm} 
    \includegraphics[scale = 0.2]{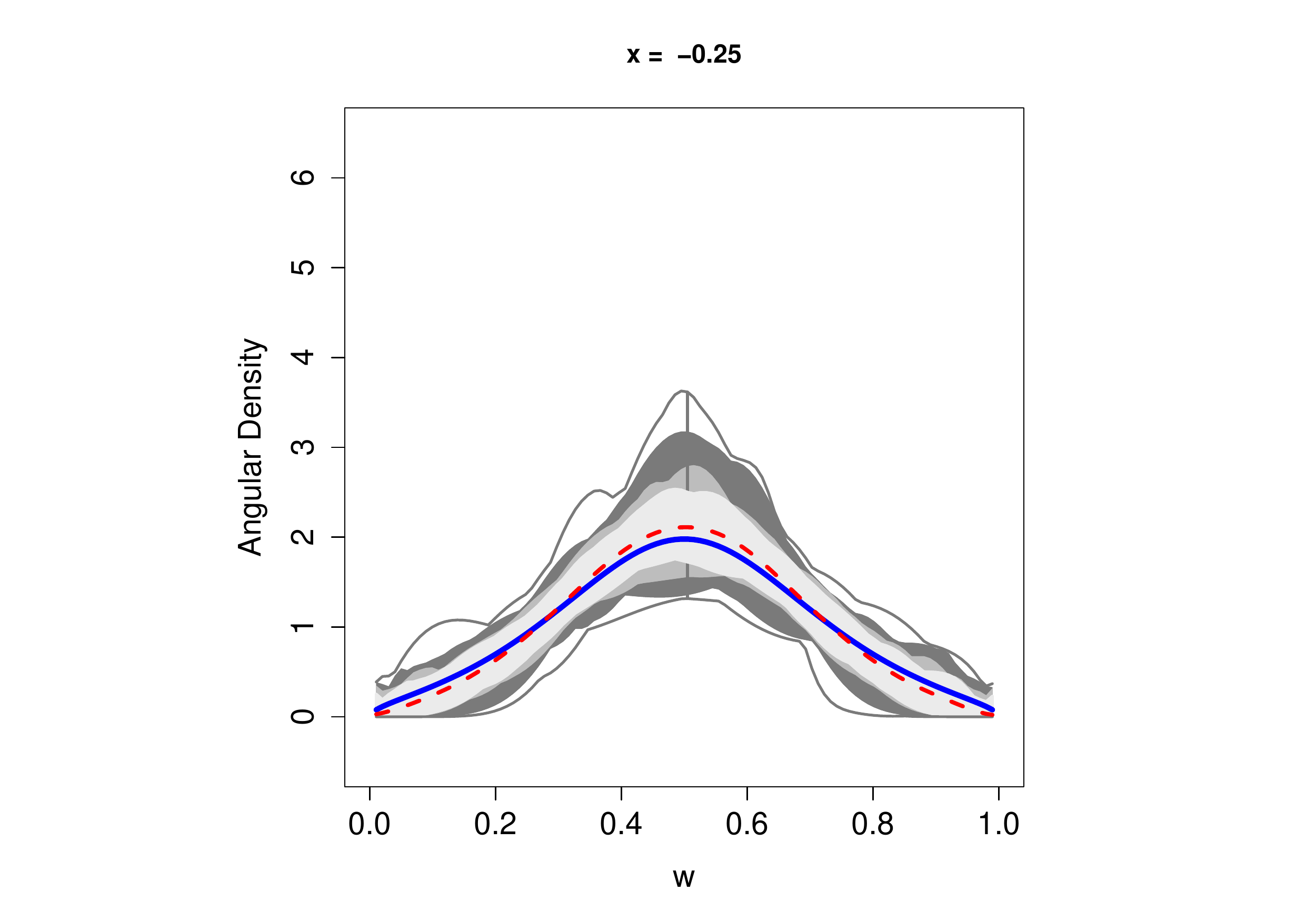}
  \end{minipage}%
  \\
    \vspace{-0.5 cm}
    \footnotesize \rotatebox{90}{\textbf{\hspace{-1cm}L--L weights}}
    \begin{minipage}[c]{0.4\linewidth}
    \hspace{-.1cm}
    \includegraphics[scale = 0.2]{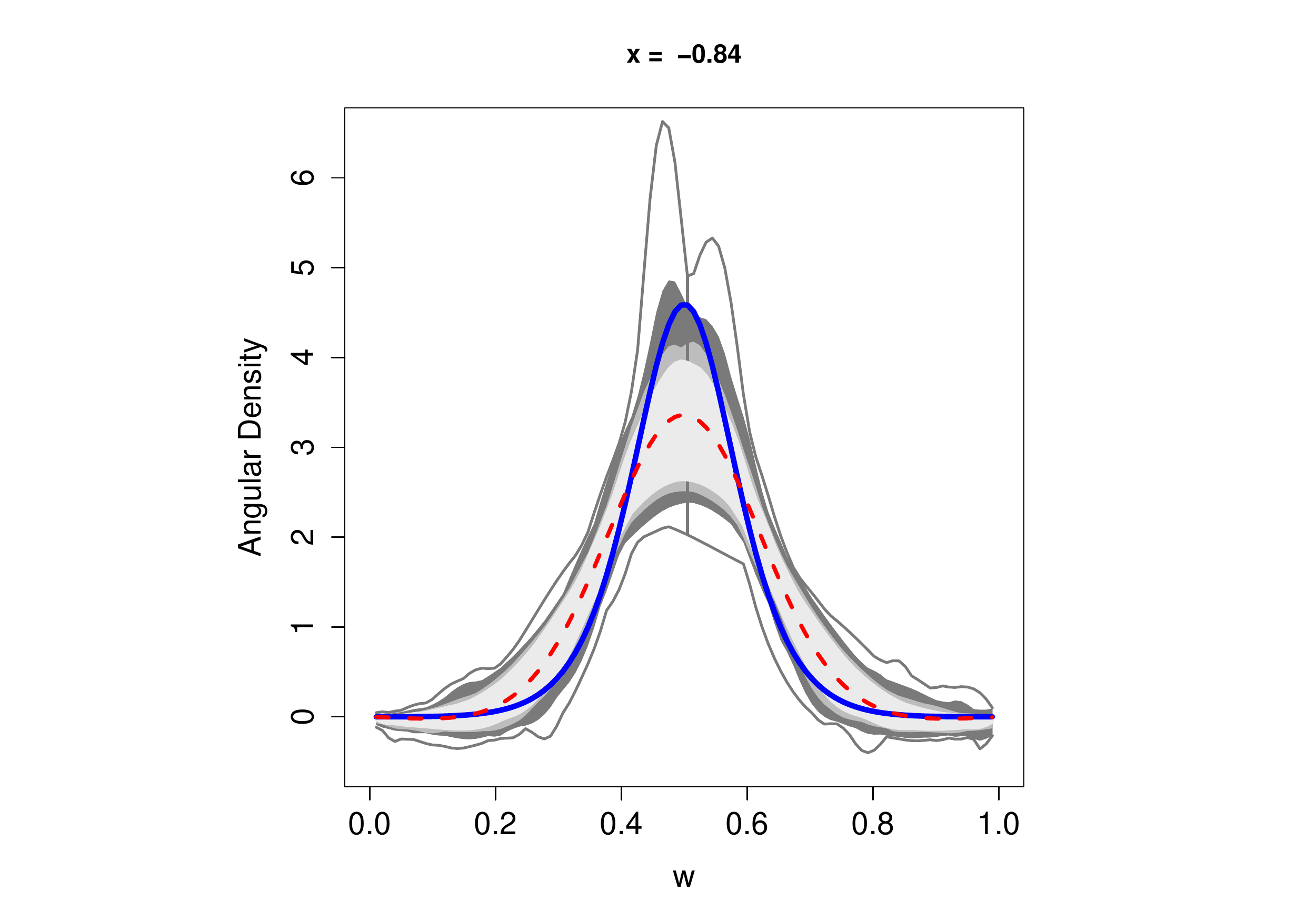}
  \end{minipage}%
  \hspace{-2cm}
  \begin{minipage}[c]{0.4\linewidth}
   \hspace{.2cm}
    \includegraphics[scale = 0.2]{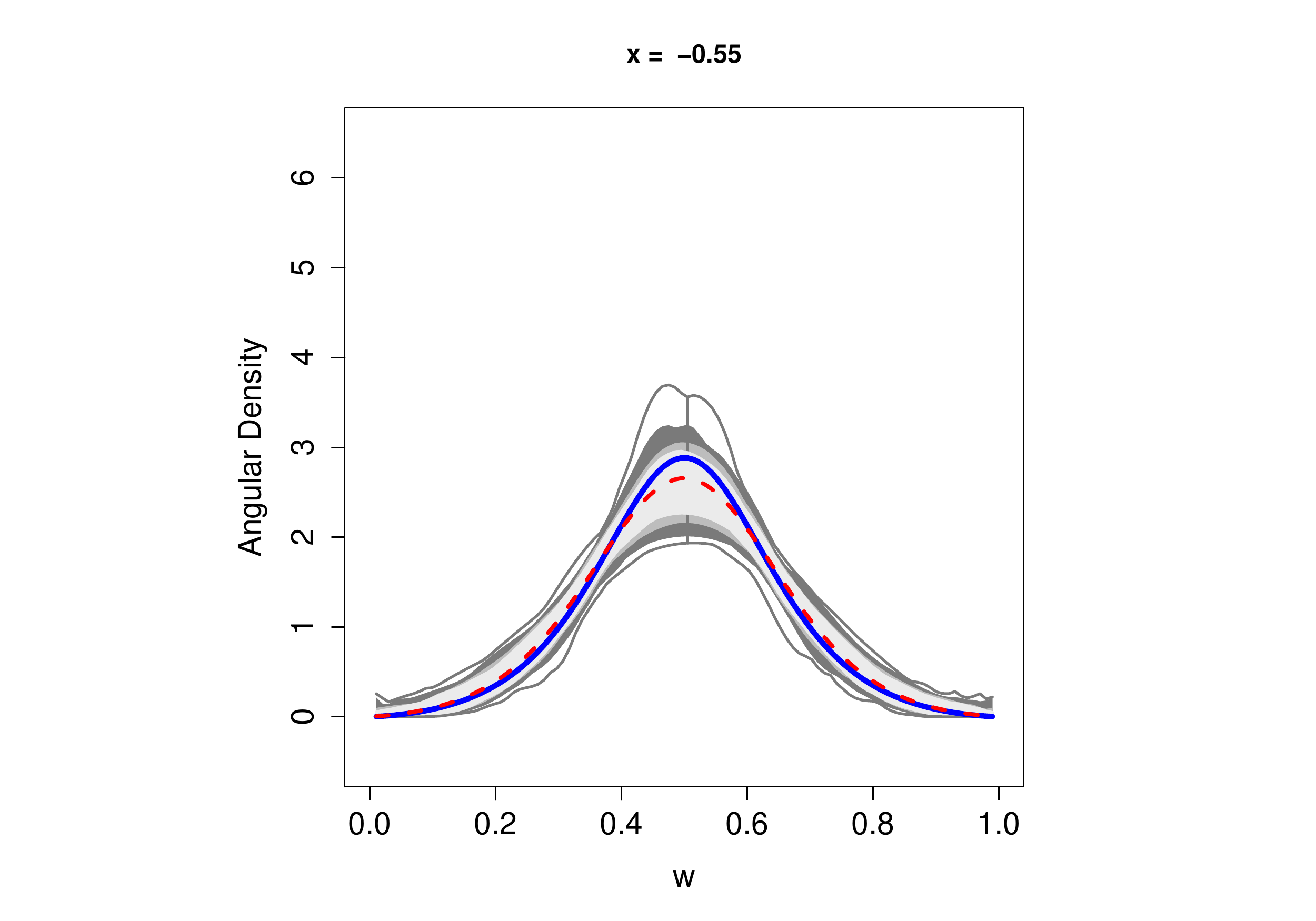}
  \end{minipage}%
  \hspace{-1.8cm}
  \begin{minipage}[c]{0.4\linewidth}
 \hspace{.2cm} 
    \includegraphics[scale = 0.2]{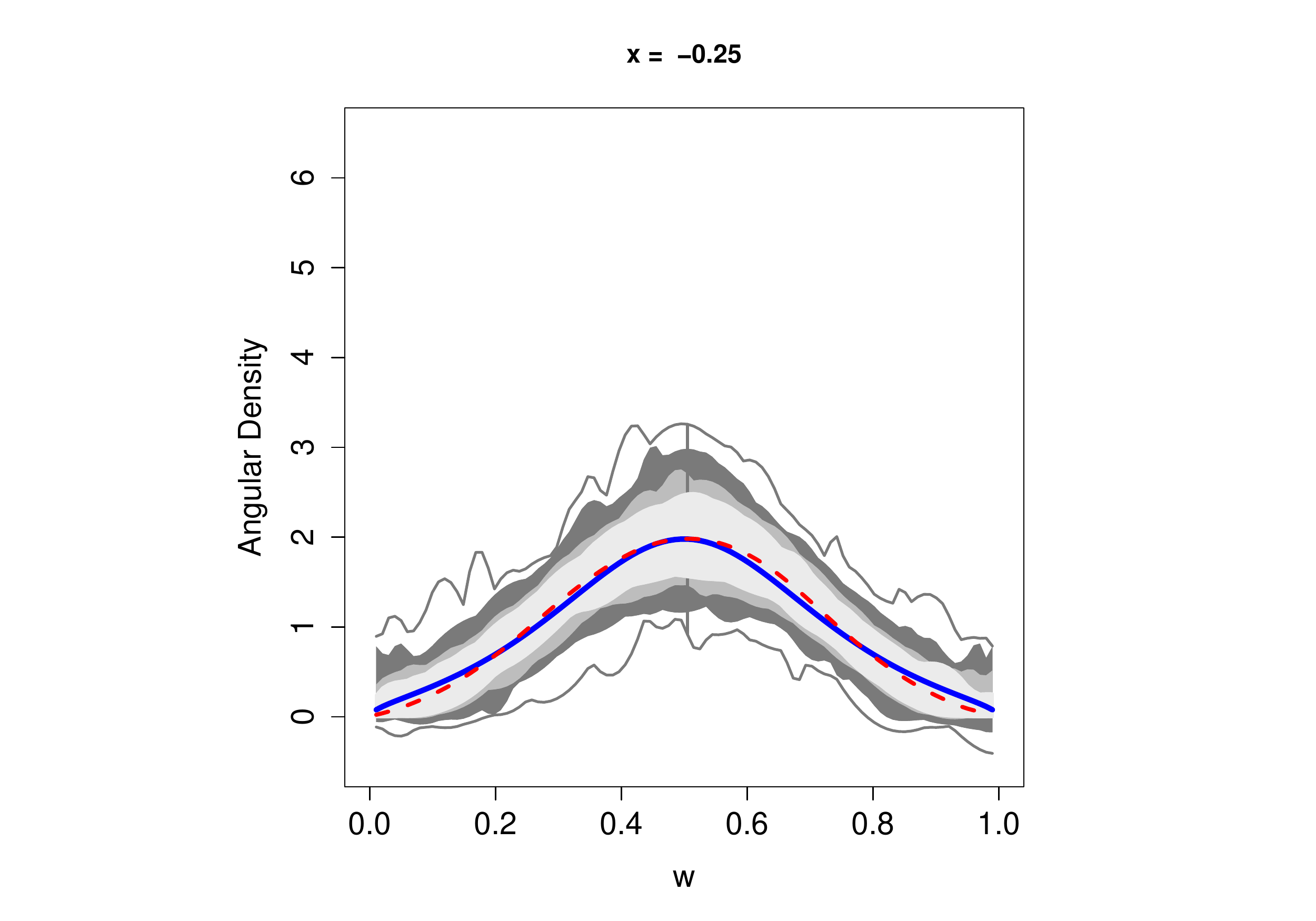}
  \end{minipage}%
  \\
  \begin{center}
    \begin{footnotesize}
      {\textbf{Symmetric Dirichlet angular surface}}
    \end{footnotesize}
  \end{center}
  \vspace{-0.2 cm}
  \footnotesize \rotatebox{90}{\textbf{\hspace{-1cm}N--W weights}}
  \begin{minipage}[c]{0.4\linewidth}
    \hspace{-.1cm}
    \includegraphics[scale = 0.2]{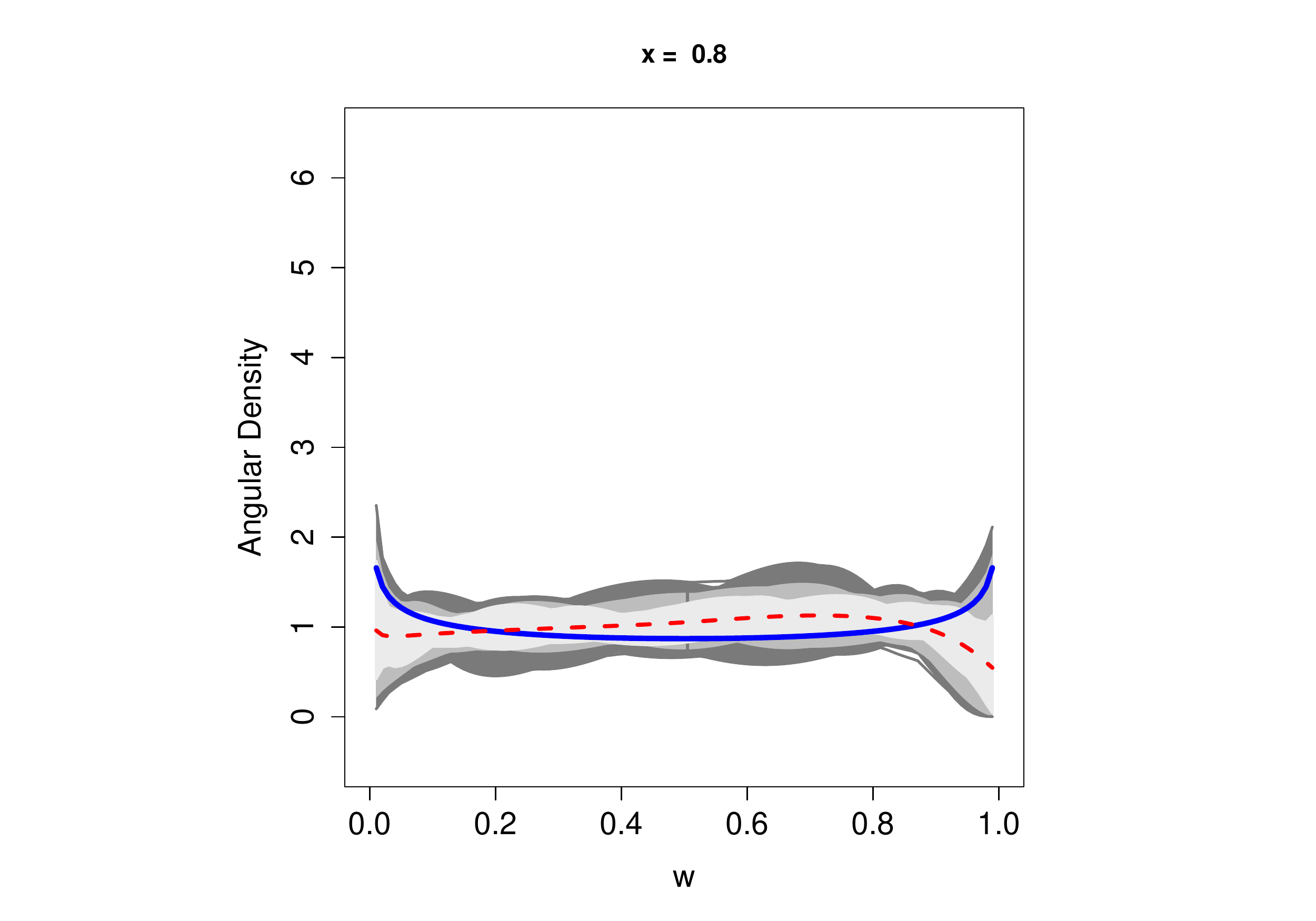}
  \end{minipage}%
  \hspace{-2cm}
  \begin{minipage}[c]{0.4\linewidth}
    \hspace{.2cm}
    \includegraphics[scale = 0.2]{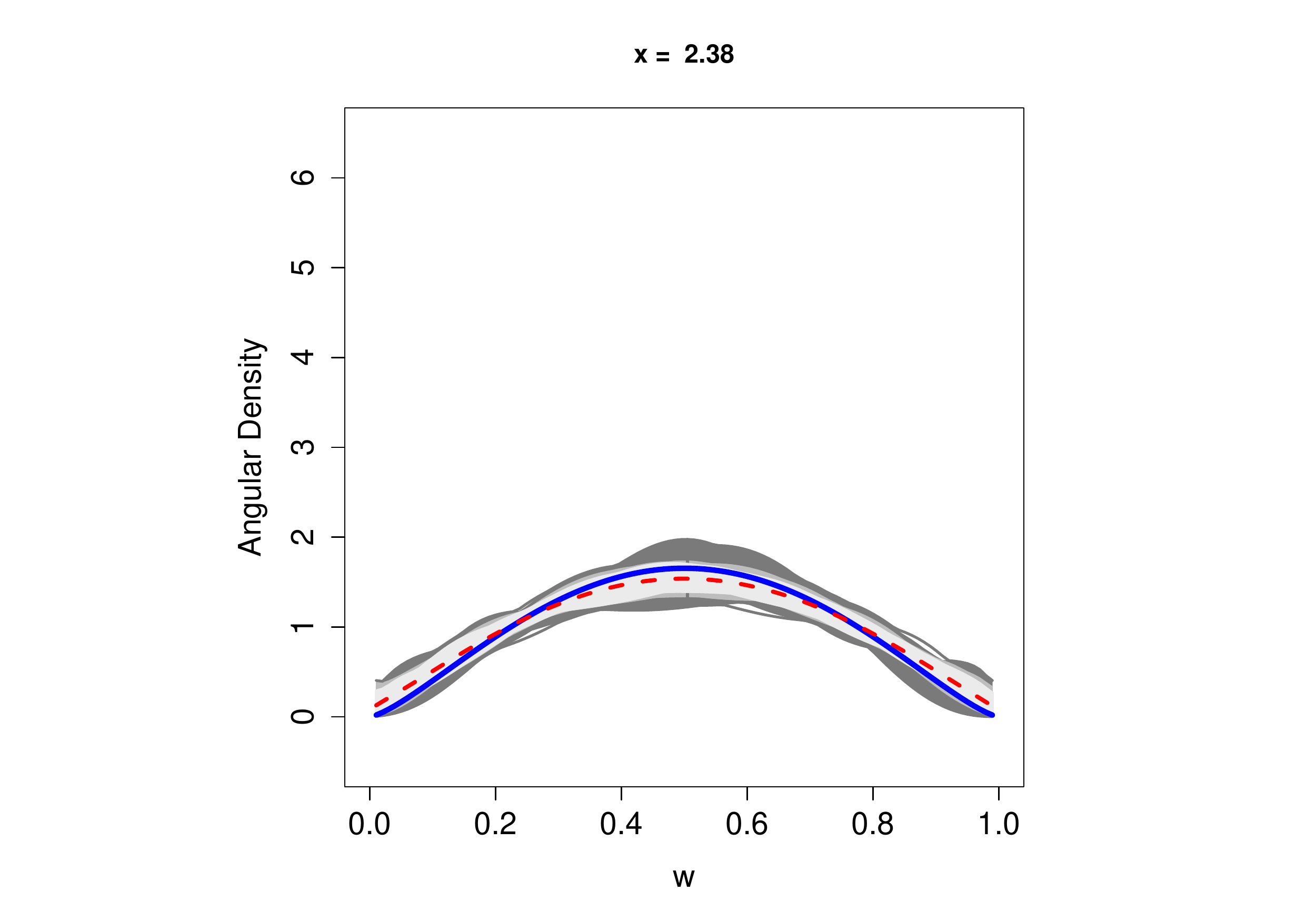}
  \end{minipage}%
  \hspace{-1.8cm}
  \begin{minipage}[c]{0.4\linewidth}
    \hspace{.2cm}
    \includegraphics[scale = 0.2]{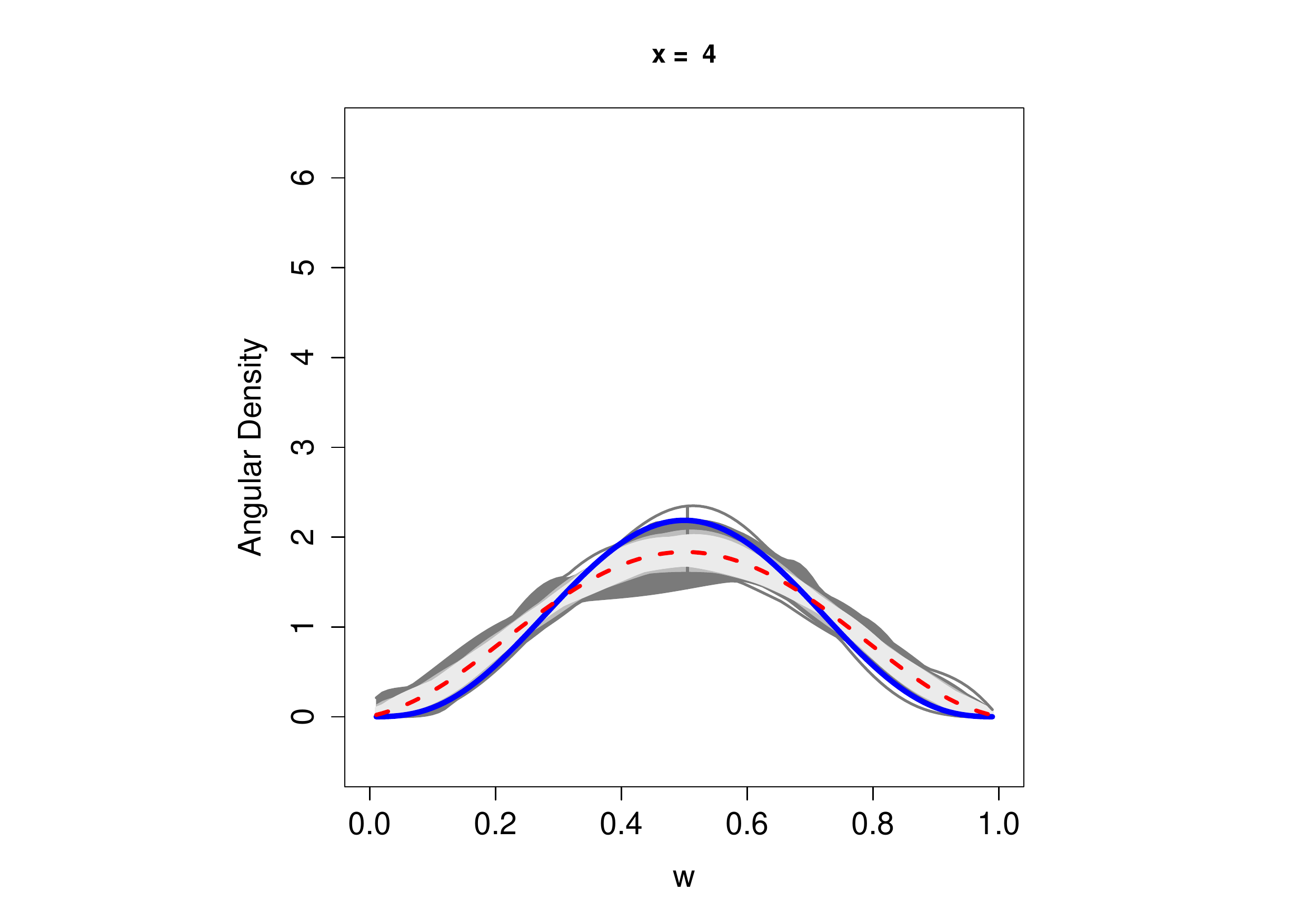}
  \end{minipage}%
  \\
  \vspace{-0.5 cm}
      \footnotesize \rotatebox{90}{\textbf{\hspace{-1cm}L--L weights}}
  \begin{minipage}[c]{0.4\linewidth}
    \hspace{-.1cm}
    \includegraphics[scale = 0.2]{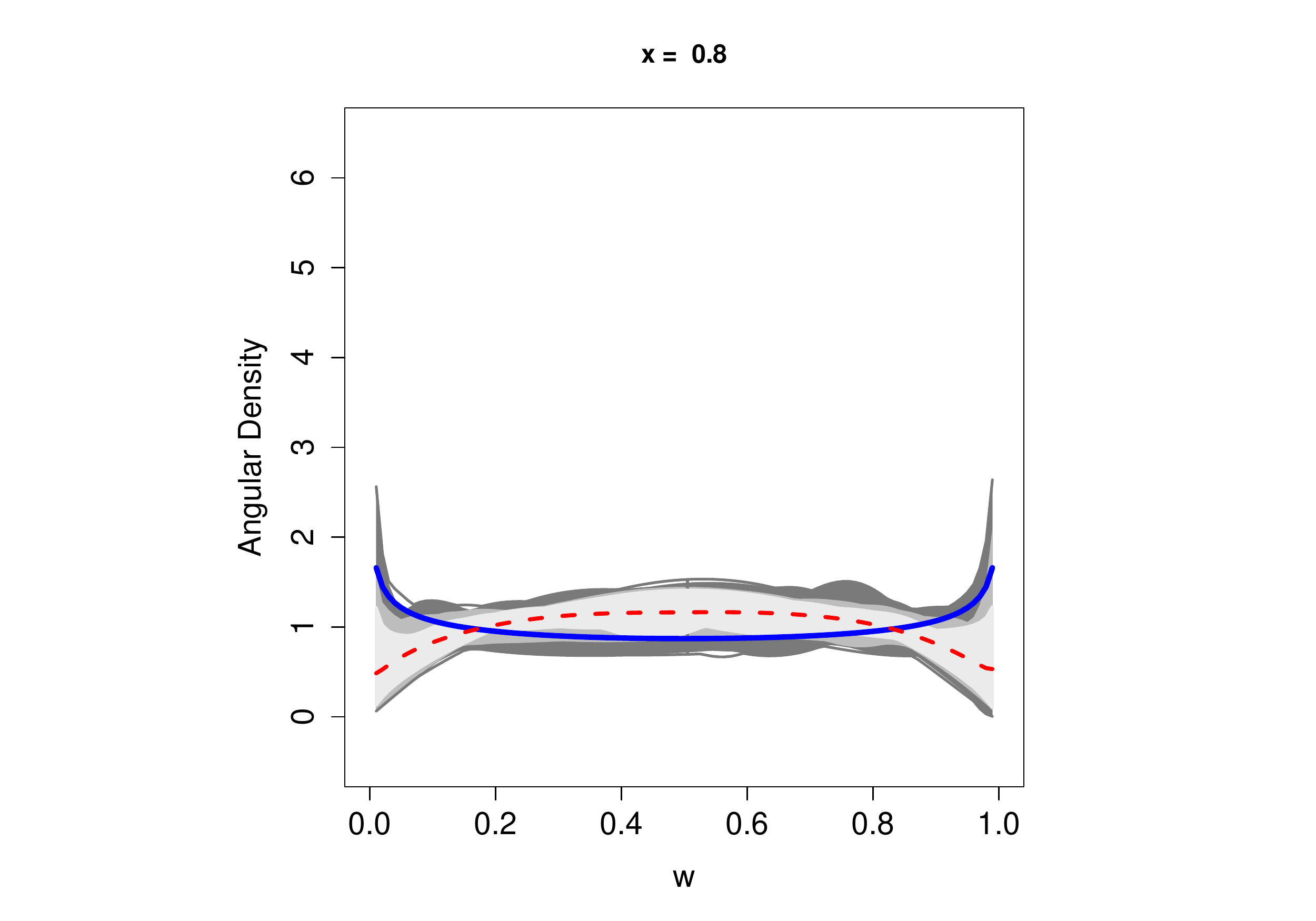}
  \end{minipage}%
  \hspace{-2cm}
  \begin{minipage}[c]{0.4\linewidth}
    \hspace{.2cm}
    \includegraphics[scale = 0.2]{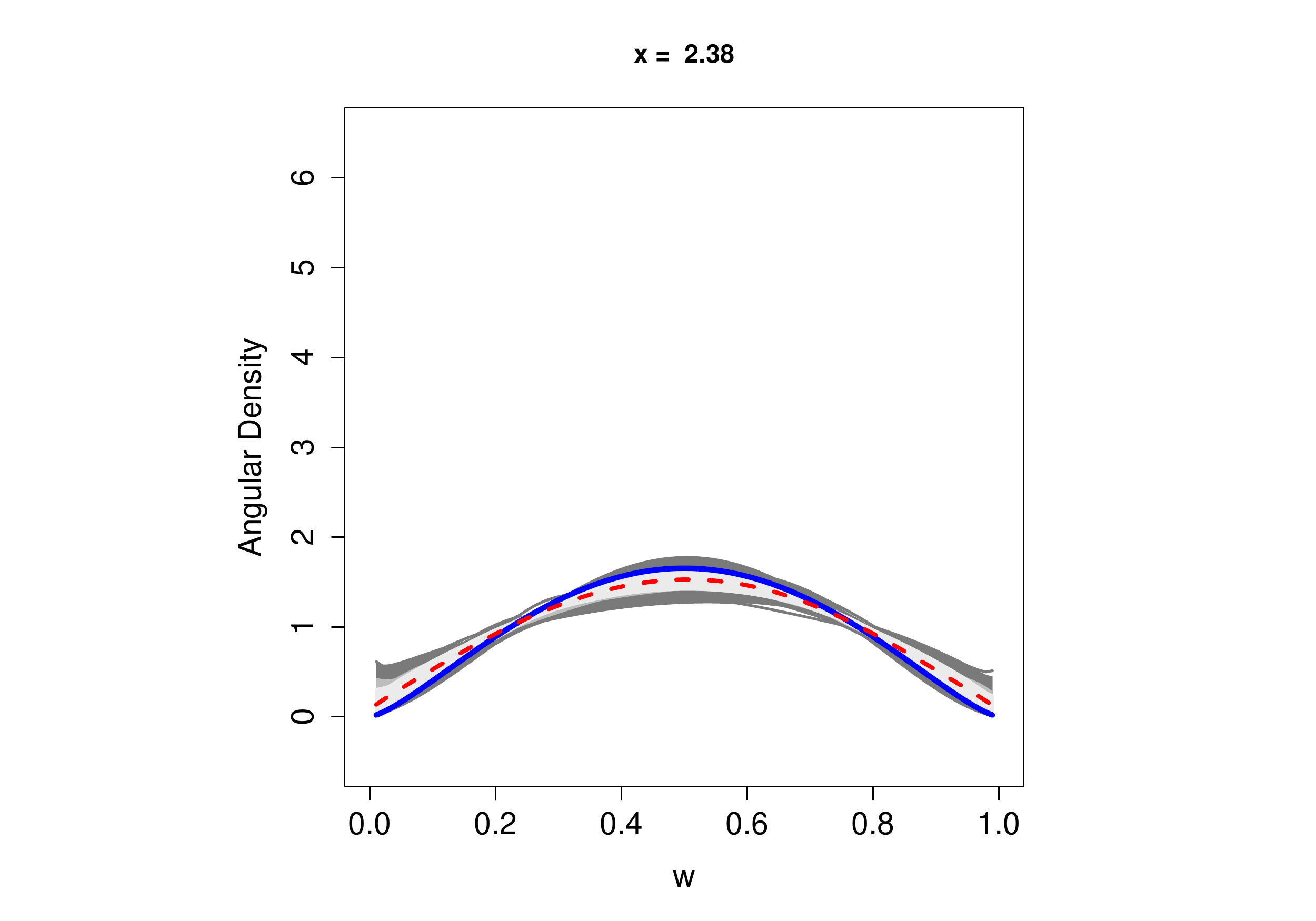}
  \end{minipage}%
  \hspace{-1.8cm}
  \begin{minipage}[c]{0.4\linewidth}
    \hspace{.2cm}
    \includegraphics[scale = 0.2]{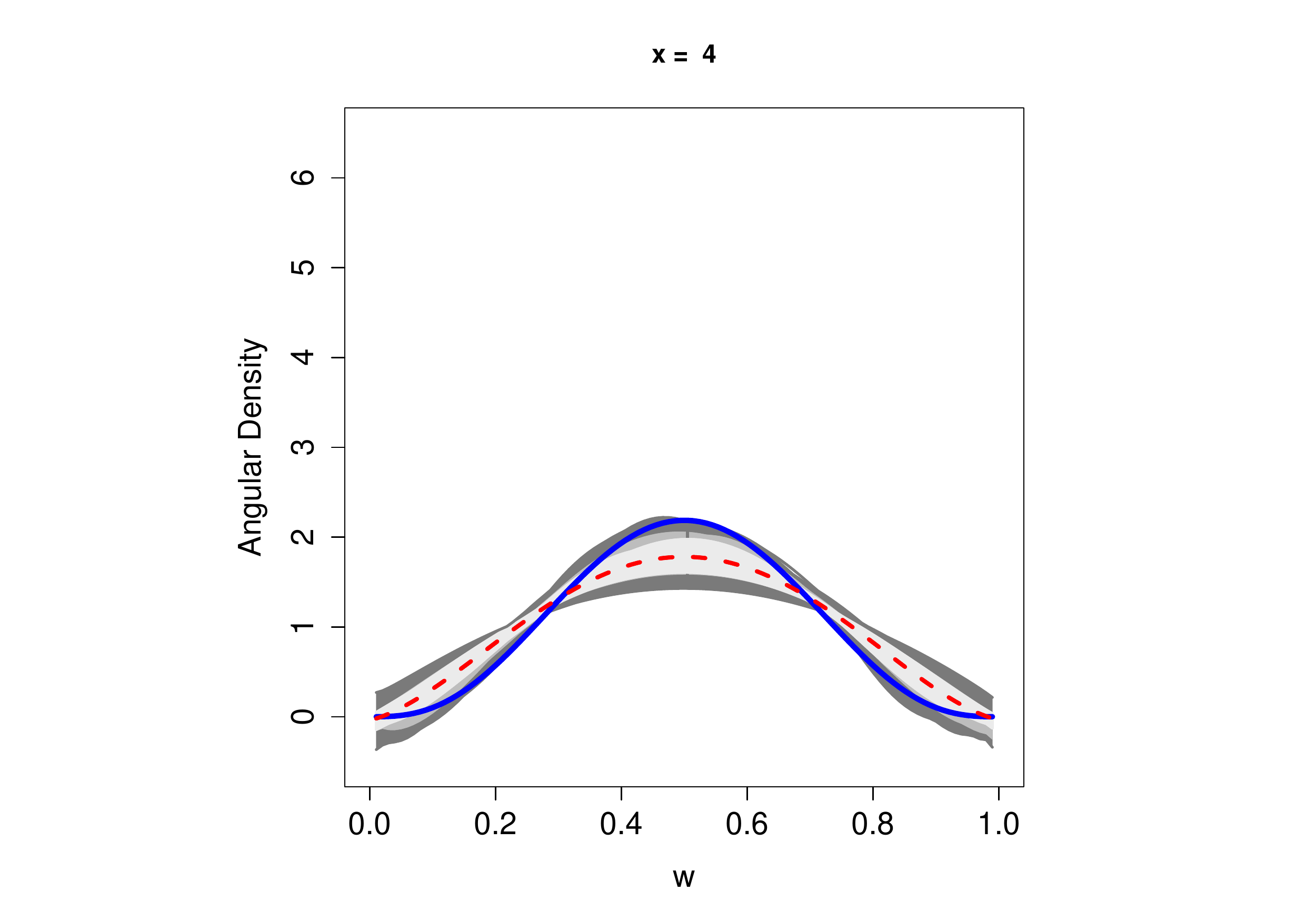}
  \end{minipage}%
  \\
  \begin{center}
    \begin{footnotesize}
      {\textbf{Asymmetric Dirichlet angular surface}}
    \end{footnotesize}
  \end{center}
  \vspace{-0.2cm}
    \footnotesize \rotatebox{90}{\textbf{\hspace{-1cm}N--W weights}}
  \begin{minipage}[c]{0.4\linewidth}
    \hspace{-.1cm}
    \includegraphics[scale = 0.2]{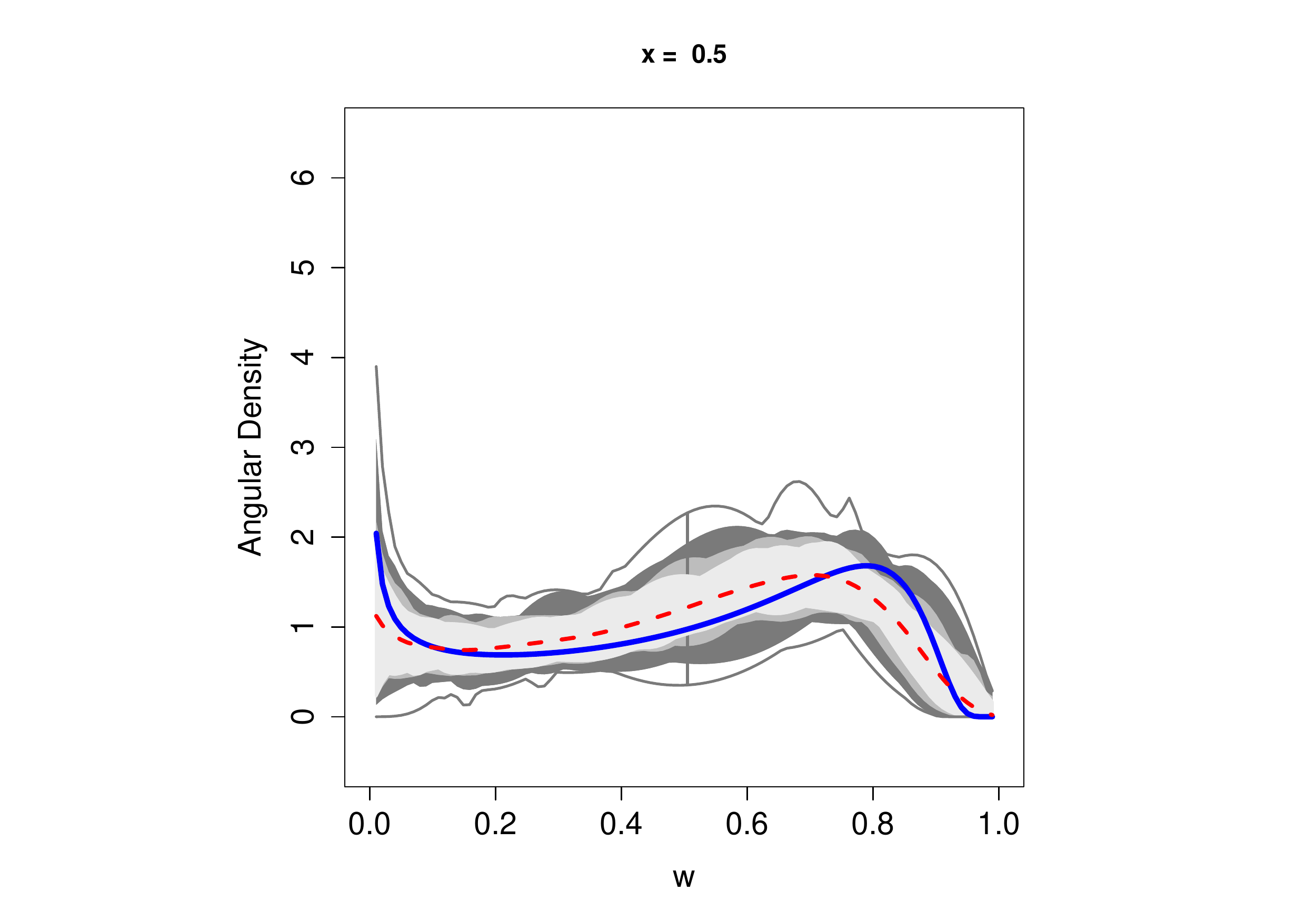}
  \end{minipage}%
  \hspace{-2cm}
  \begin{minipage}[c]{0.4\linewidth}
    \hspace{.2cm}
    \includegraphics[scale = 0.2]{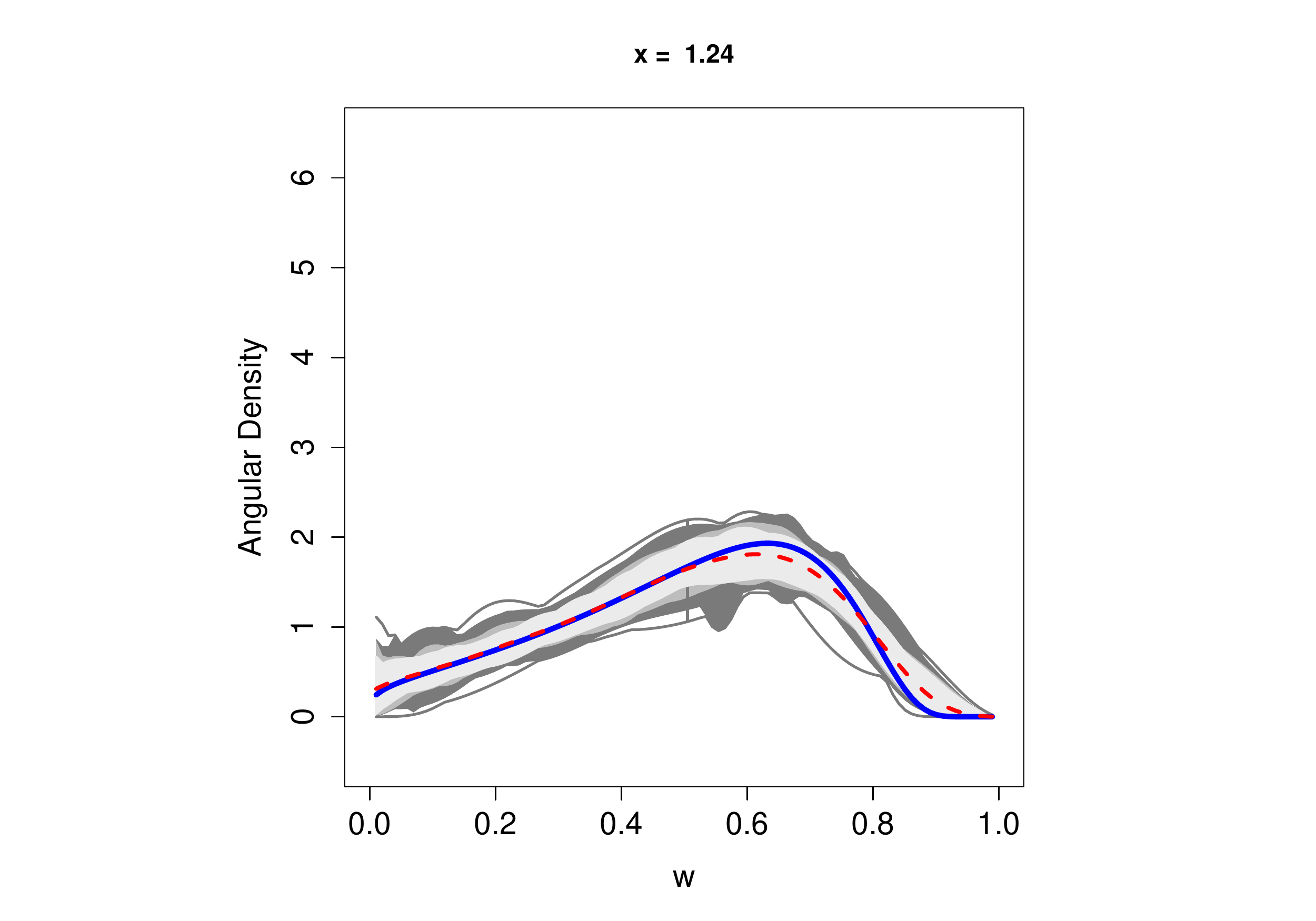}
  \end{minipage}%
  \hspace{-1.8cm}
  \begin{minipage}[c]{0.4\linewidth}
    \hspace{.2cm}
    \includegraphics[scale = 0.2]{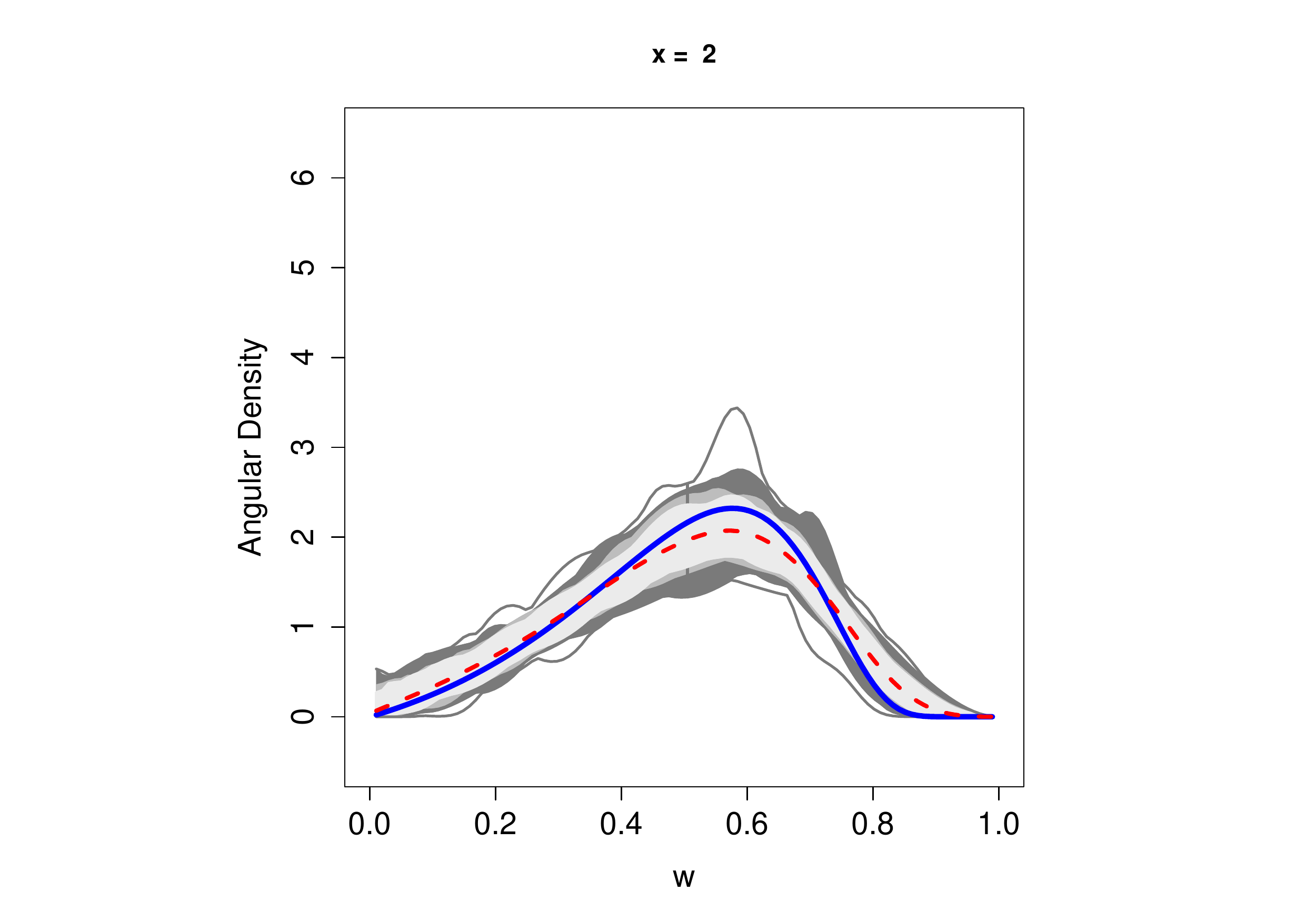}
  \end{minipage}%
  \\
  \vspace{-0.5 cm}
      \footnotesize \rotatebox{90}{\textbf{\hspace{-1cm}L--L weights}}
  \begin{minipage}[c]{0.4\linewidth}
    \hspace{-.1cm}
    \includegraphics[scale = 0.2]{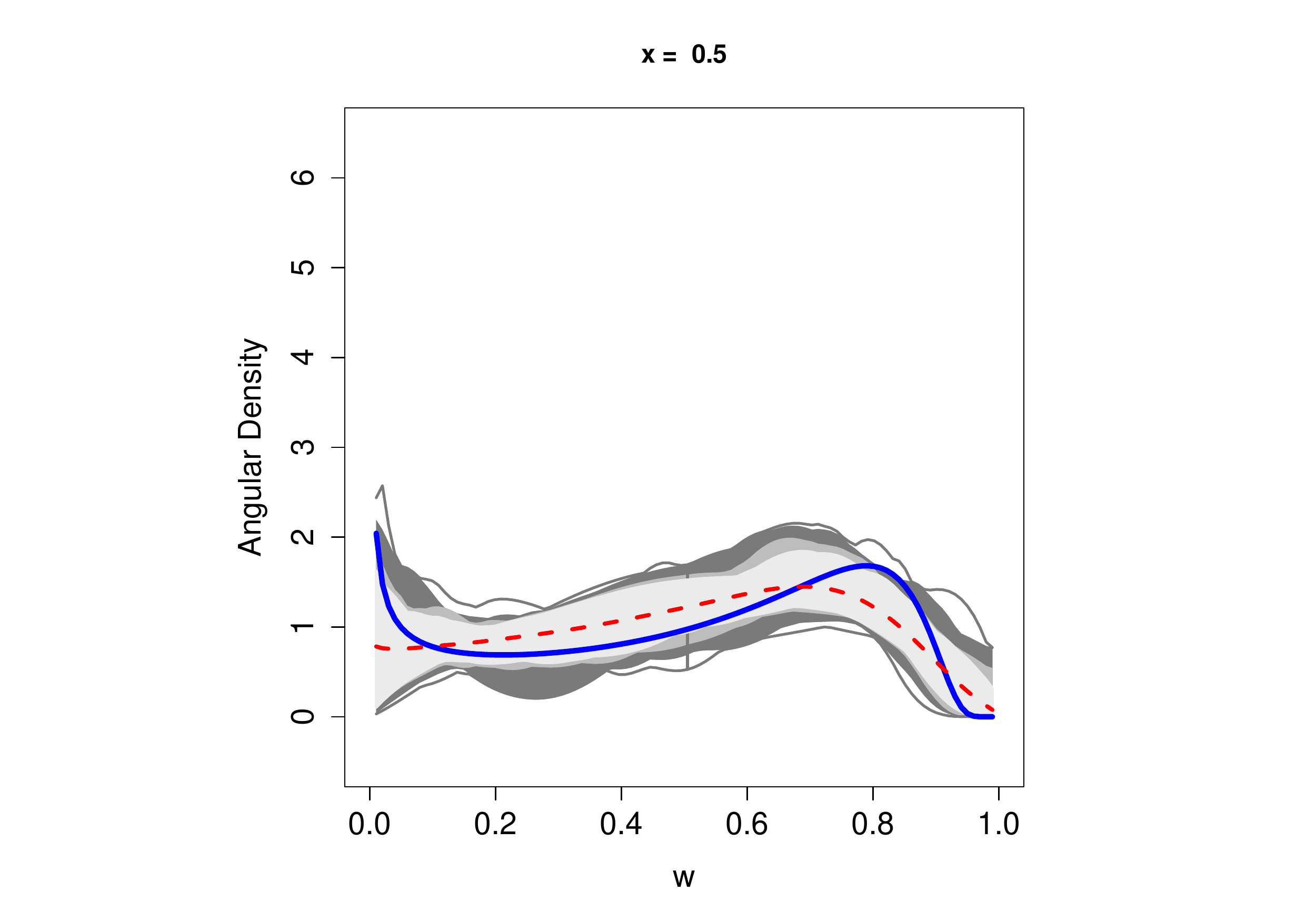}
  \end{minipage}%
  \hspace{-2cm}
  \begin{minipage}[c]{0.4\linewidth}
    \hspace{.2cm}
    \includegraphics[scale = 0.2]{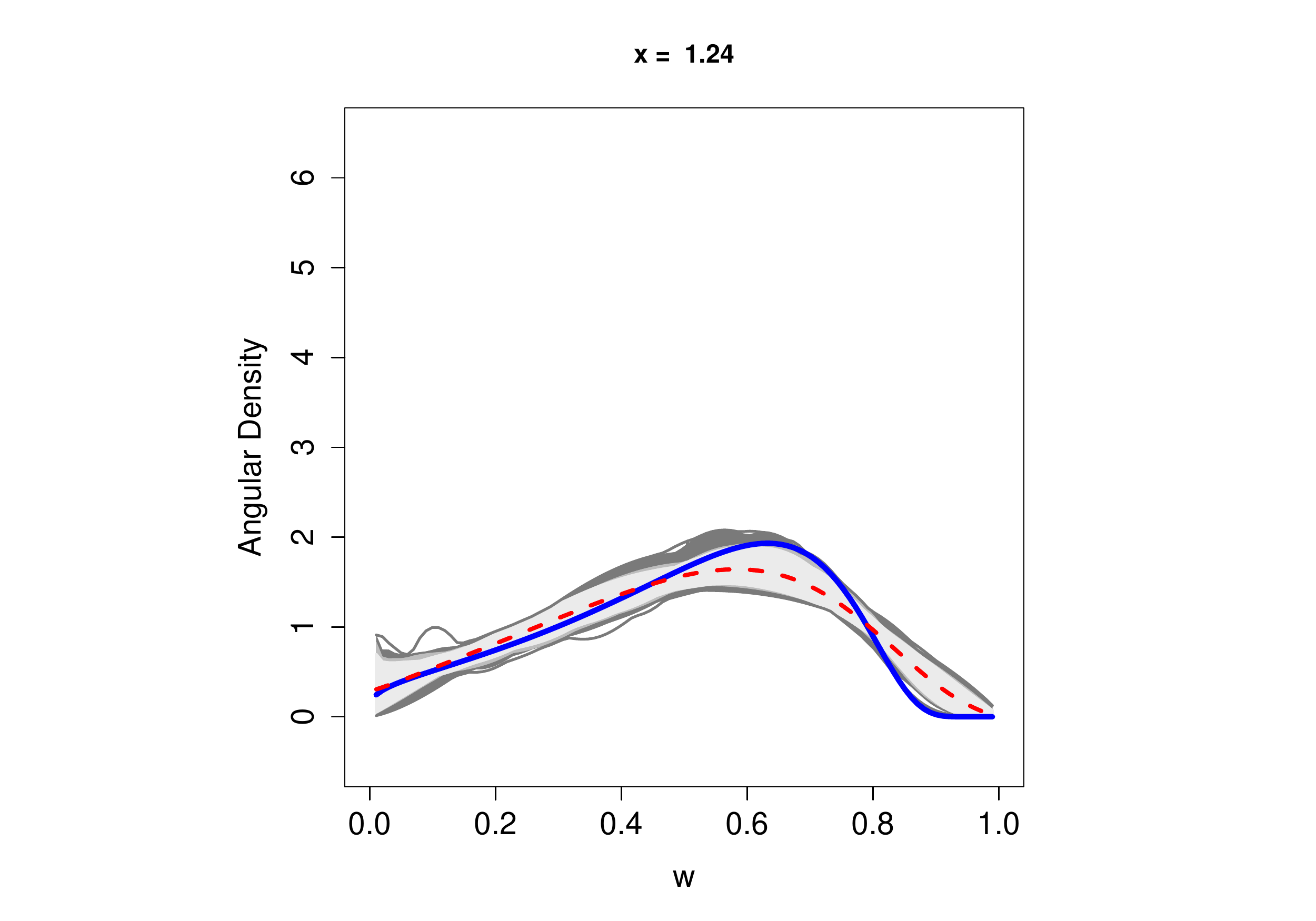}
  \end{minipage}%
  \hspace{-1.8cm}
  \begin{minipage}[c]{0.4\linewidth}
    \hspace{.2cm}
    \includegraphics[scale = 0.2]{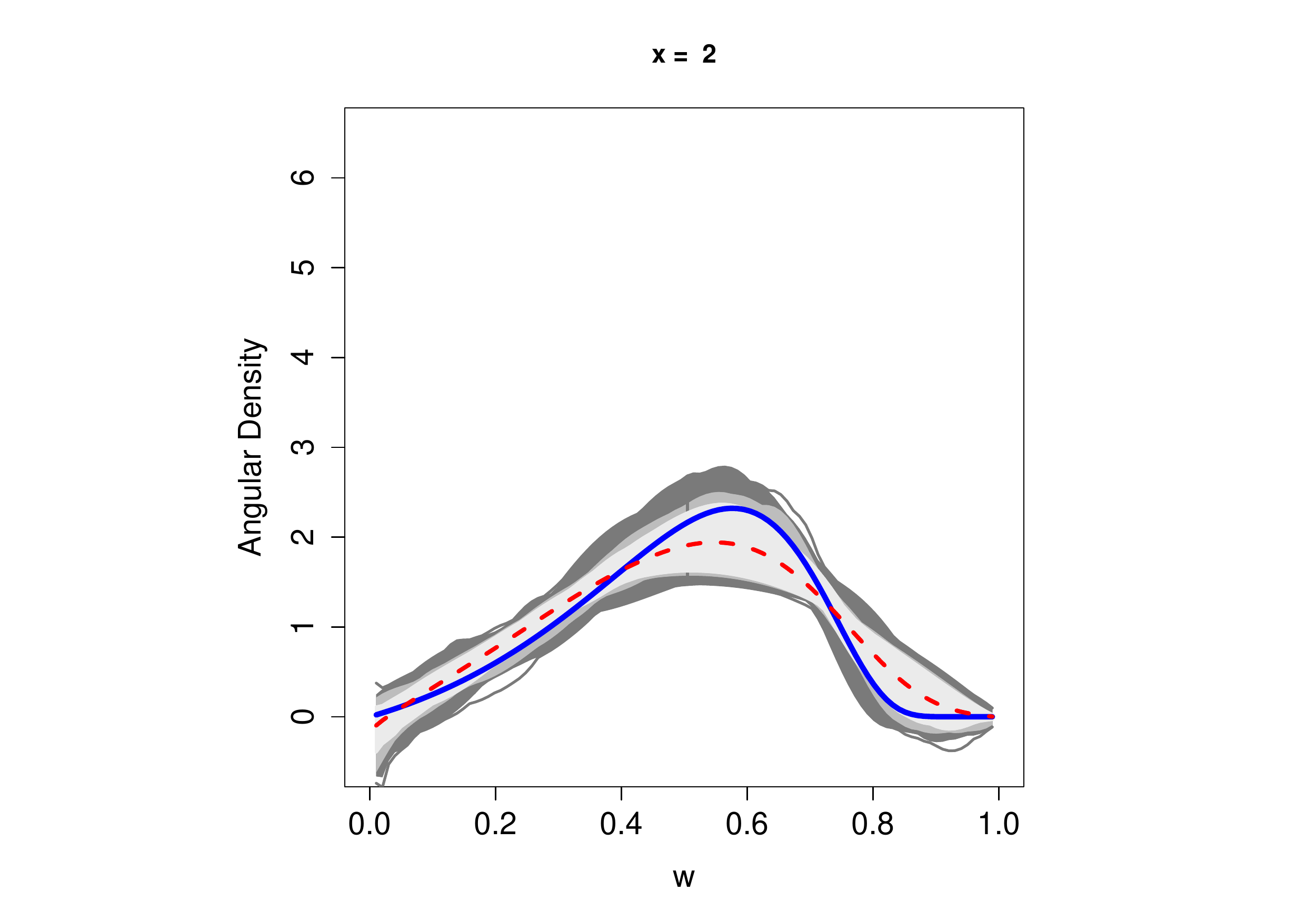}
  \end{minipage}%
  \caption{\footnotesize Functional boxplots (gray shadow) showing the 50\%, 75\%, and 95\% central regions (as defined by~\cite{sun2012functional}) of 1000 samples of size 500 for the conditional models presented in Section~\ref{Data configuration}, as well as their corresponding true values (solid blue line) and Monte Carlo means (dashed red line).}
  \label{trajectories}
\end{figure}

\subsection{Simulation results}\label{Finite sample performance}
To construct the simulation studies, we took 1000 samples of sizes 300 and 500 for the three conditional models presented in Section~\ref{Data configuration}. For the samples of size 500, Figure~\ref{trajectories} displays functional boxplots~\citep{sun2012functional} of cross sections of the angular surface (conditional angular density estimates at fixed values of $x$) along with their Monte Carlo means. {Functional boxplots are constructed introducing measures to define functional quantiles and the centrality or outlyingness of a curve. Specifically, \cite{sun2012functional} use band depths to order a sample of curves from the center outwards, defining $100\alpha\%$ central regions ($0<\alpha<1$). These central regions can be estimated using the $\alpha$ proportion of deepest curves; a formal definition of these regions can be found in~\citet[][Section~3]{sun2012functional}. The gray areas in Figure~\ref{trajectories} show the sample 50\%, 75\%, and 95\% central regions of the sampled curves}. These plots allow us to illustrate the performance of our estimator in terms of variability, under different dependence dynamics. For example, the logistic model estimates presented in the top panel of Figure~\ref{trajectories} turn out to be the most dispersed over all three scenarios. We argue that there are two related reasons for this: the limitations due to boundary bias that were discussed in Section~\ref{Data configuration}, and the fact that the range of extremal dependence in the logistic conditional surface is greater compared to the two other scenarios. Both estimators seems to perform similarly, although the local--linear estimator is more variable when the extremal dependence is stronger. The center panel corresponds to the symmetric Dirichlet angular model, which displays a good mean performance in the last two cases, but some bias when the true angular density is U-shaped. Estimates using the Nadaraya--Watson weights seem to be slightly more variable than the ones using the local linear weights. Finally, the asymmetric Dirichlet angular model presented in the bottom panel, displays more dispersed estimates than its symmetric counterpart for both estimators (and between them it seems that the Nadaraya--Watson estimates are again more variable than the ones using local linear weights), although the Monte Carlo mean produces suitable approximations. The asymmetry does not seem to be a major issue. Overall, estimates for the three models display reasonable performance in recovering the different shapes of the densities, and Monte Carlo means produce reliable estimates. Monte Carlo mean surfaces for the three models and the two estimators can be found in the supplementary material.

We assess the performance of our estimator using the mean integrated absolute error (MIAE), 
\begin{eqnarray}
\text{MIAE} = \E\bigg(\int_{\mathcal{X}}\int_0^1
|\widehat{h}_x(w)-h_x(w)| \, \dif w \, \dif x\bigg),
\label{miae}
\end{eqnarray} 
and report the results in Table~\ref{tableMIAE}. As mentioned before, we can see that the estimator using the Nadaraya--Watson weights outperforms the one using local linear weights in the logistic and symmetric Dirichlet models, but the local linear weights seem to be a better choice for the asymmetric Dirichlet model. In any case and except for the logistic model with sample size 300, the improvements of one estimator over the other are fairly modest. As we should expect, the results show that performance increases with sample size. Overall, simulations confirm that our methods produce acceptably accurate estimates of the angular surface.

\begin{table}
  \caption{ \label{tableMIAE}\footnotesize Mean integrated absolute error 
    estimates computed over 1000 samples for the data-generating configurations 
    discussed in Section~\ref{Data configuration} for the Nadaraya--Watson (N--W) and the local lineal (L--L) weights.}
  \fbox
  {
    \begin{tabular}{l| l | l| l| ll}
    $n$&conditional Model& Specification & \multicolumn{2}{c}{MIAE}\\ 
        \cline{4-5}
     & & &  N--W weights & L--L weights\\
      \midrule
      300& Logistic & $\alpha_{x}=\Phi(x)$ & 0.09& 0.92\\
         & Symmetric Dirichlet  & $(a_x,b_x) = (x, x)$ &0.42&0.60 \\	
         & Asymmetric Dirichlet & $(a_x, b_x)=(x,100)$ &0.63& 0.59  \\
      \hline
      500& Logistic & $\alpha_{x}=\Phi(x)$ & 0.08 & 0.14\\
         & Symmetric Dirichlet  & $(a_x,b_x) = (x, x)$ &0.39& 0.55\\
         & Asymmetric Dirichlet & $(a_x, b_x)=(x,100)$ & 0.62& 0.55  \\
    \end{tabular}
  }
\end{table}

We conclude this section providing some comments on implementation of the tuning parameter selection (Section~\eqref{tuning}). Since in some cases optimization over $\mathcal{R}_{\mathcal{X},n}$ (defined in Eq.~\eqref{Rset}) can be computationally expensive, our experiments suggest that optimization over $\mathcal{R}_{n}$ defined as
\begin{equation*}
\mathcal{R}_{n} = \{(b,\nu,\tau)\in(0, \infty)^{3}: \nu \{1-W_i\theta_b(X_{j})\}+\tau>0,\text{for } i,\text{ }j=1,\dots, n\},
\end{equation*} 
performs reasonably well. Note that $\mathcal{R}_{n}$ is a version of $\mathcal{R}_{\mathcal{X},n}$ determined only by the observed covariate values, and not by the entire covariate space $\mathcal{X}$. Furthermore, for large $n$, unconstrained optimization over $(0, \infty)^3$ typically also performs well. We thus recommend the user to initially try unconstrained optimization for large $n$, or optimization over $\mathcal{R}_{n}$ for moderate $n$. Only if the resulting parameter values do not yield a valid estimator over the study region of interest does one then need to implement the constrained optimization over $\mathcal{R}_{\mathcal{X},n}$.

\section{Dynamics of joint extremal losses in leading European stock markets}
\label{Application}
\subsection{Background and motivation for empirical analysis}\label{background}
In 1999, 11 European Union (EU) countries formed the Economic and Monetary Union (EMU), which led them to adopt a common currency and monetary policy as well as the conduction of coordinated economic policies. 

The process of creation of the EMU was the outcome of three stages of development,  further details of which can be found on the European Central Bank website: 
\begin{center}
  \url{https://www.ecb.europa.eu/ecb/history}
\end{center}
\noindent see also \cite{J12}. To join the Eurozone (countries who adopted the Euro as their common currency) member states had to qualify by meeting the criteria of the Maastricht Treaty in terms of budget deficits, inflation, interest rates, and other monetary requirements. At the moment the Euro is the single currency shared by 19 of the 28 EU members. The remaining 9 countries, including the UK, are endowed with `opt-out' clauses which exempts them from using the Euro as their currency. In recent years there have been several studies providing evidence for an increased integration of European stock markets, and the EMU has been frequently put forward as the causal driver for this increase, along with some other determinants \cite[][and the references therein]{F02, KAL05, HAL06,BH11}. \cite{HAL06} found however that the UK, who chose not to enter the eurozone, showed no increase in stock market integration by that time. 

Although there is a wealth of studies analyzing stock market integration over time, {few attempts have been made to ascertain the dynamics governing extreme value dependence of stock market returns over time. The huge literature looking into dependence of financial markets \citep[see for example][]{KAL94, LS95, LS01, KS96, FR02, BN04, BN05, RN09} has collected evidence compatible with the hypothesis that the comovement of returns has not remained constant over time. Yet, none of these papers has focused on tracking the dynamics of extremal dependence of returns, which is the object of the current inquiry.} An exception in this respect is the seminal paper of \citet{PAL03}, which provides evidence of increasing levels of extremal dependence for three major stock markets within Europe [CAC (France), DAX (Germany), and FTSE (UK)]. The subperiod analysis of \citet[][Section~3.3.2]{PAL03} is however exploratory, in the sense that they arbitrarily partitioned the sample period into three periods, and thus estimation of extremal dependence on each period only takes data from that period into account. 

\begin{figure}
\begin{minipage}[c]{0.4\linewidth}
  \hspace{-1cm}
  \includegraphics[scale=0.30]{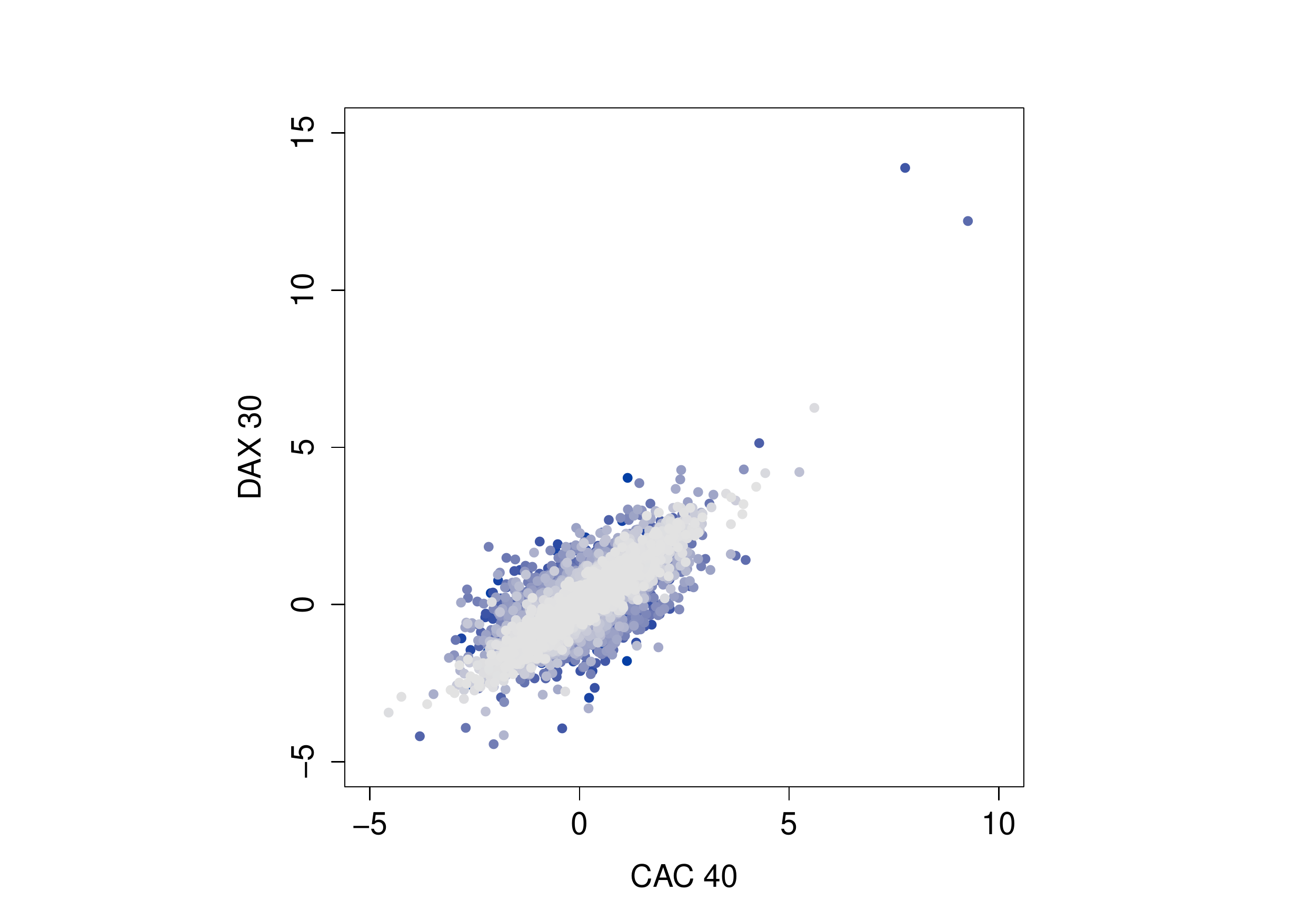}
 \end{minipage}%
\hspace{-1.8cm}
\begin{minipage}[c]{0.4\linewidth}
  \hspace{-.6cm}
 \includegraphics[scale=0.30]{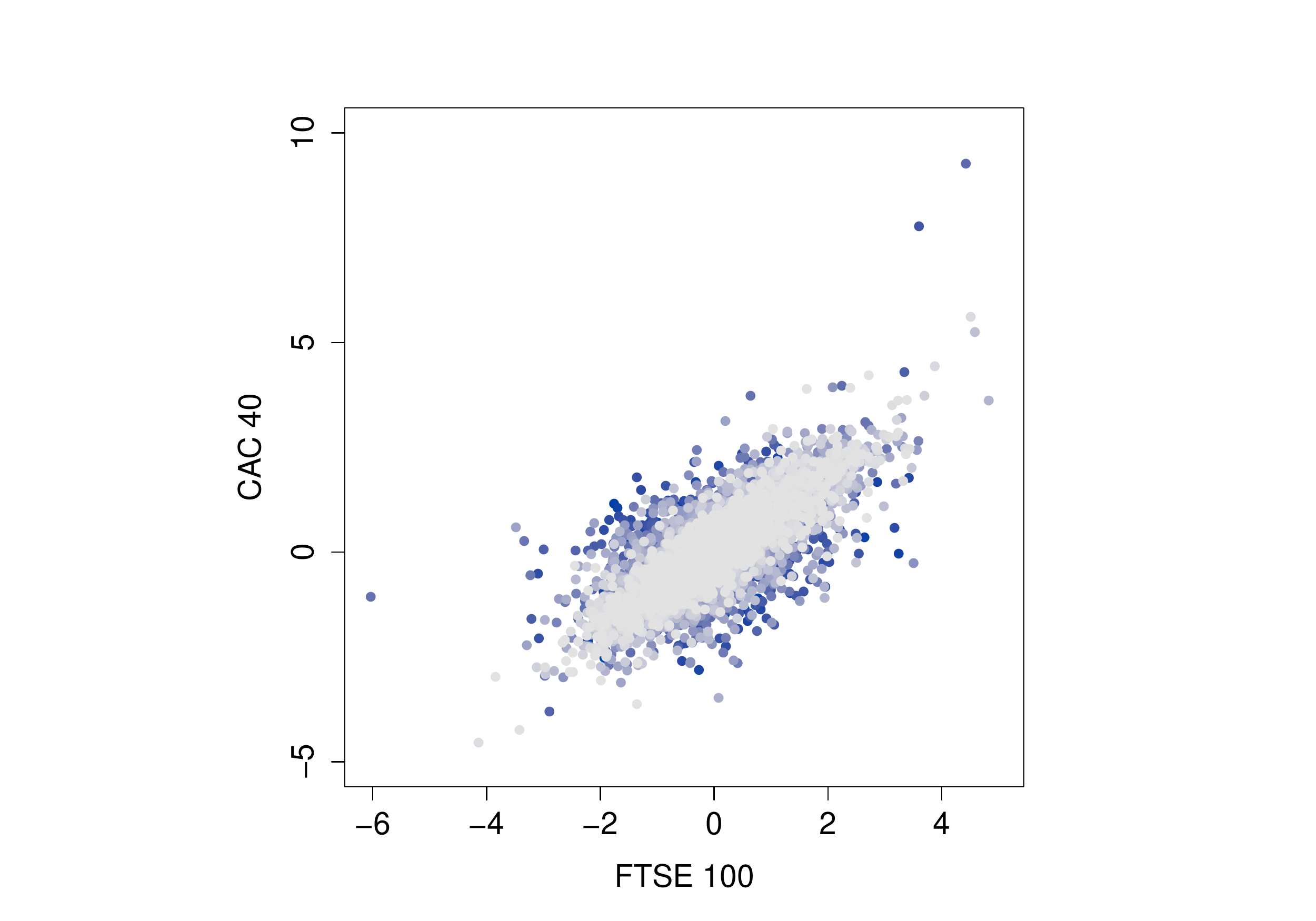}
\end{minipage}%
\hspace{-1.6cm}
\begin{minipage}[c]{0.4\linewidth}
\hspace{-.4cm}
 \includegraphics[scale=0.30]{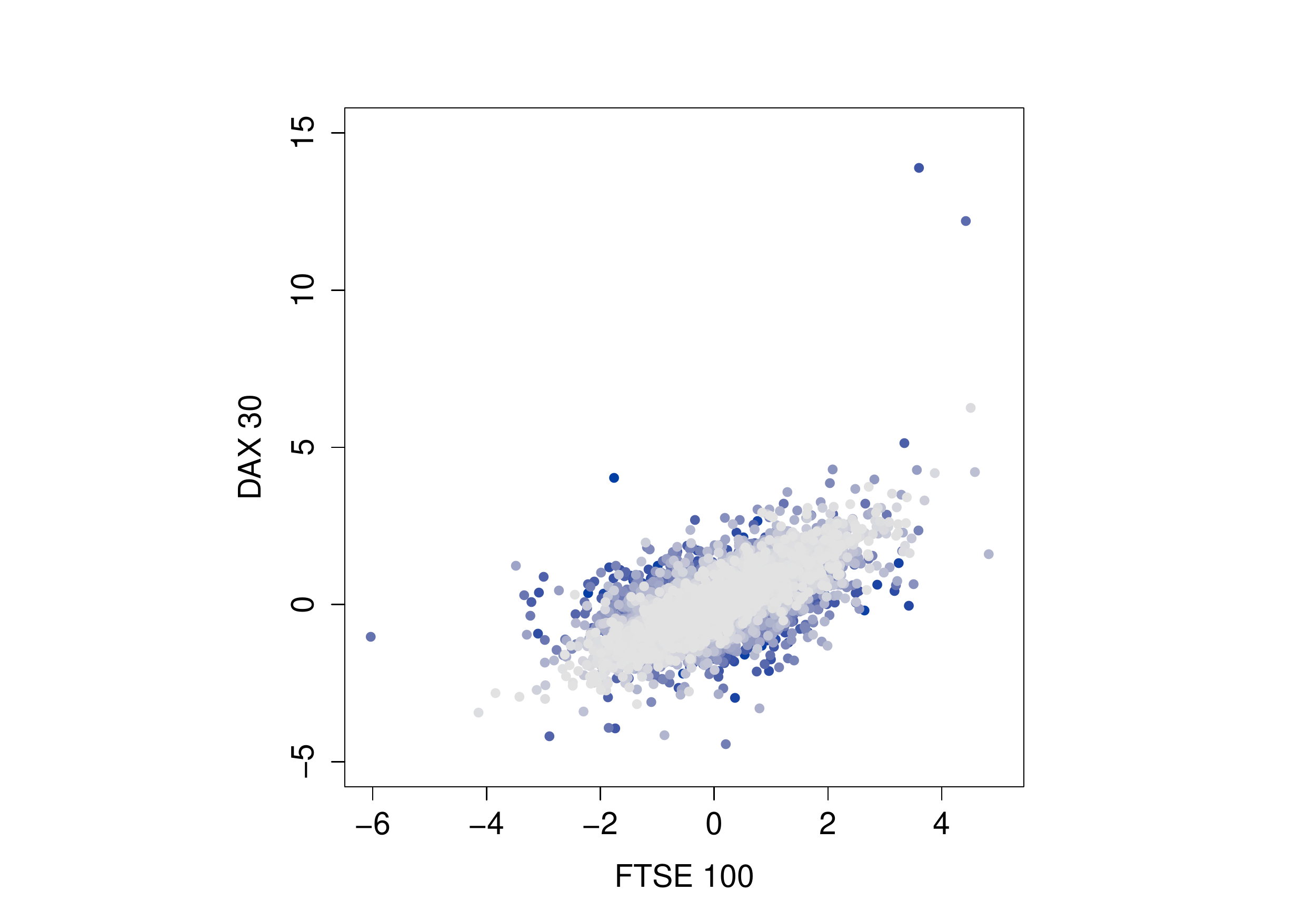}
\end{minipage}%
\\
\begin{minipage}[c]{0.4\linewidth}
\hspace{.3cm}
 \includegraphics[scale=0.25]{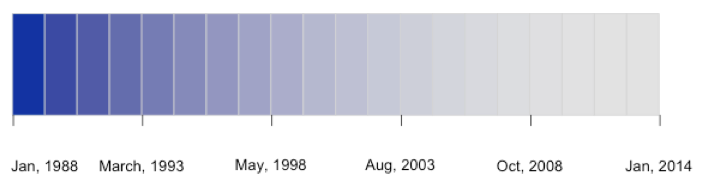}
\end{minipage}%
 \caption{\footnotesize Scatterplots using a time-varying color palette for GARCH-filtered residuals for CAC 40 (FR), DAX 30  (DE) and FTSE 100 (UK) spanning the period from January 1, 1988 to January 1, 2014.}
 \label{scatter.pdf}
\end{figure}

Below, we apply our methods to address a similar question to that of \cite{PAL03,PAL04}. Specifically, one of our main interests is disentangling the dynamics governing the dependence of extreme losses on three leading European stock markets---using CAC, DAX, and FTSE---in recent years. The motivation for choosing these markets is twofold: these are the stock markets of the European members of G5; these are also the same European stock markets considered by \cite{PAL03,PAL04}. Moreover they display a stronger type of extremal dependence than some of the other markets studied by \cite{PAL03,PAL04}, i.e., asymptotic dependence as defined in Section~\ref{bse}.
\vspace{-.1 cm}
\subsection{Data description, preprocessing, and exploratory considerations}\label{exploratory}
Our data were gathered from Datastream and consist of daily closing stock index levels of three leading European stock markets: CAC 40, DAX 30, and FTSE 100 (henceforth CAC, DAX, and FTSE). The sample period spans from January 1, 1988 to January 1, 2014 ($N = 6784$ observations), and hence it includes the Great Moderation and Great Recession which are by all standards challenging modeling issues. Since we want to focus on extreme losses, we use daily negative returns as a unit of analysis. Daily negative returns are computed by taking the negative of the first differences of the logarithmic indices. Following the bivariate analysis in~\cite{PAL04} both observations of a particular day are removed if at least one of the two observations is a zero return (plots of the data and summary statistics can be found in the supplementary material). The \emph{Engle} (1) statistic of~\cite{engle1982autoregressive} (not reported here) is large and significant for all three stock return series, indicating strong heteroskedasticity which can be removed by fitting volatility filters. In the spirit of~\cite{PAL04} we fit three different filters: GARCH(1,1) assuming $t-$distributed errors for CAC and normal for FTSE and DAX, NGARCH (also known as nonlinear asymmetric GARCH) with normal innovations, and the stochastic volatility model (SV) of~\cite{kim1998stochastic} with hyperparameters chosen according to the latter paper. Diagnostic plots (not shown here) suggest that the GARCH fits are superior than the NGARCH fits for the three stock markets, and heteroskedasticity is successfully removed with the GARCH and NGARCH filters, but not with the SV filter. The results shown below correspond to the GARCH-filtered residuals, but similar conclusions can be drawn using the NGARCH filter (angular surfaces based on the NGARCH-filtered residuals can be found in the supplementary material). 
Scatterplots of possible combinations of pairs of filtered residual series are displayed in Figure~\ref{scatter.pdf}, depicted using a time-varying color palette which allows us to uncover the nonstationary nature of joint extremes. This is in line with the findings of \cite{PAL03,PAL04}. 

To verify that our methods are a sensible approach for modeling these data, we need to assess whether the filtered residuals are asymptotically dependent. As mentioned in Section~\ref{bse}, in the modeling of extreme events two different classes of extreme value dependence can arise: asymptotic dependence and asymptotic independence. Dependence between moderately large values can arise in both cases, but the very largest values from each variable can occur together only under asymptotic dependence. To make ideas concrete, let $Y_1$ and $Y_2$ be any two filtered residuals of interest, transformed to have unit Fr\'echet margins. Under an exploratory setting, two measures of tail dependence can be obtained to summarize the strength of extremal dependence: 
\begin{align*}
\chi = \lim_{u\to\infty} \pr(Y_1>u \mid Y_2>u), \quad 
\overline{\chi} =\lim_{u\to\infty}\frac{2\log \pr(Y_1>u)}{\log \pr(Y_1>u,Y_2>u)}-1.
\end{align*}
Here, $\chi\in[0,1]$ measures the strength of dependence within the class of asymptotically dependent variables, whereas $\overline{\chi}\in[-1,1]$ is often used to measure the strength of dependence within the class of asymptotically independent variables. Taken together, the pair $(\chi,\overline{\chi})$ provides a summary of extremal dependence for the vector ($Y_1,Y_2$). For asymptotically dependent variables we have $\overline{\chi} = 1$ and the value of $\chi>0$ increases with the strength of dependence at extreme levels. For asymptotically independent variables we have $\chi=0$ and $\overline{\chi}\leqslant1$ increases with the strength of dependence at extreme levels. Roughly speaking, if $\overline{\chi}>0$ then we often speak about `positive extremal dependence,' whereas if $\overline{\chi}<0$ we use the expression `negative extremal dependence'. Indeed, for the bivariate normal dependence structure $\overline{\chi}$ corresponds to Pearson correlation; see \cite{H00} for further examples.

\begin{figure}[h]
\begin{center}\hspace{-.1cm} {\fontsize{6}{5}\selectfont{\textbf{CAC 40--DAX 30}}} \hspace{2.2cm} {\fontsize{6}{5}\selectfont{\textbf{FTSE 100--CAC 40}}} \hspace{2.4cm}{\fontsize{6}{5}\selectfont{\textbf{FTSE 100--DAX 30}}} \end{center}
\begin{minipage}[c]{0.4\linewidth}
  \vspace{-.5cm}\hspace{-1cm}
  \includegraphics[scale=0.28]{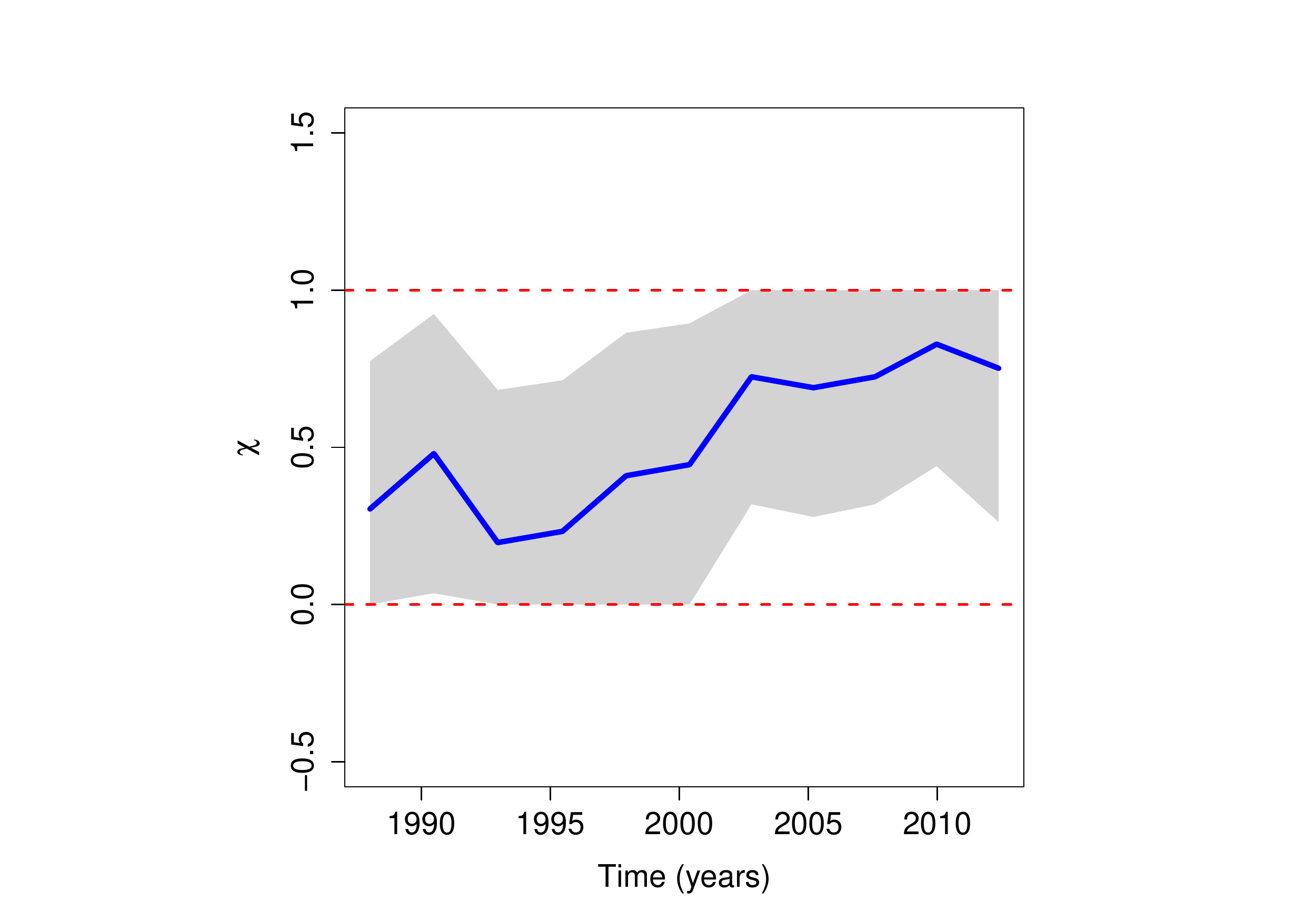}
 \end{minipage}%
\hspace{-1.8cm}
\begin{minipage}[c]{0.4\linewidth}
    \vspace{-.5cm}\hspace{-.8cm}
   \includegraphics[scale=0.28]{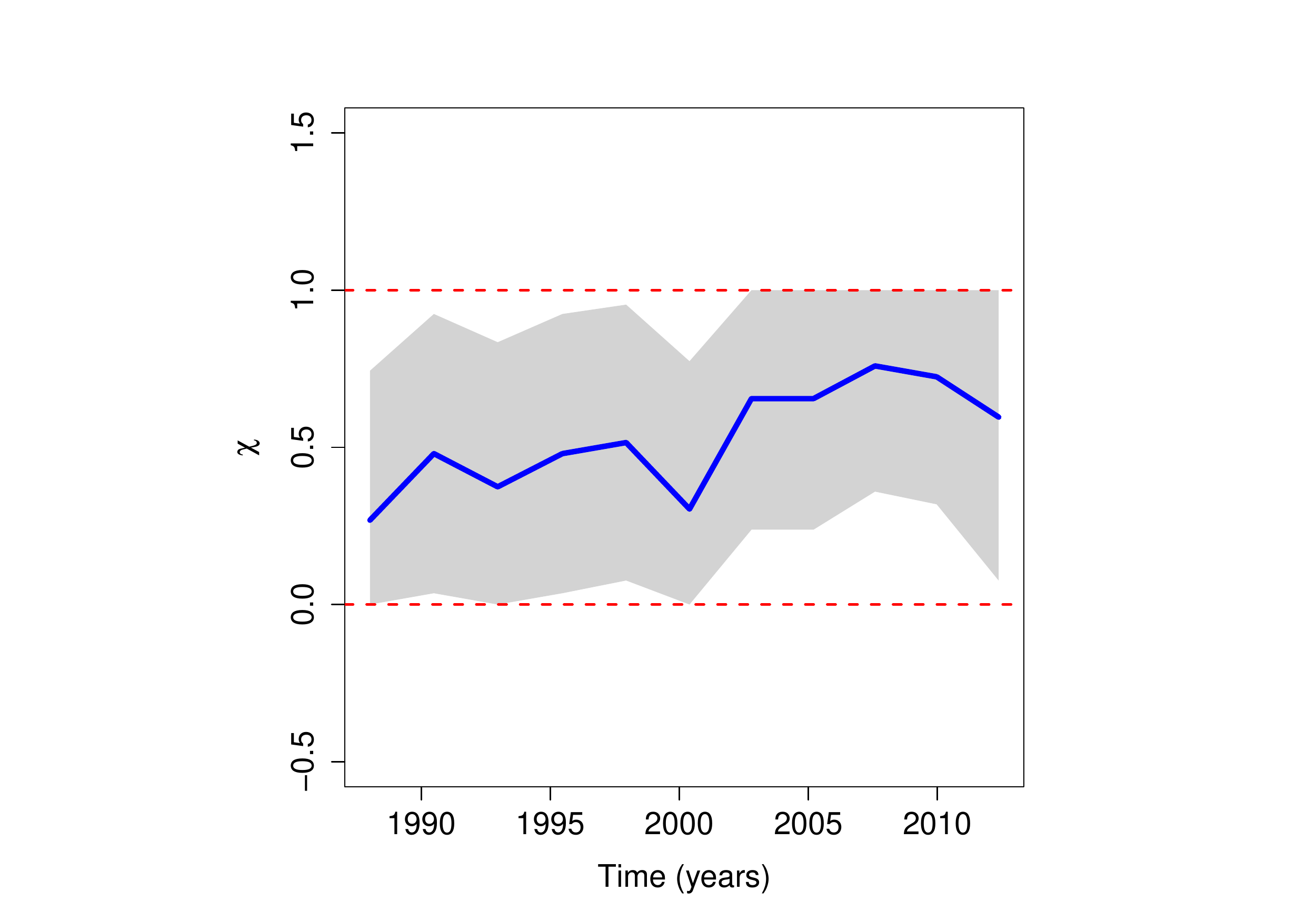}
\end{minipage}%
\hspace{-1.8cm}
\begin{minipage}[c]{0.4\linewidth}
  \vspace{-.5cm}\hspace{-.6cm}
 \includegraphics[scale=0.28]{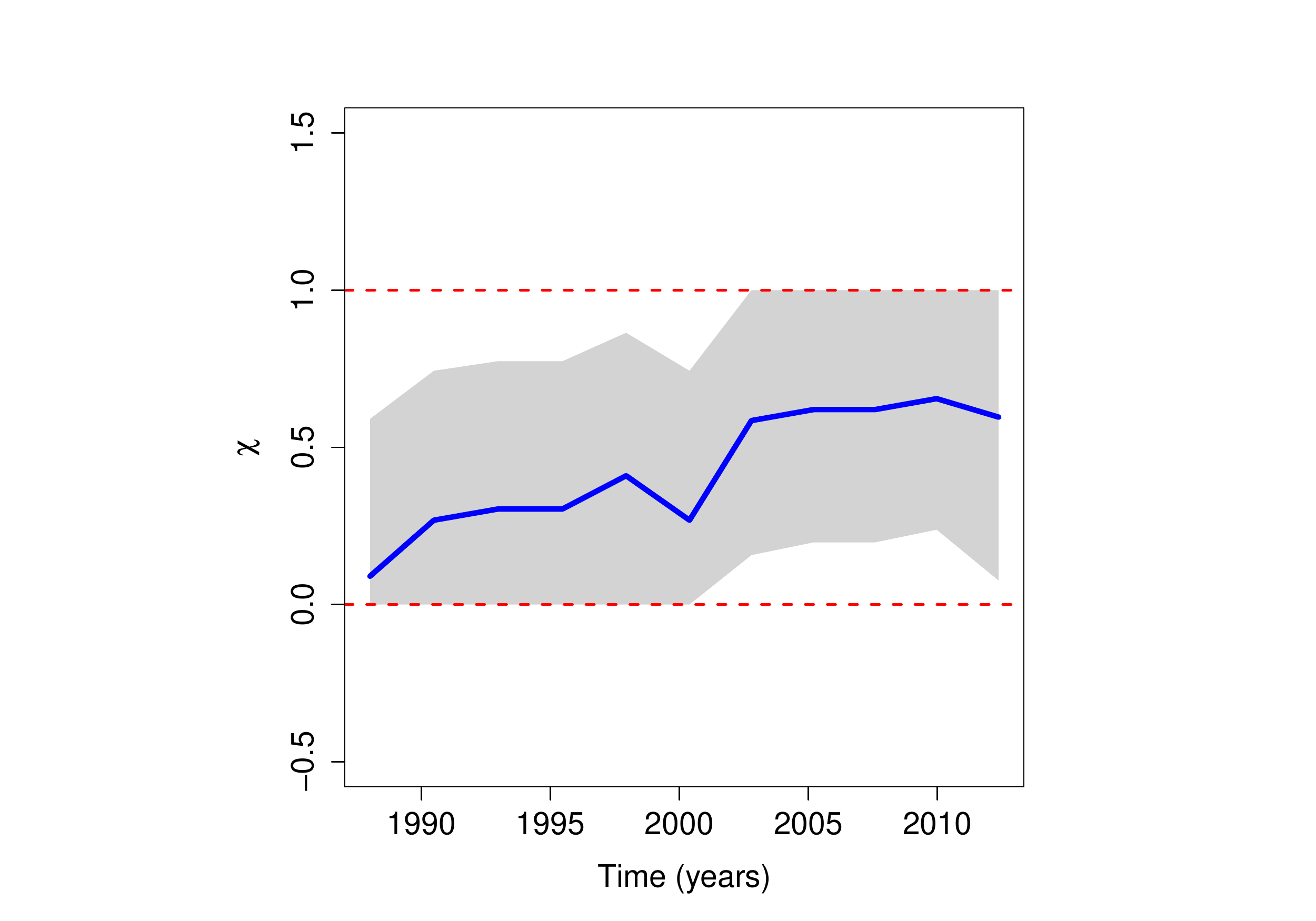}
\end{minipage}%
\\
\begin{minipage}[c]{0.4\linewidth}
    \vspace{-.5cm}\hspace{-1cm}
   \includegraphics[scale=0.28]{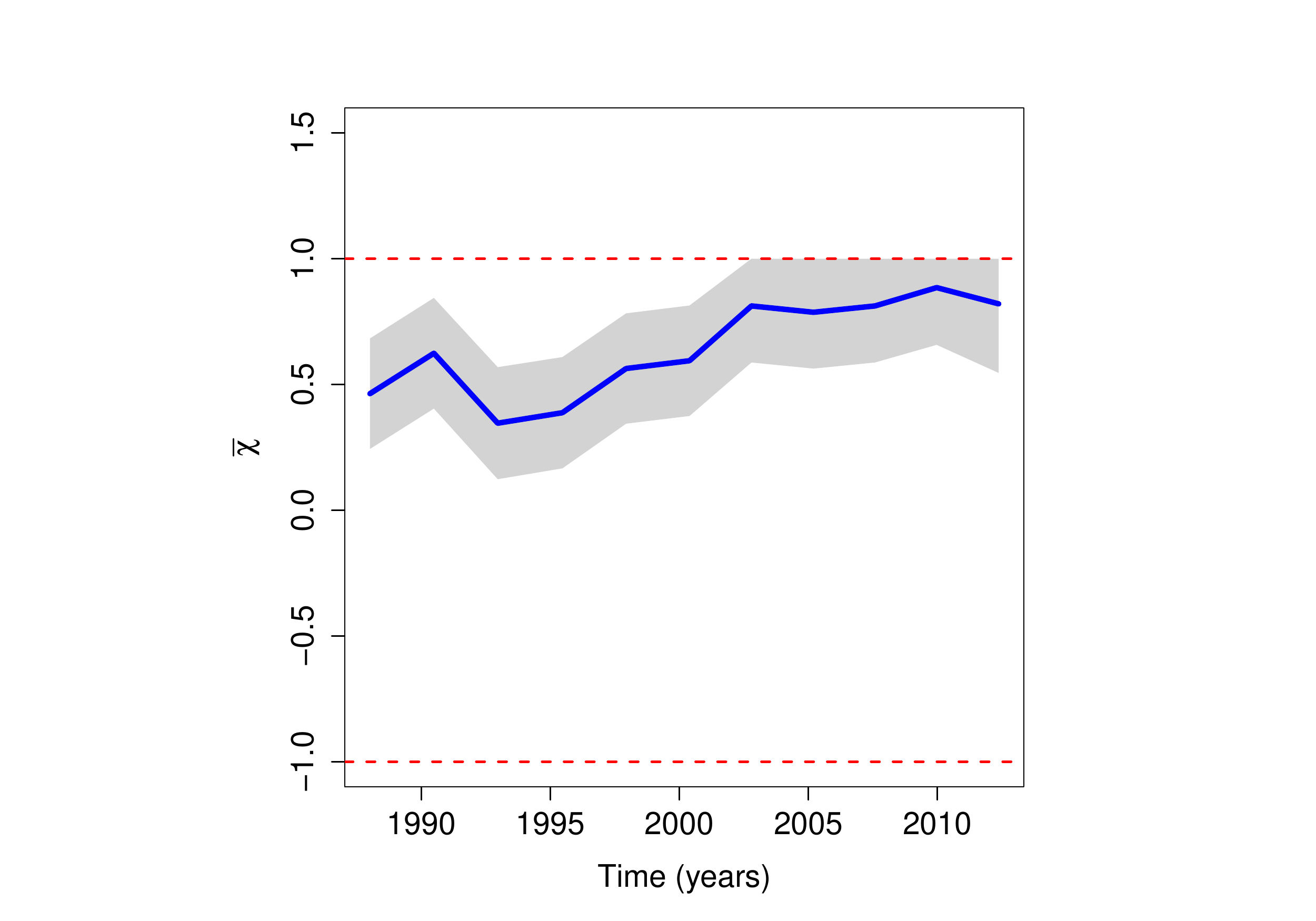}
 \end{minipage}%
\hspace{-1.8cm}
\begin{minipage}[c]{0.4\linewidth}
    \vspace{-.5cm}\hspace{-.8cm}
    \includegraphics[scale=0.28]{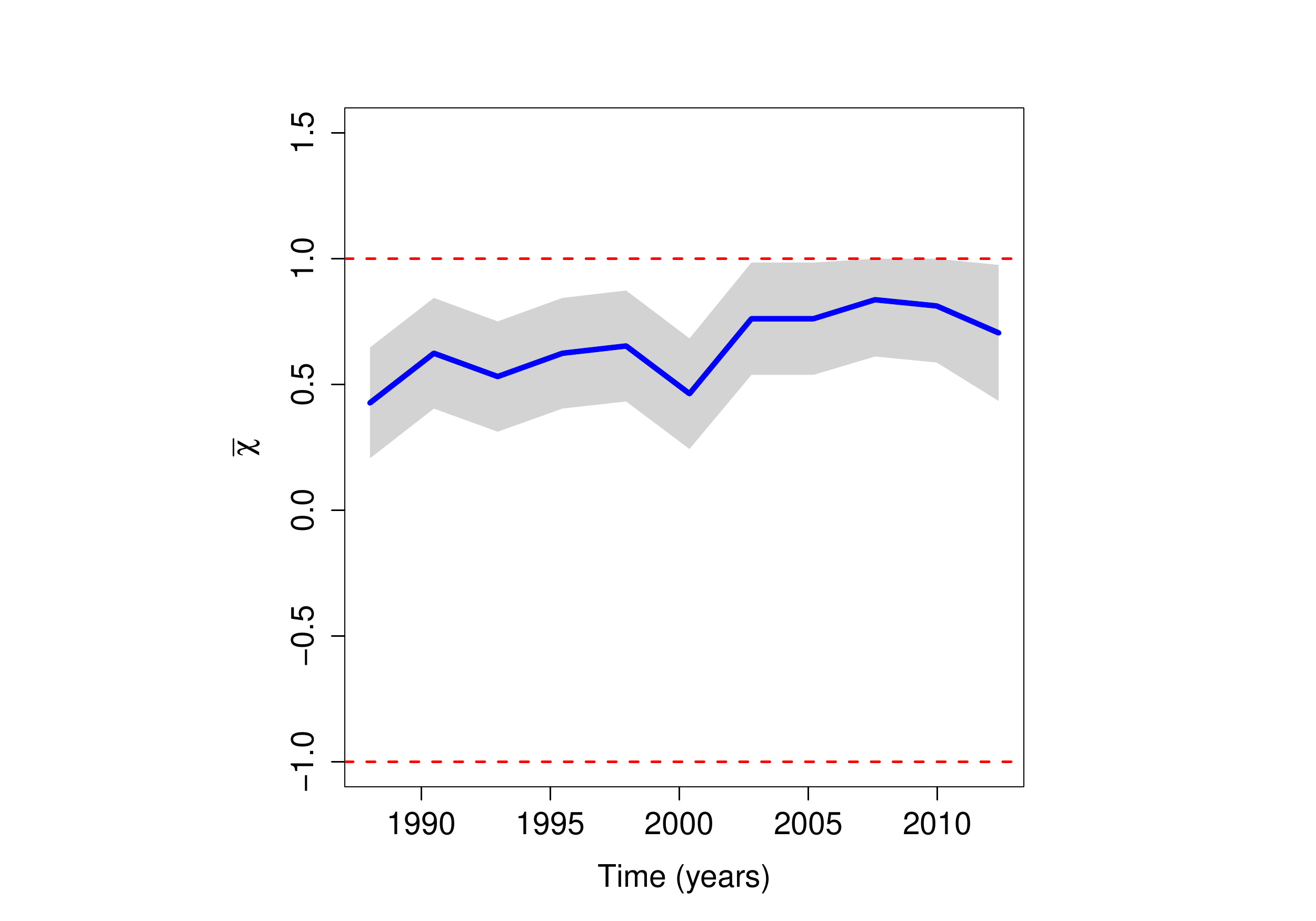}
\end{minipage}%
\hspace{-1.8cm}
\begin{minipage}[c]{0.4\linewidth}
  \vspace{-.5cm}\hspace{-.6cm}
 \includegraphics[scale=0.28]{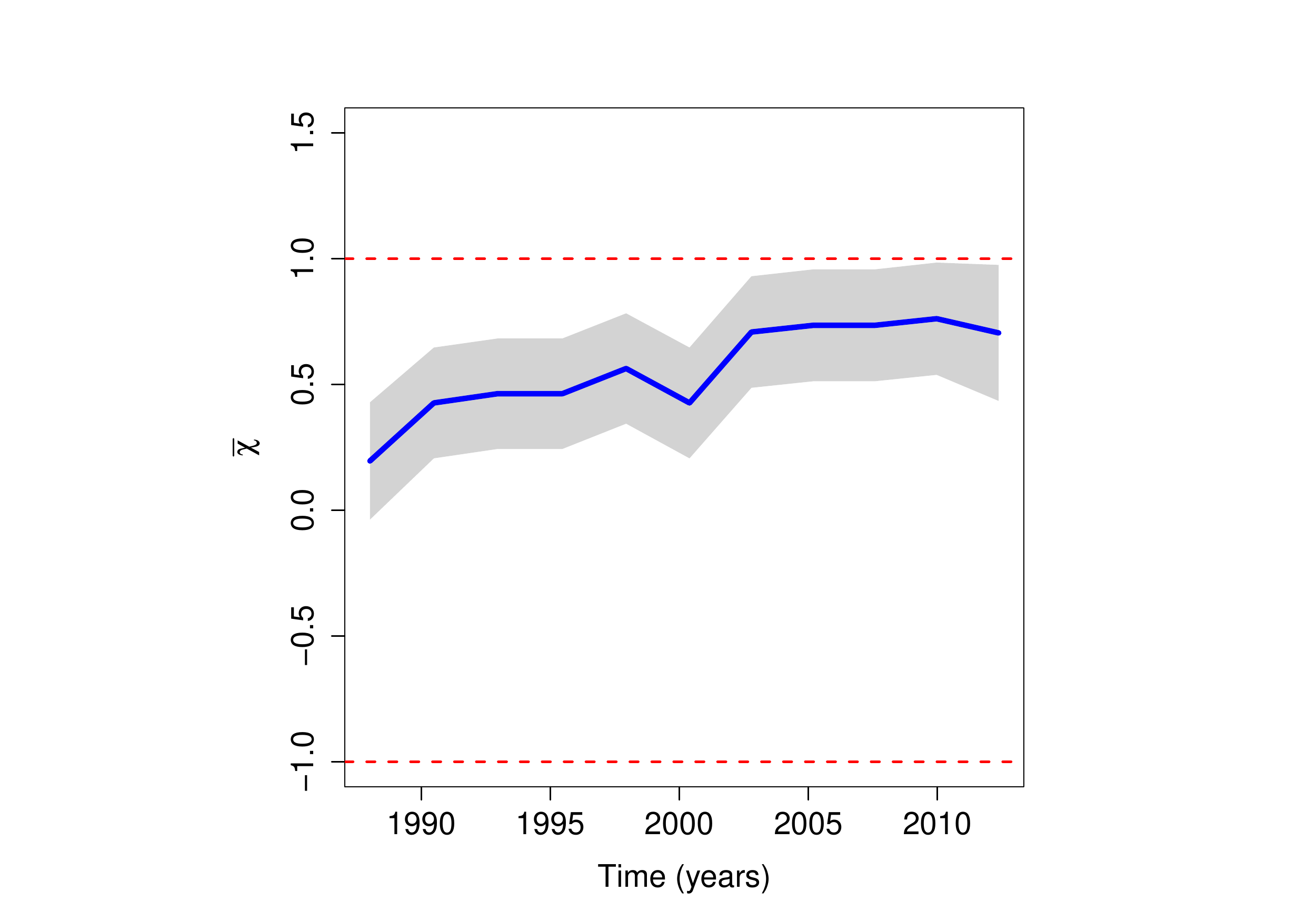}
\end{minipage}%
  \caption{\footnotesize Rolling window estimates of $\chi$ (top) and $\overline{\chi}$ (bottom) at the 95\% quantile using moving windows of size 600, applied to the three pairs under study.}
  \label{chi.pdf}
\end{figure}

In Figure~\ref{chi.pdf} we present rolling window estimates of $\chi$ and $\overline{\chi}$ with approximate 95\% confidence intervals, which is tantamount to the subperiod analysis of \citet[][Section~3.3.2]{PAL03}. The rolling window estimates were computed using the empirical estimators of $\chi$ and $\overline{\chi}$~\citep[][p.~348]{BAL04} at the 95\% quantile for moving windows of 600 observations. Given the large uncertainty entailed in the estimation of $\chi$, interpretation of these plots is far from straightforward. Nevertheless, pointwise estimation for $\chi$ seems reasonably different from 0 for the three pairs under study, and despite some drops around 1992 and 2000, there seems to be an increasing trend for the three cases. Moreover values for $\overline{\chi}$ are closer to 1 as time passes. This combined information indicates that the assumption of asymptotic dependence is certainly plausible for the later years, and might be adequate for earlier years. We discuss the asymptotic independence issue again in Section~\ref{Discussion}. 

\subsection{Modeling time-varying extremal dependence}\label{tmed}
\begin{figure}[h]
\begin{tabular}{ccc}
   \footnotesize \rotatebox{90}{\textbf{\hspace{-1cm}N--W weights}}
\begin{minipage}[c]{0.4\linewidth}
 \hspace{-1.5cm}
  \includegraphics[width=1.3\textwidth]{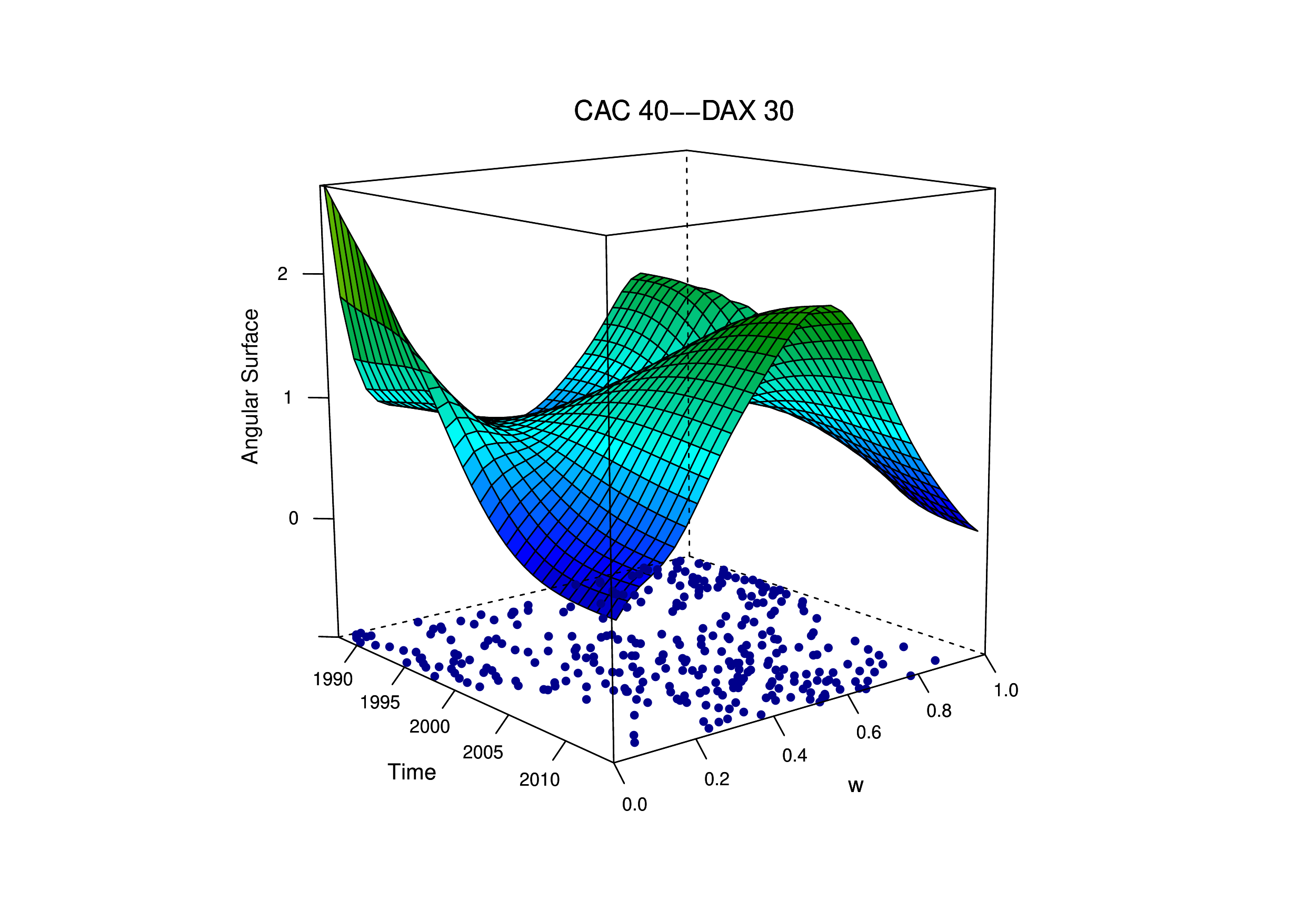}
 \end{minipage}&
\begin{minipage}[c]{0.4\linewidth}
\hspace{-3cm}
 \includegraphics[width=1.3\textwidth]{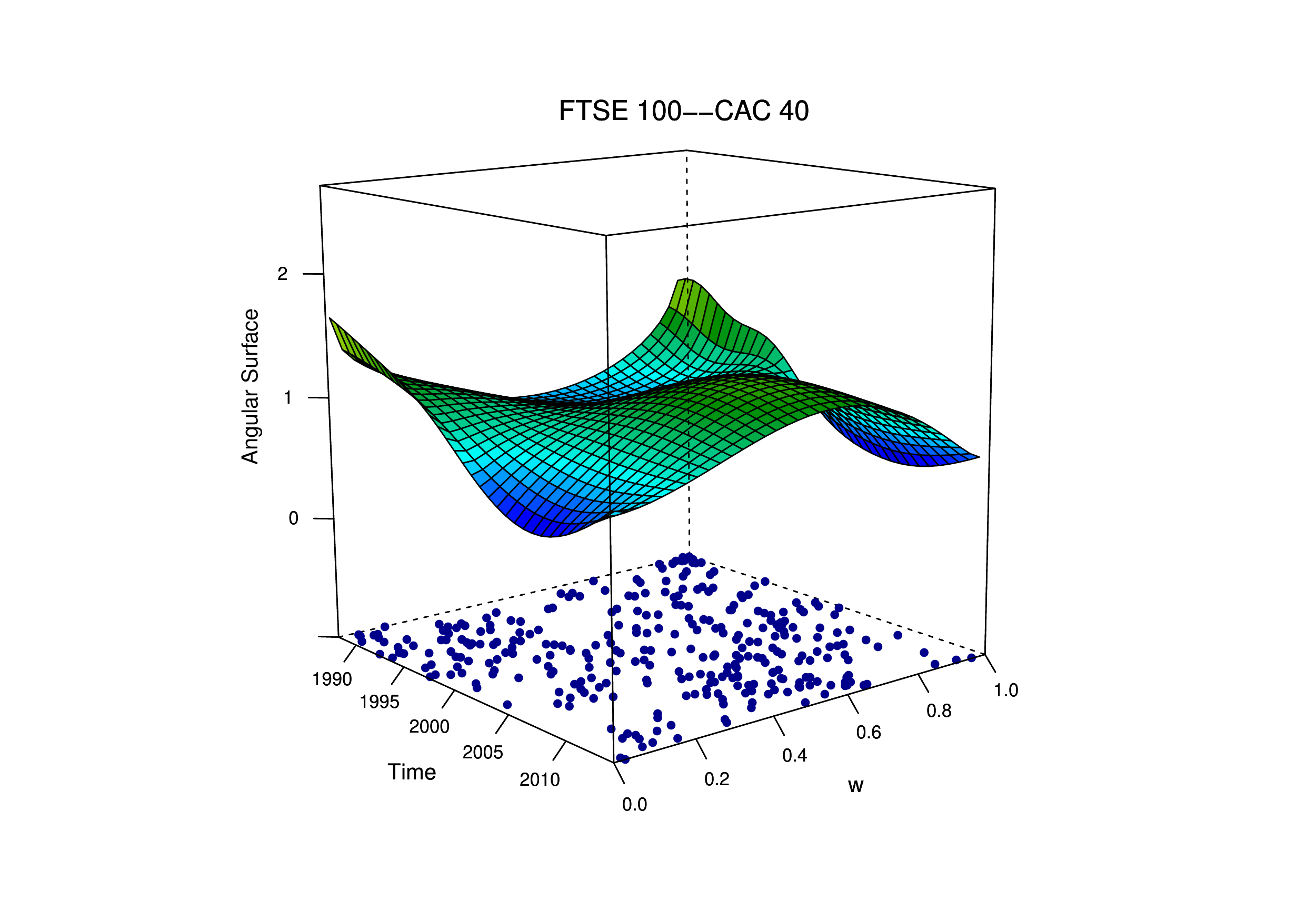}
\end{minipage}&
\begin{minipage}[c]{0.4\linewidth}
\hspace{-4.8cm}
 \includegraphics[width=1.3\textwidth]{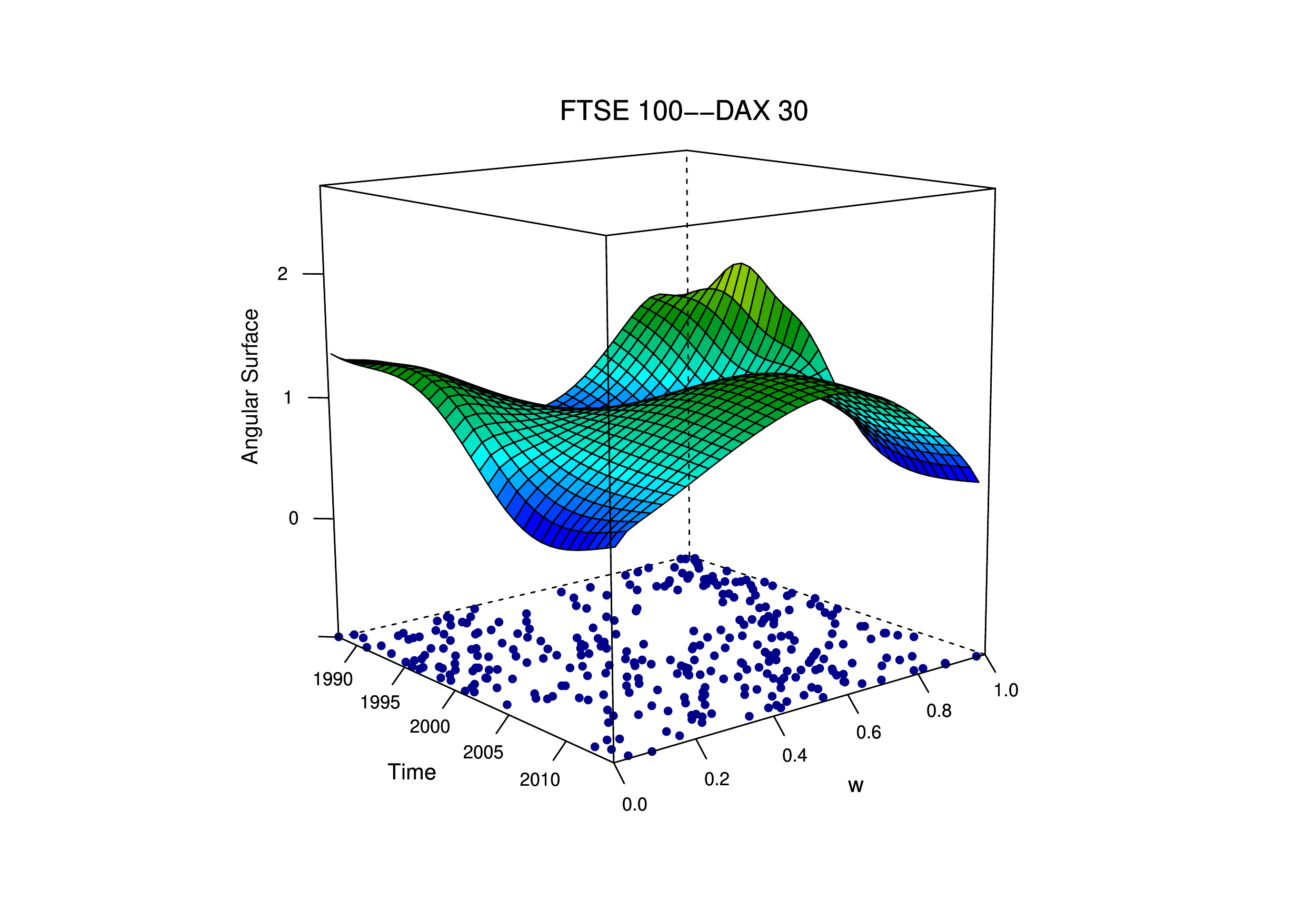}
\end{minipage}\\
      \footnotesize \rotatebox{90}{\textbf{\hspace{-1cm}L--L weights}}
\begin{minipage}[c]{0.4\linewidth}
 \hspace{-1.3cm}
  \includegraphics[width=1.3\textwidth]{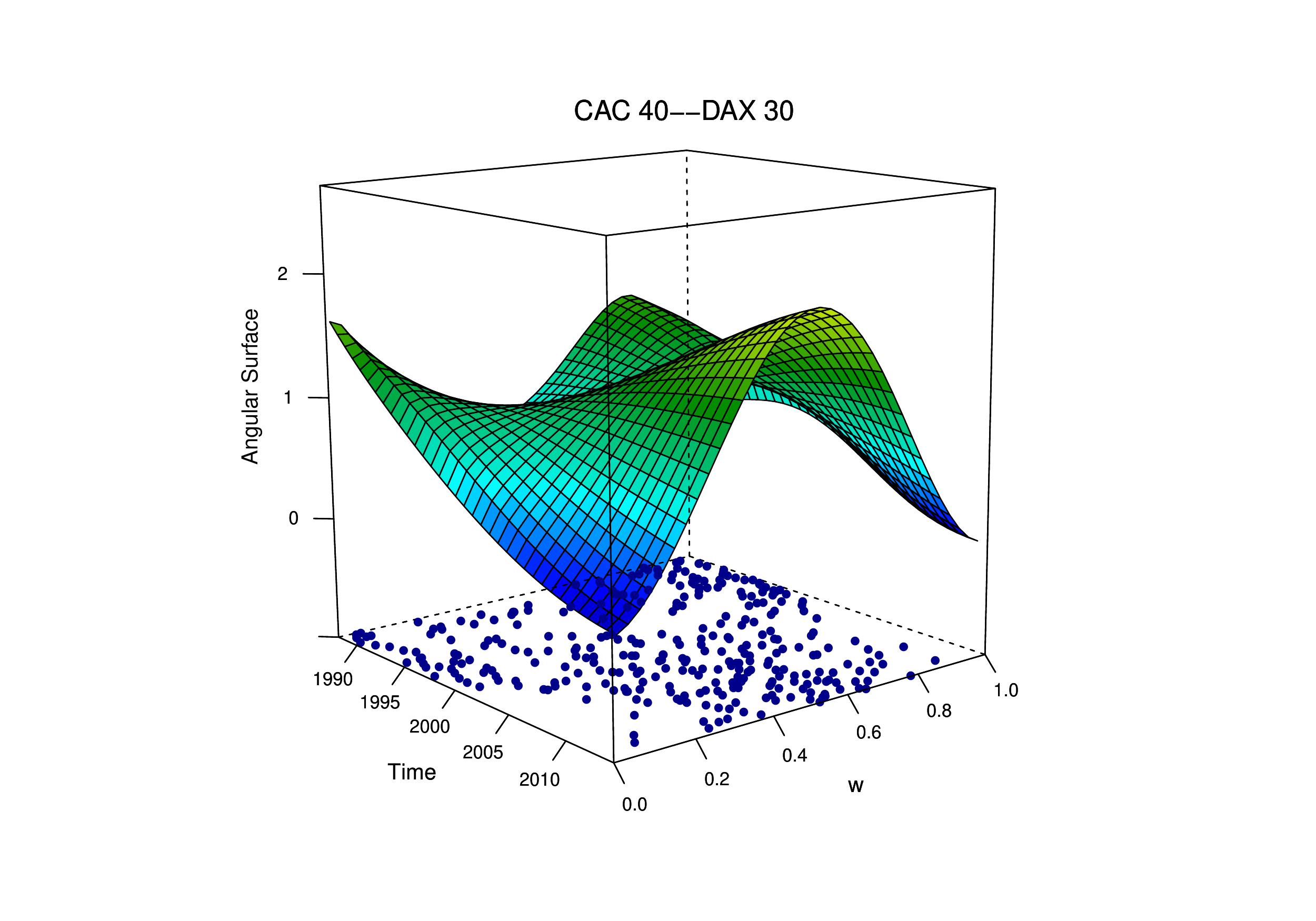}
 \end{minipage}&
\begin{minipage}[c]{0.4\linewidth}
\hspace{-3cm}
 \includegraphics[width=1.3\textwidth]{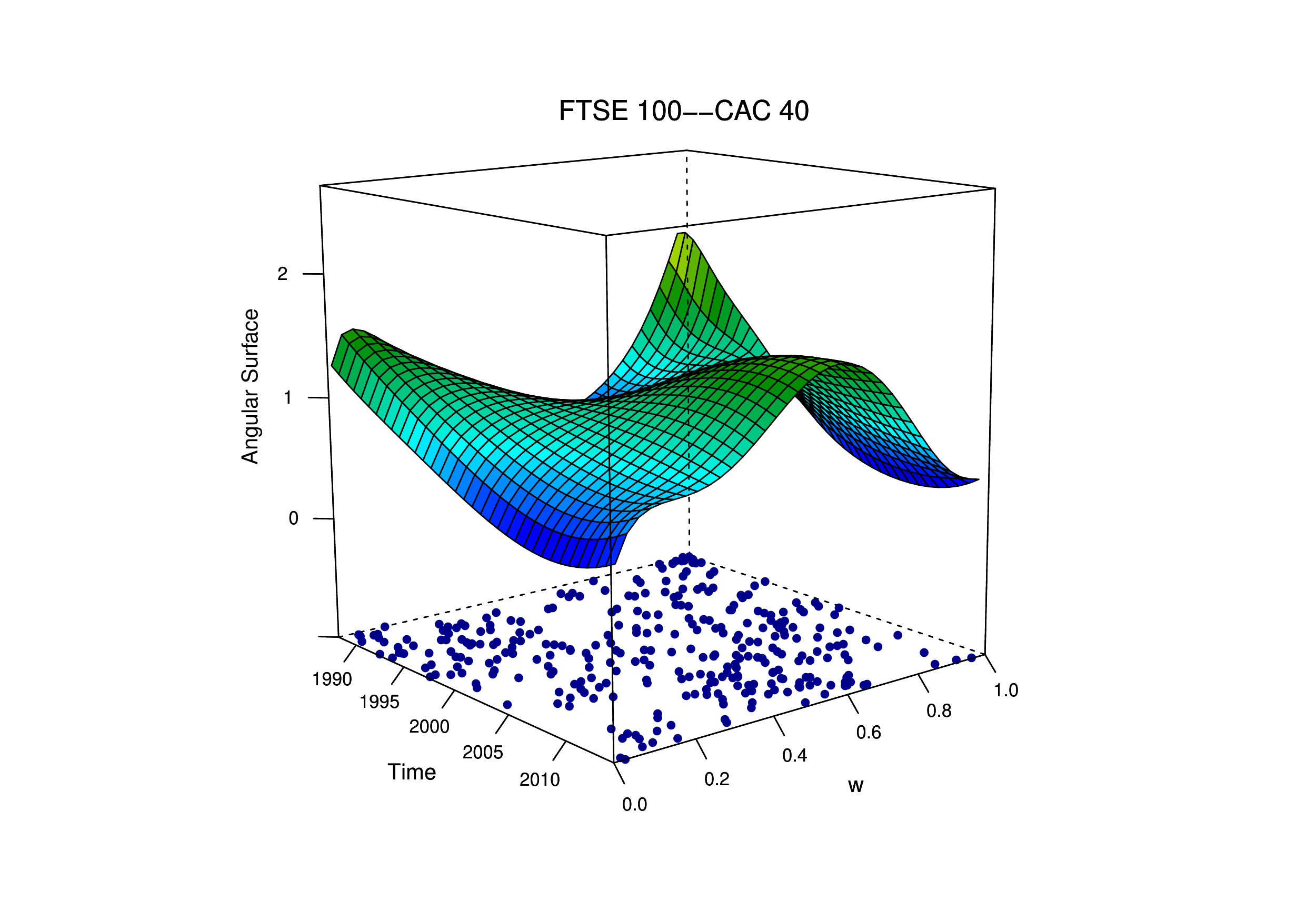}
\end{minipage}&
\begin{minipage}[c]{0.4\linewidth}
\hspace{-4.8cm}
 \includegraphics[width=1.3\textwidth]{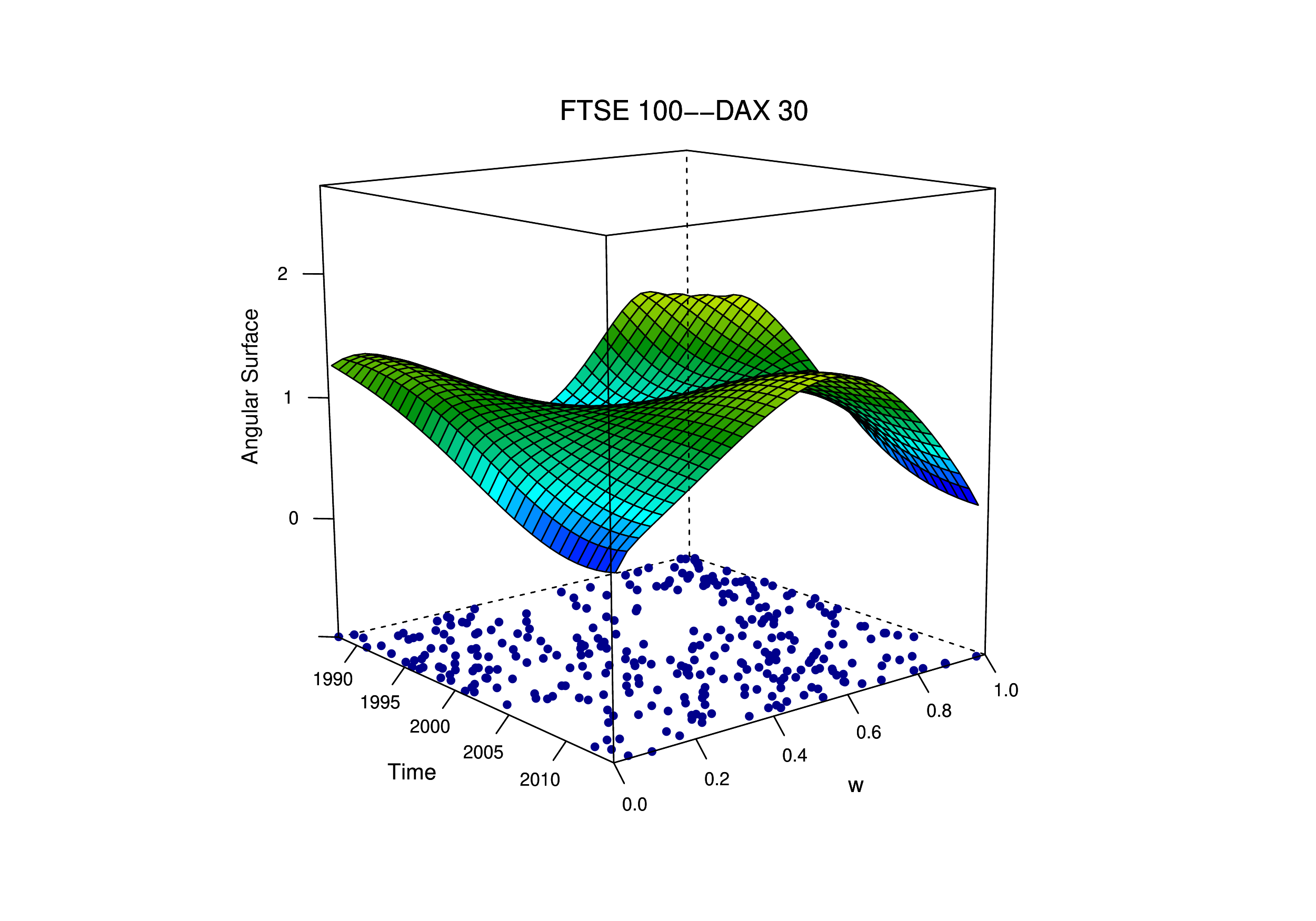}
\end{minipage}\\
 \end{tabular}
 \vspace{-.5cm}
 \caption{\footnotesize Angular surfaces estimates for CAC--DAX, FTSE--CAC and FTSE--DAX using Nadaraya--Watson (top) and local linear (bottom) weights, with pseudo-angles overlaid on the bottom of the box.}
 \label{CacDaxFtse.pdf}
\end{figure}

The time-varying color palette scatterplots in Figure~\ref{scatter.pdf} and the rolling window estimates in Figure~\ref{chi.pdf} provide evidence of nonstationary extremal dependence, but they are only exploratory. In this section we complete the analysis from Section~\ref{exploratory} by applying our conditional modeling approach to assess how the dependence structure of bivariate extreme losses in the three pairs has been evolving over recent years. Before we proceed any further, some comments regarding implementation are in order. As mentioned in Section~\ref{modeling}, the data were transformed to have standard Fr\'echet margins. This was done as follows. Given a sample of pairs of filtered residuals $(r_{1,1}, r_{1,2}), \ldots, (r_{N,1}, r_{N,2})$, we construct proxies for the unobservable pseudo-angles $W_i$ by setting
\begin{equation*}
  W_i = \widehat{Y}_{i,1} / ( \widehat{Y}_{i,1} + \widehat{Y}_{i,2} ), \quad
  R_i = \widehat{Y}_{i,1} + \widehat{Y}_{i,2},
\end{equation*}
where $\widehat{Y}_{i,1} = -1 / \log\{\widehat{F}_{r_1}(r_{i,1})\}$ and $\widehat{Y}_{i,2} = -1 / \log\{\widehat{F}_{r_2}(r_{i,2})\}$ and where $\widehat{F}_{r_1}$ and $\widehat{F}_{r_2}$ are estimates of the marginal distribution functions $F_{r_1}$ and $F_{r_2}$. A robust choice for $\widehat{F}_{r_1}$ and $\widehat{F}_{r_2}$ is the pair of univariate empirical distribution functions, normalized by $N+1$ rather than by $N$ to avoid division by zero. Following Section~\ref{dpa}, after fitting a spline--based nonparametric quantile regression we found evidence of dependence of the pseudo-radii $\{R_1,\ldots,R_N\}$ on time, and so we proceed under a nonstationary assumption. Specifically, we model the $95\%$ quantile of the pseudo-radii through nonparametric quantile regression and threshold the pseudo-radii according to the fit. The tail region to study the extreme losses is therefore defined through the pseudo-angles associated with the threshold exceedances of the pseudo-radii. After thresholding, the number of pseudo-angles is 312 for CAC--DAX and FTSE--CAC and 314 for FTSE--DAX. The pseudo-angles corresponding to these observations are plotted in the two-dimensional bottom plane in Figure~\ref{CacDaxFtse.pdf}. The tuning parameters $(b, \nu, \tau)$ were computed as discussed in Sections~\ref{tuning} and \ref{Finite sample performance}.

\begin{figure}
  \begin{center}
    \begin{footnotesize}
      {\textbf{CAC 40--DAX 30}}
    \end{footnotesize}
  \end{center}
  \vspace{-0.2 cm}
\footnotesize \rotatebox{90}{\textbf{\hspace{-1cm}N--W weights}}
  \begin{minipage}[c]{0.4\linewidth}
    \hspace{-.1cm}
    \includegraphics[scale=\sizeA]{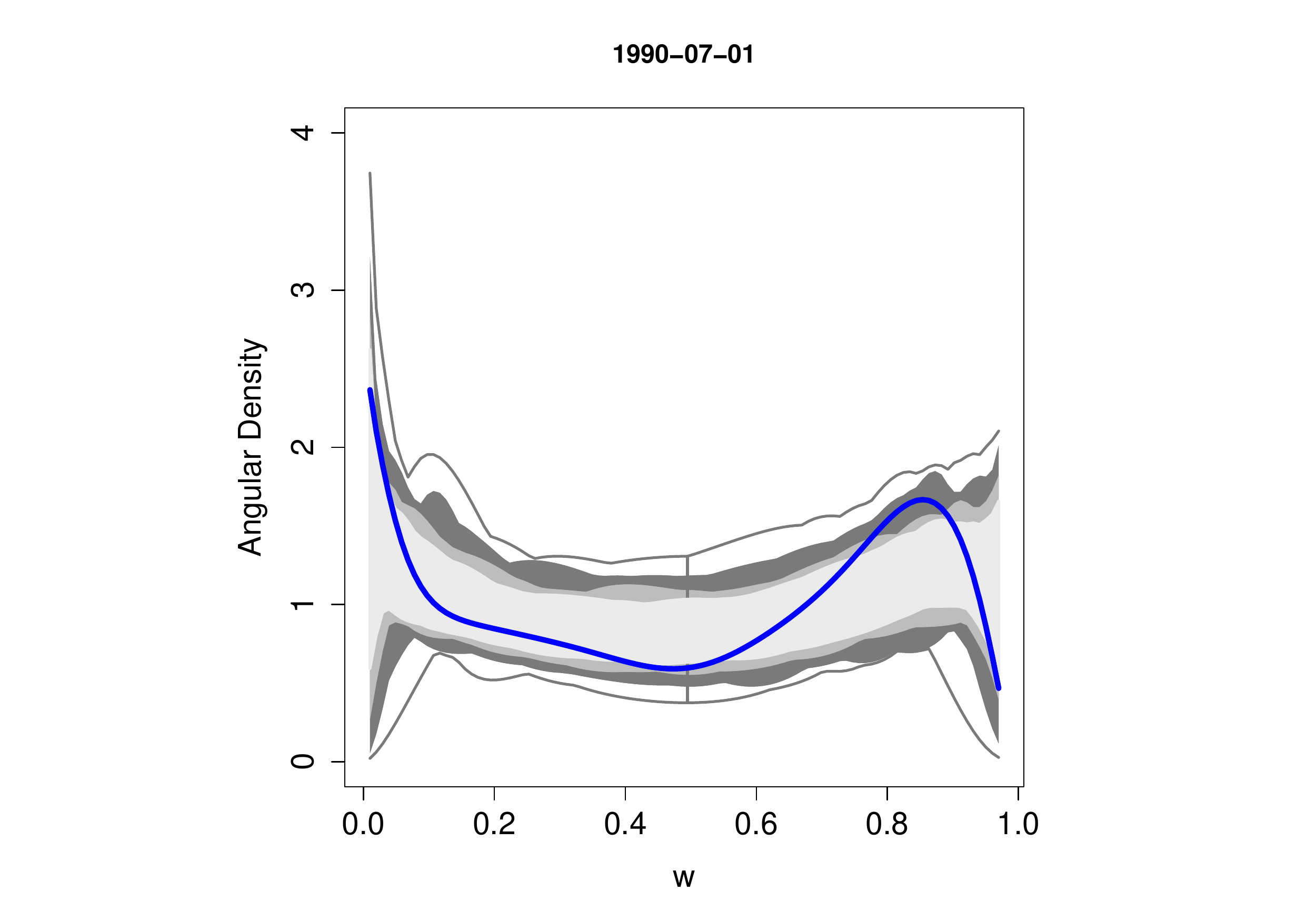}
  \end{minipage}%
  \hspace{-2cm}
  \begin{minipage}[c]{0.4\linewidth}
   \hspace{.2cm}
    \includegraphics[scale=\sizeA]{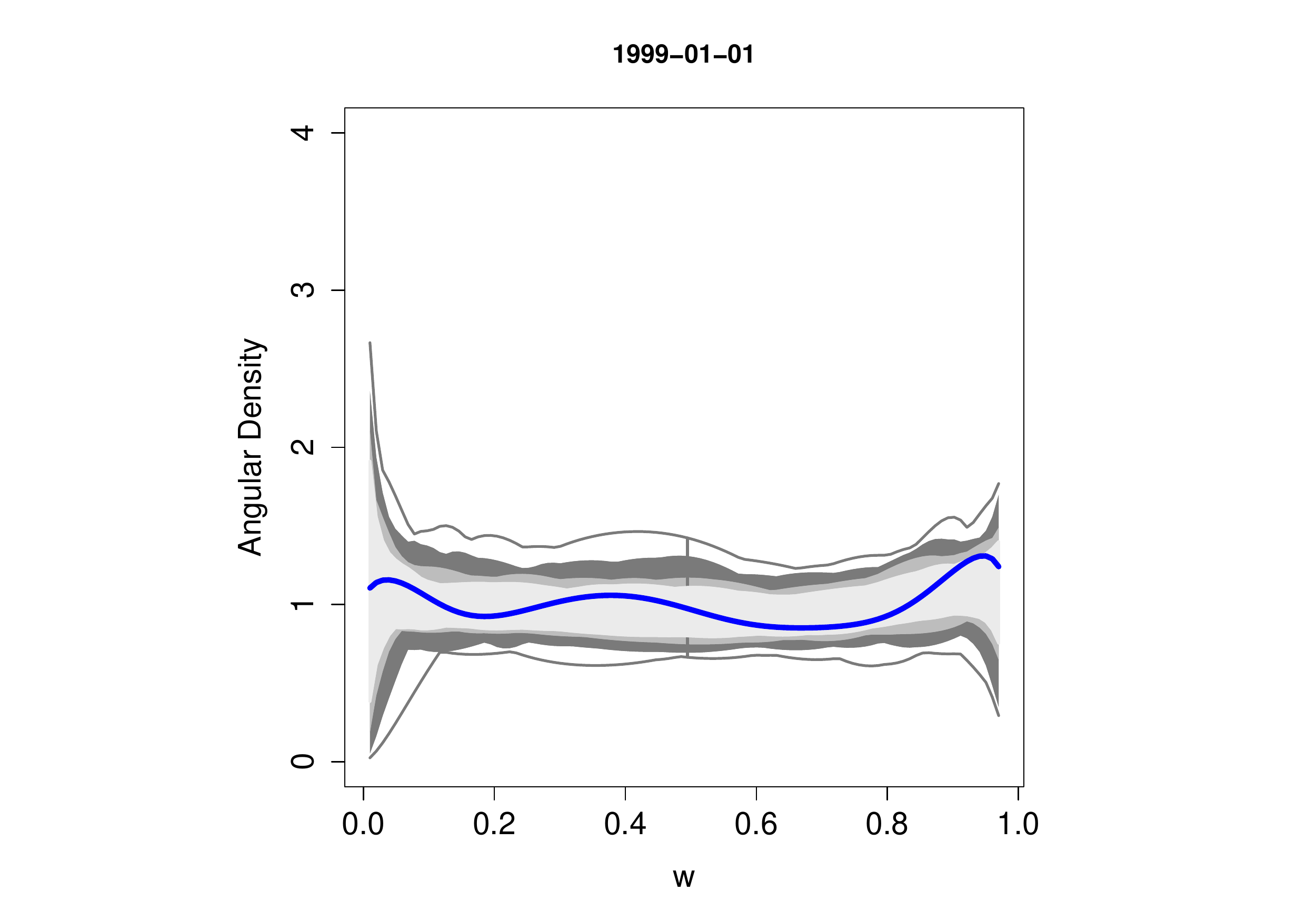}
  \end{minipage}%
  \hspace{-1.8cm}
  \begin{minipage}[c]{0.4\linewidth}
 \hspace{.2cm} 
    \includegraphics[scale=\sizeA]{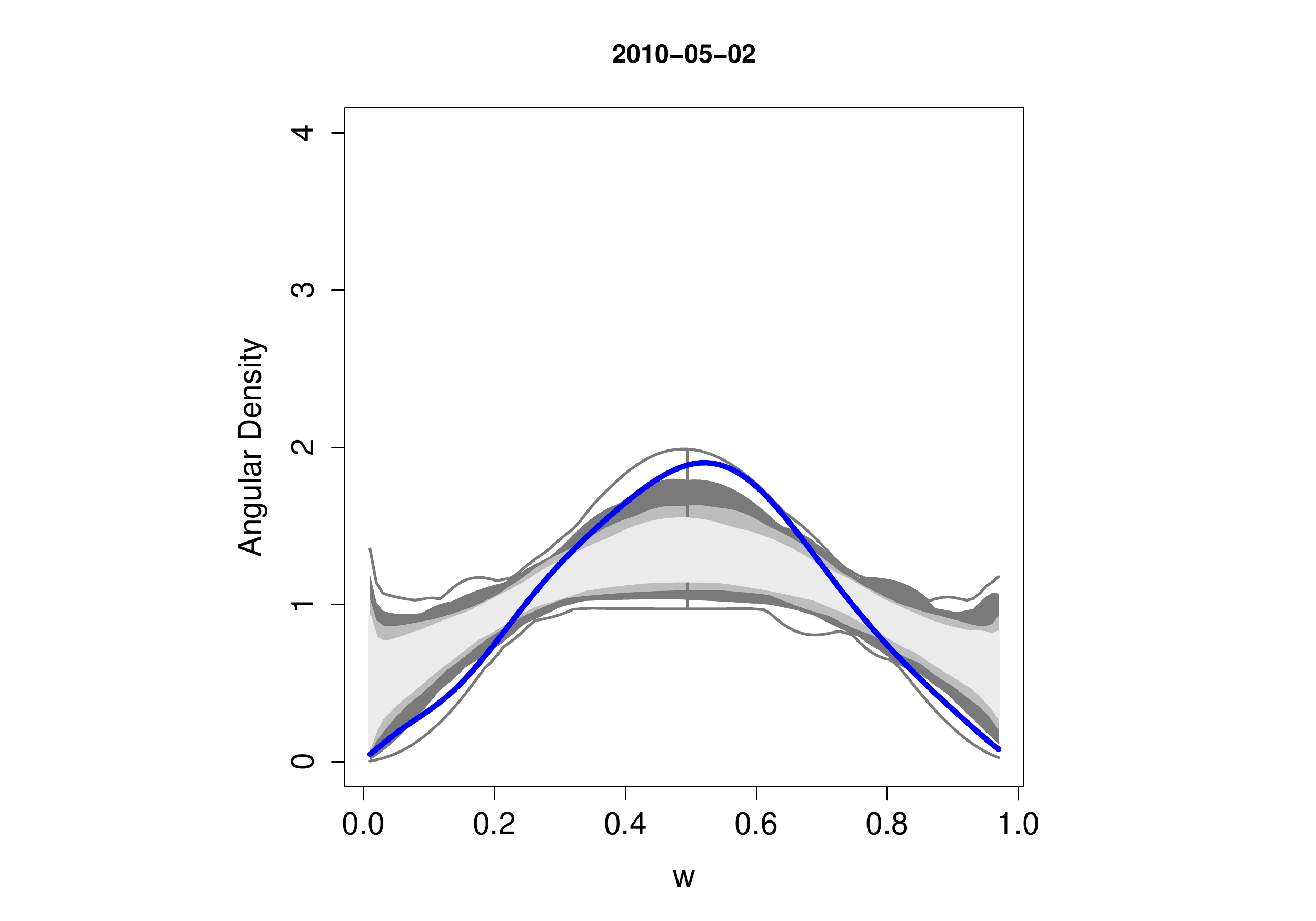}
  \end{minipage}%
  \\
    \vspace{-0.5 cm}
    \footnotesize \rotatebox{90}{\textbf{\hspace{-1cm}L--L weights}}
    \begin{minipage}[c]{0.4\linewidth}
    \hspace{-.1cm}
    \includegraphics[scale=\sizeA]{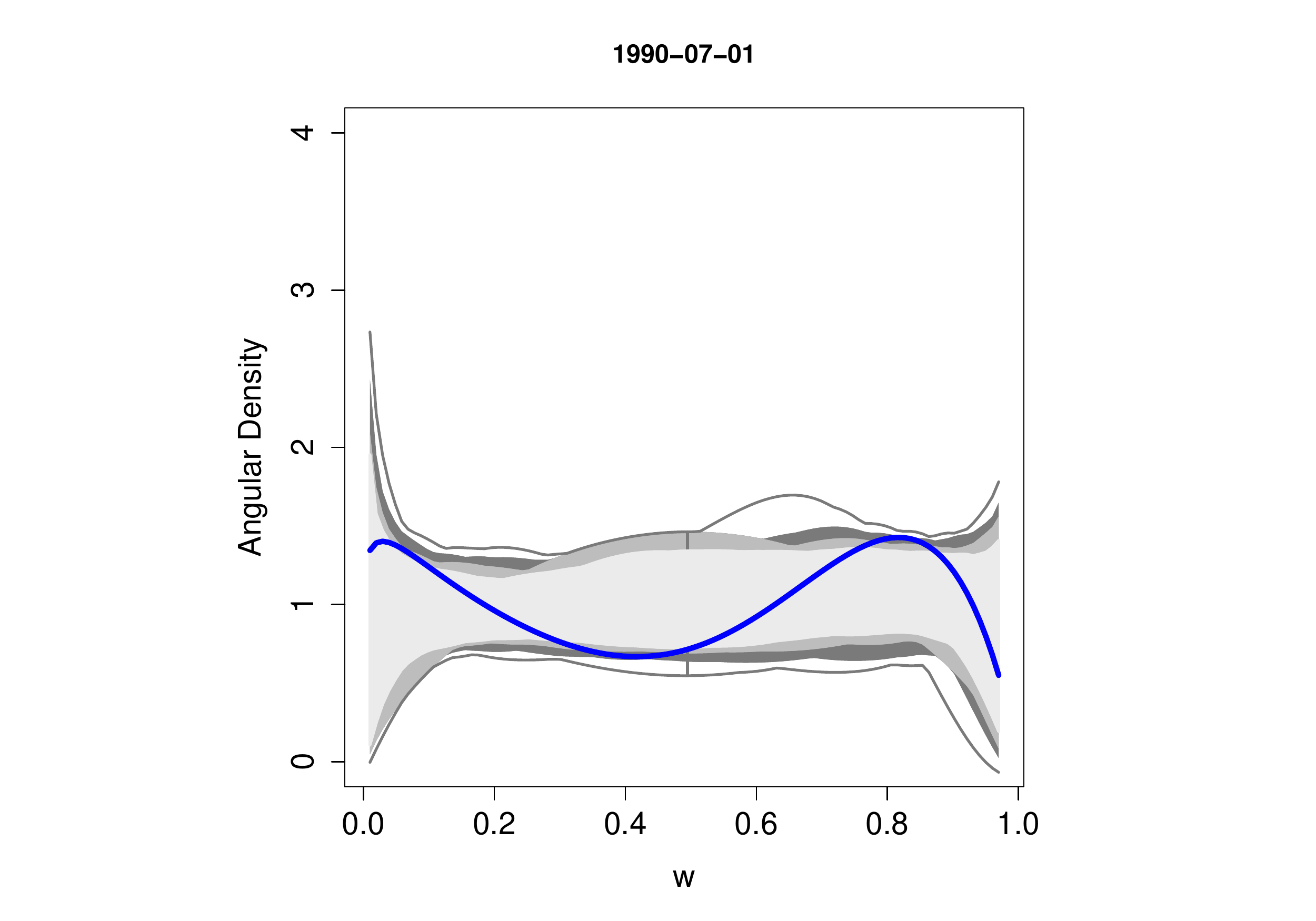}
  \end{minipage}%
  \hspace{-2cm}
  \begin{minipage}[c]{0.4\linewidth}
   \hspace{.2cm}
    \includegraphics[scale=\sizeA]{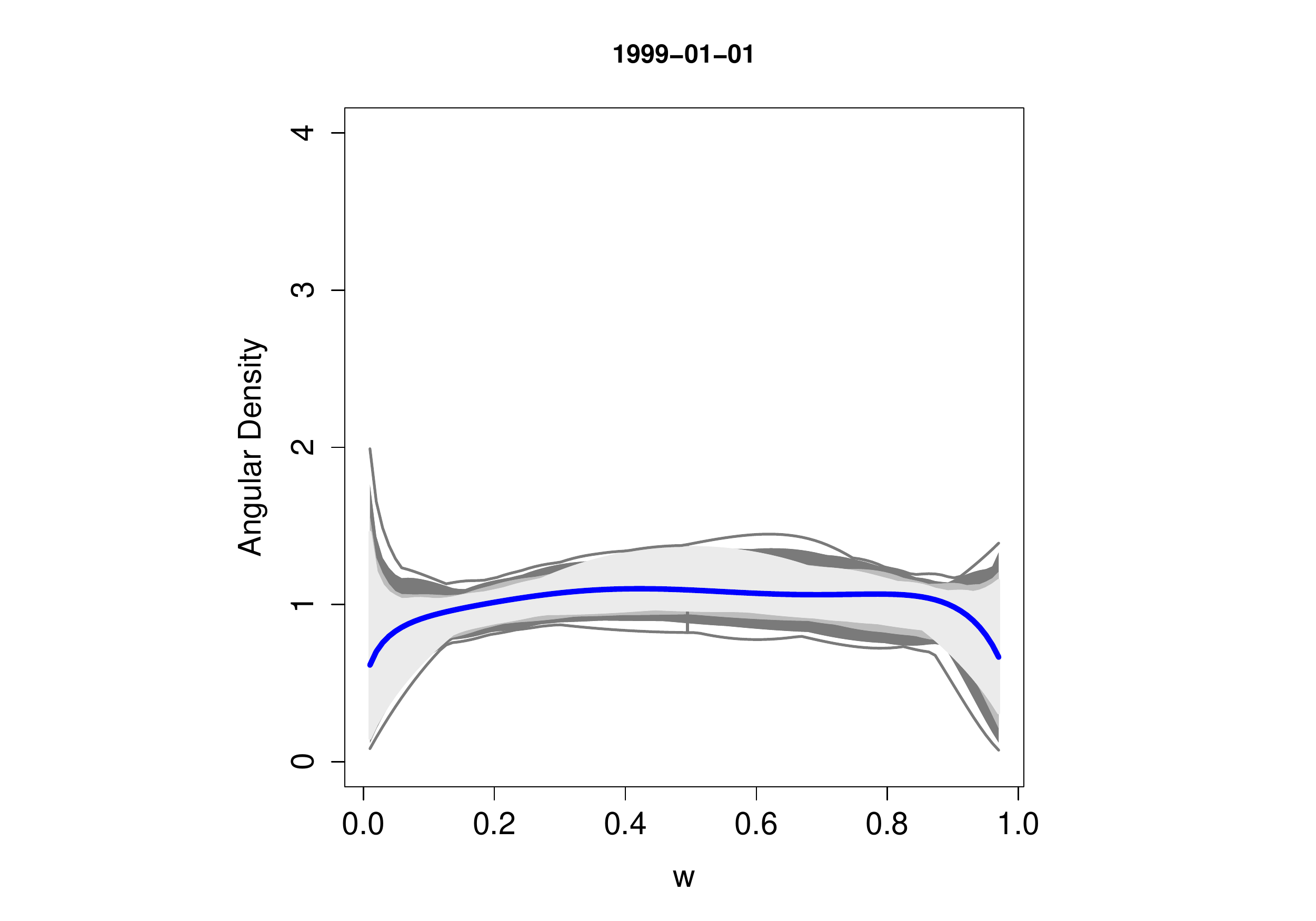}
  \end{minipage}%
  \hspace{-1.8cm}
  \begin{minipage}[c]{0.4\linewidth}
 \hspace{.2cm} 
    \includegraphics[scale=\sizeA]{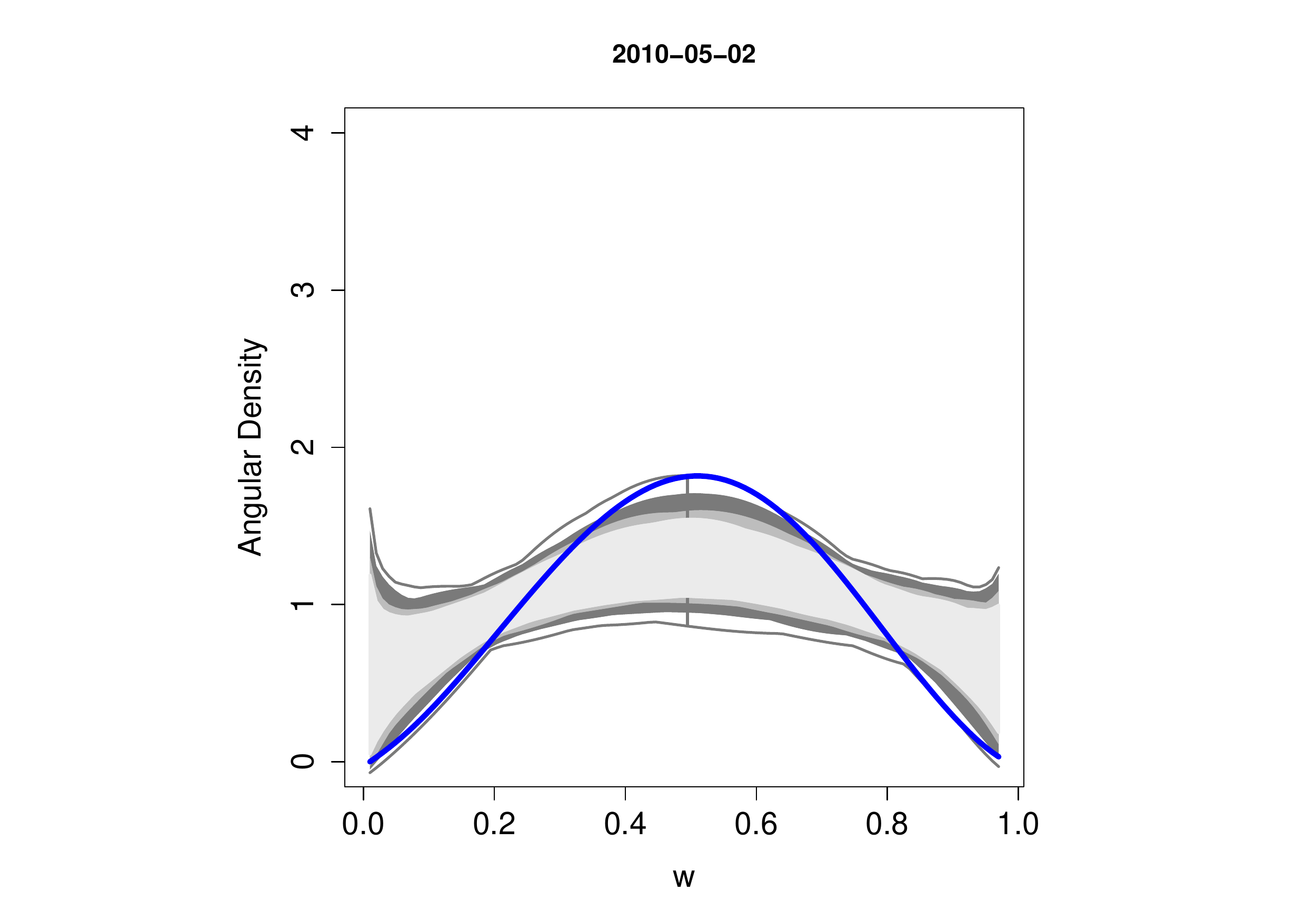}
  \end{minipage}%
  \\
  \begin{center}
    \begin{footnotesize}
      {\textbf{FTSE 100--CAC 40}}
    \end{footnotesize}
  \end{center}
  \vspace{-0.2 cm}
  \footnotesize \rotatebox{90}{\textbf{\hspace{-1cm}N--W weights}}
  \begin{minipage}[c]{0.4\linewidth}
    \hspace{-.1cm}
    \includegraphics[scale=\sizeA]{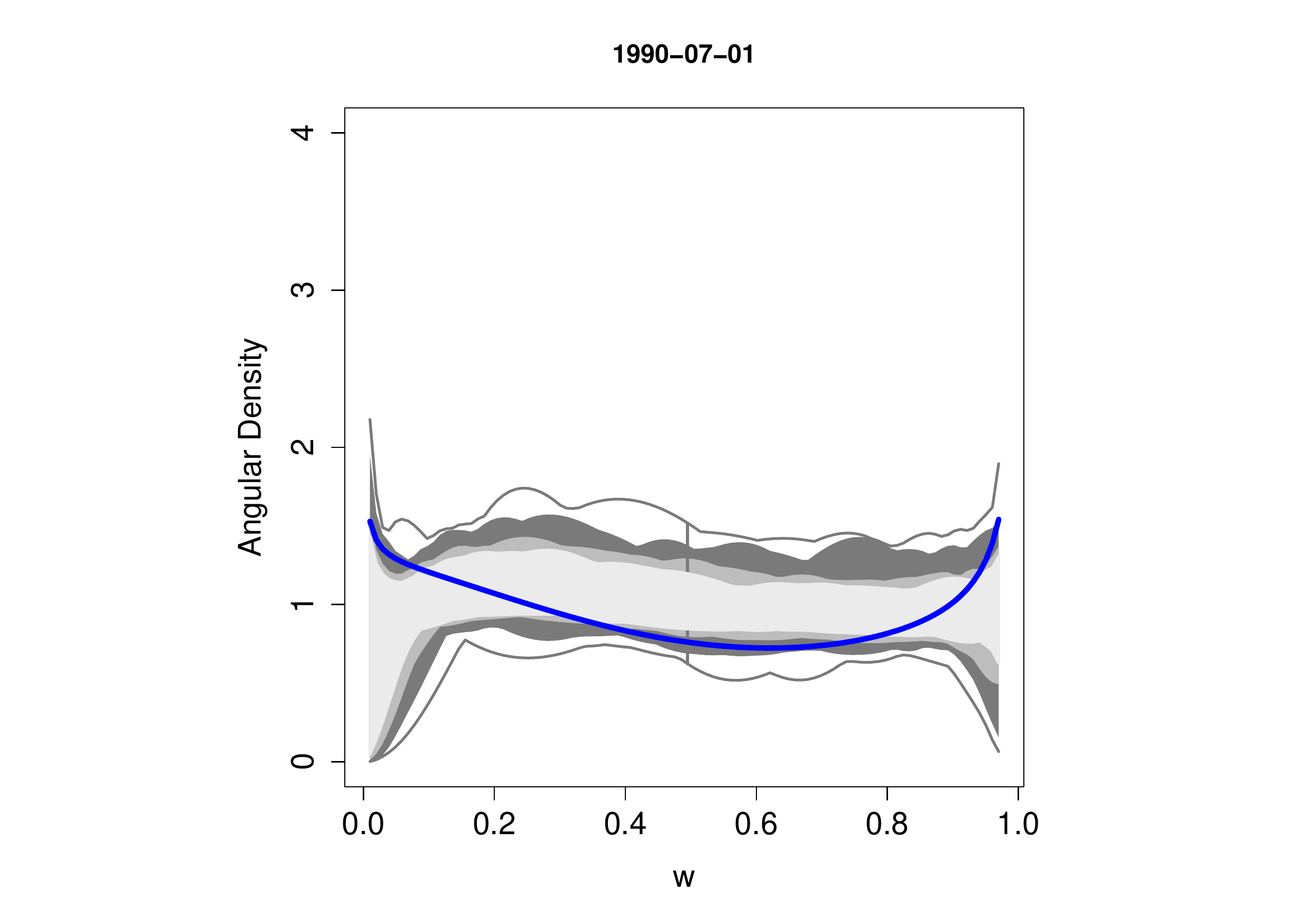}
  \end{minipage}%
  \hspace{-2cm}
  \begin{minipage}[c]{0.4\linewidth}
    \hspace{.2cm}
    \includegraphics[scale=\sizeA]{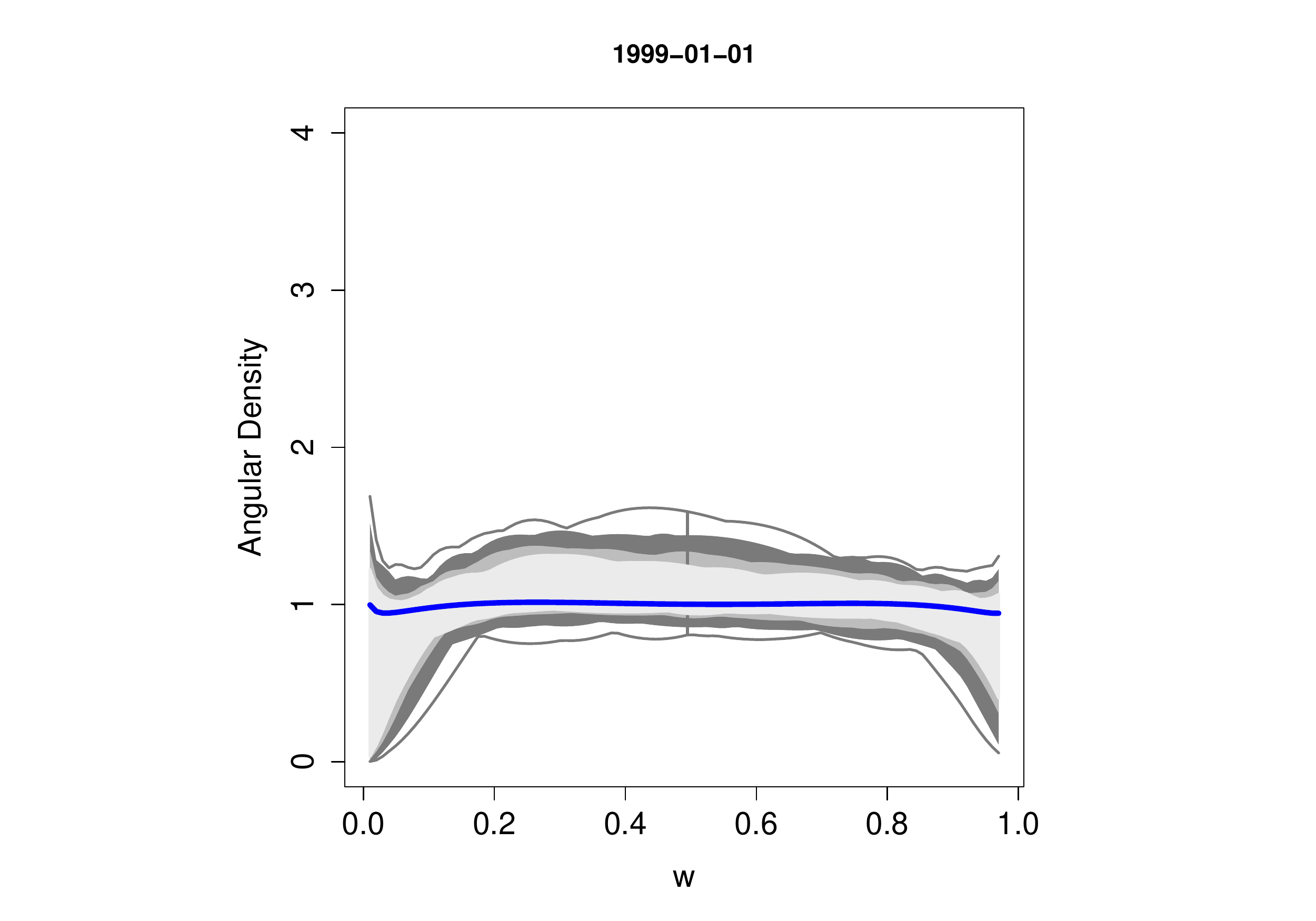}
  \end{minipage}%
  \hspace{-1.8cm}
  \begin{minipage}[c]{0.4\linewidth}
    \hspace{.2cm}
    \includegraphics[scale=\sizeA]{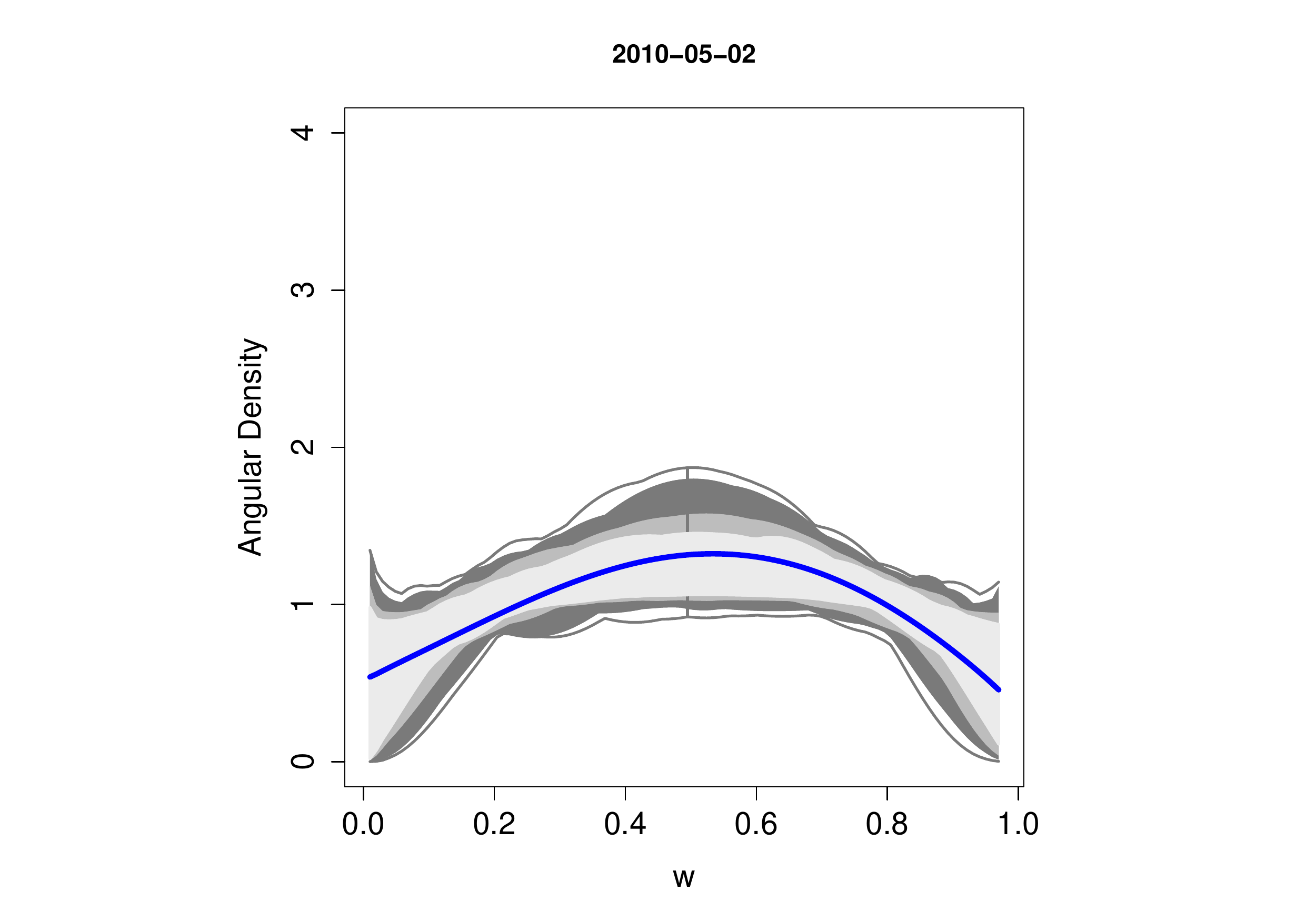}
  \end{minipage}%
  \\
  \vspace{-0.5 cm}
      \footnotesize \rotatebox{90}{\textbf{\hspace{-1cm}L--L weights}}
  \begin{minipage}[c]{0.4\linewidth}
    \hspace{-.1cm}
    \includegraphics[scale=\sizeA]{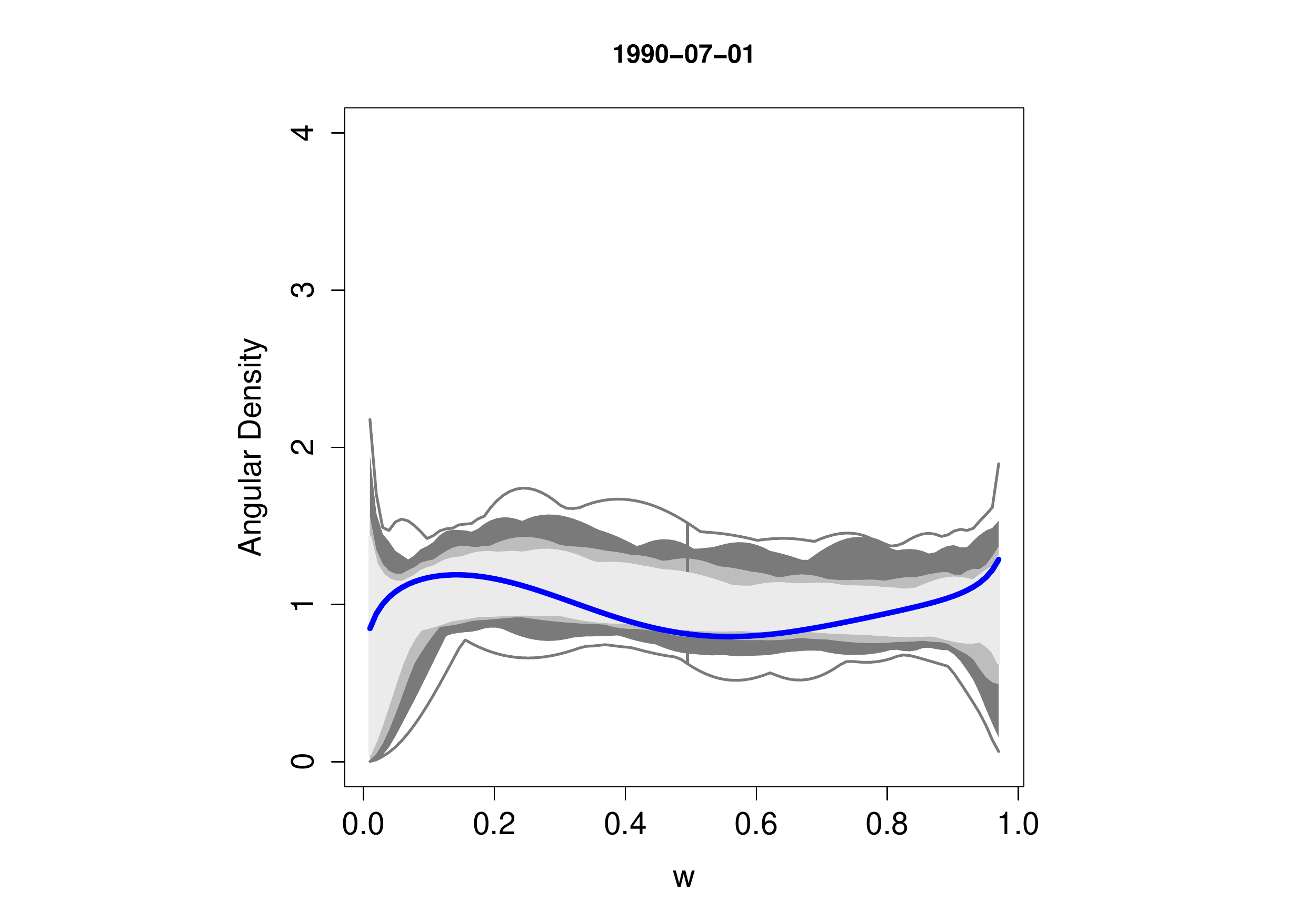}
  \end{minipage}%
  \hspace{-2cm}
  \begin{minipage}[c]{0.4\linewidth}
    \hspace{.2cm}
    \includegraphics[scale=\sizeA]{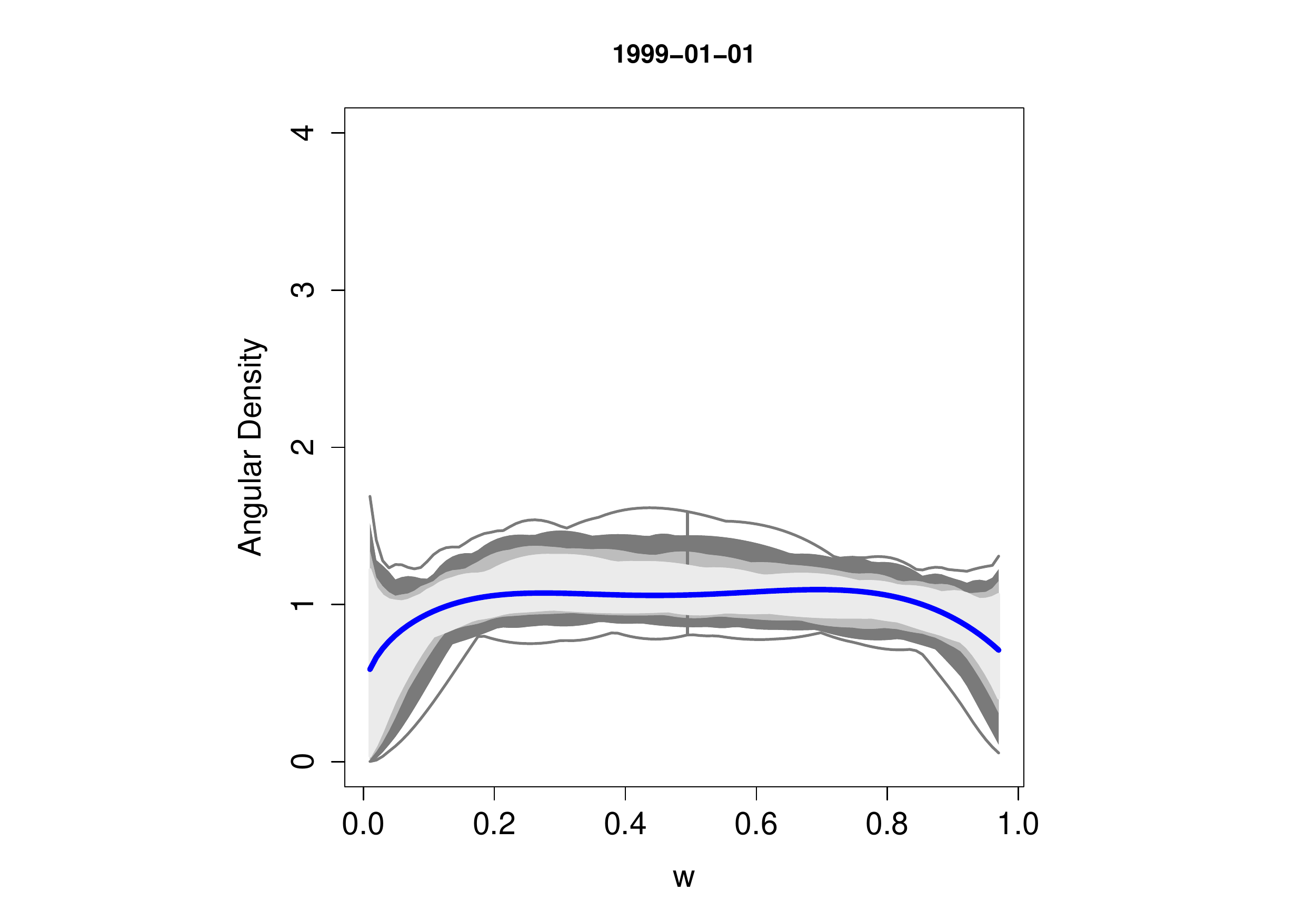}
  \end{minipage}%
  \hspace{-1.8cm}
  \begin{minipage}[c]{0.4\linewidth}
    \hspace{.2cm}
    \includegraphics[scale=\sizeA]{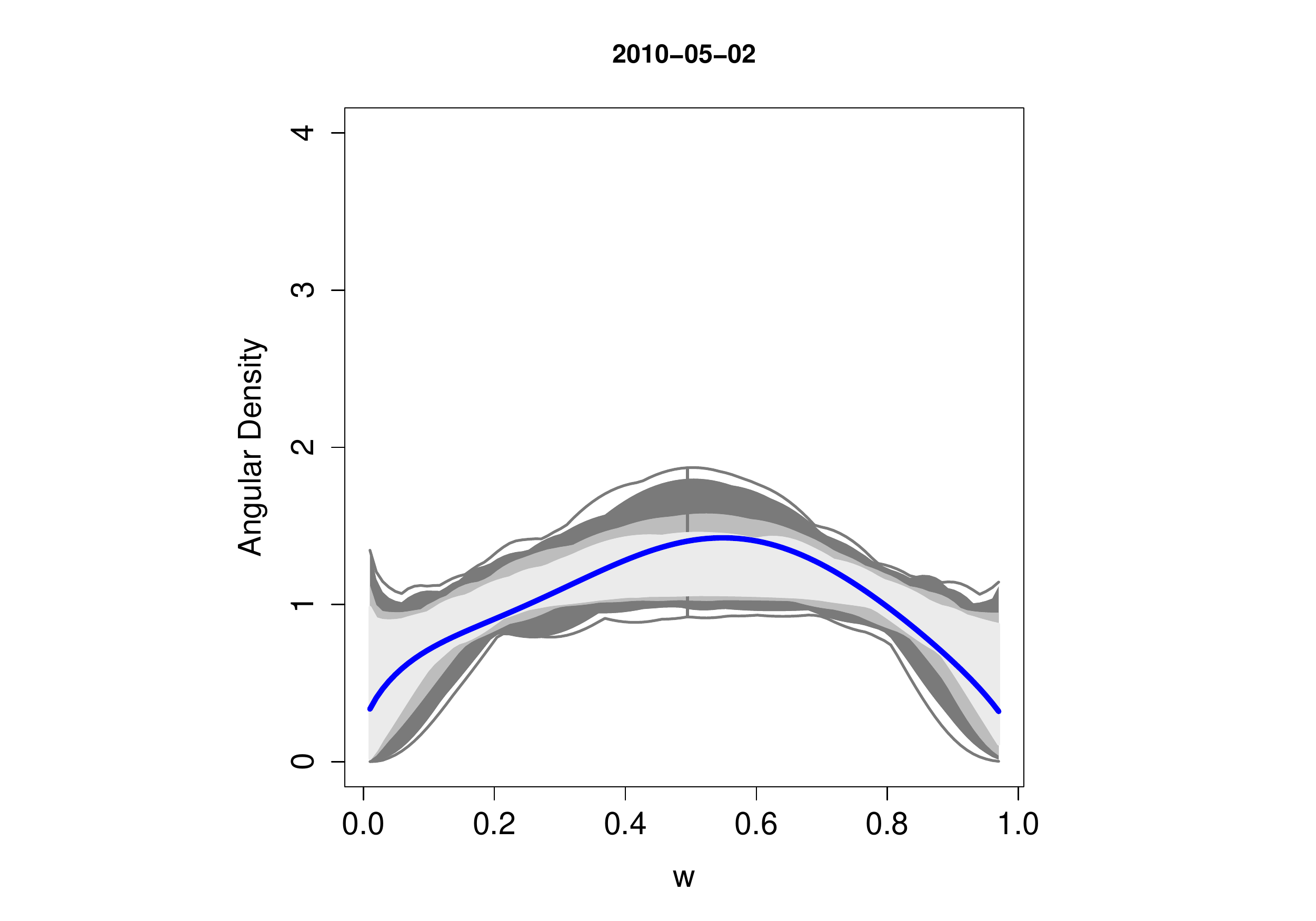}
  \end{minipage}%
  \\
  \begin{center}
    \begin{footnotesize}
      {\textbf{FTSE 100--DAX 30}}
    \end{footnotesize}
  \end{center}
  \vspace{-0.2cm}
    \footnotesize \rotatebox{90}{\textbf{\hspace{-1cm}N--W weights}}
  \begin{minipage}[c]{0.4\linewidth}
    \hspace{-.1cm}
    \includegraphics[scale=\sizeA]{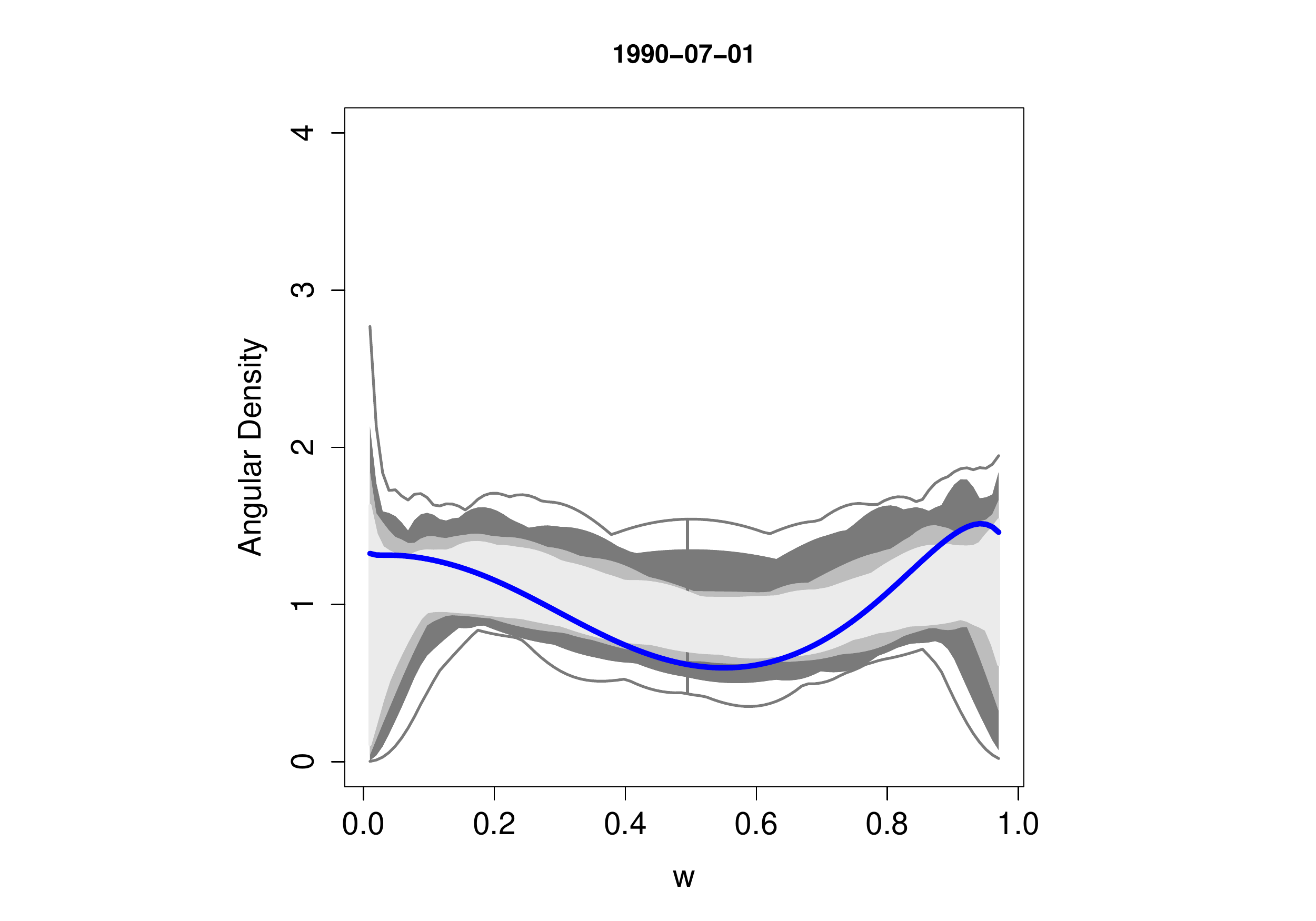}
  \end{minipage}%
  \hspace{-2cm}
  \begin{minipage}[c]{0.4\linewidth}
    \hspace{.2cm}
    \includegraphics[scale=\sizeA]{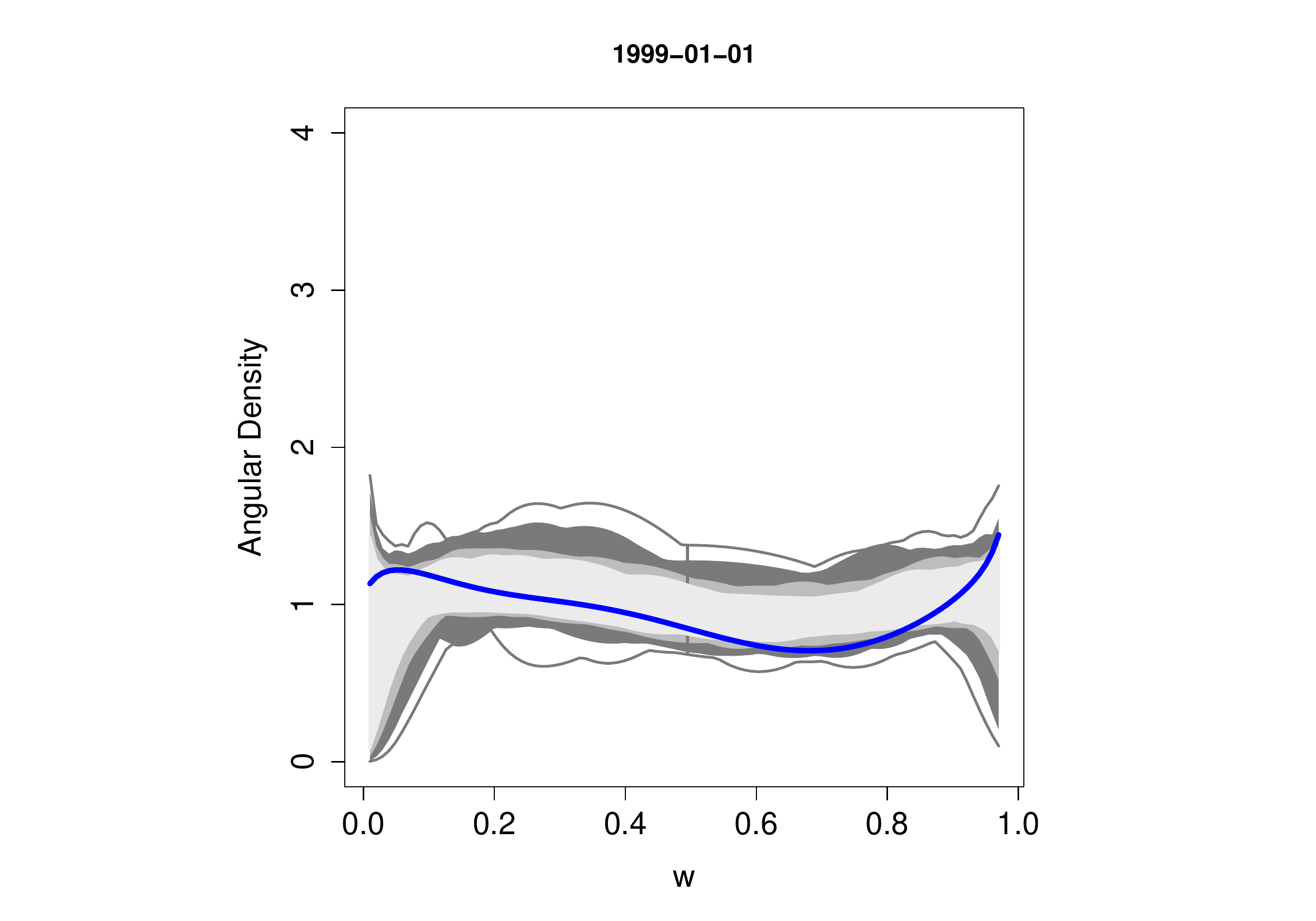}
  \end{minipage}%
  \hspace{-1.8cm}
  \begin{minipage}[c]{0.4\linewidth}
    \hspace{.2cm}
    \includegraphics[scale=\sizeA]{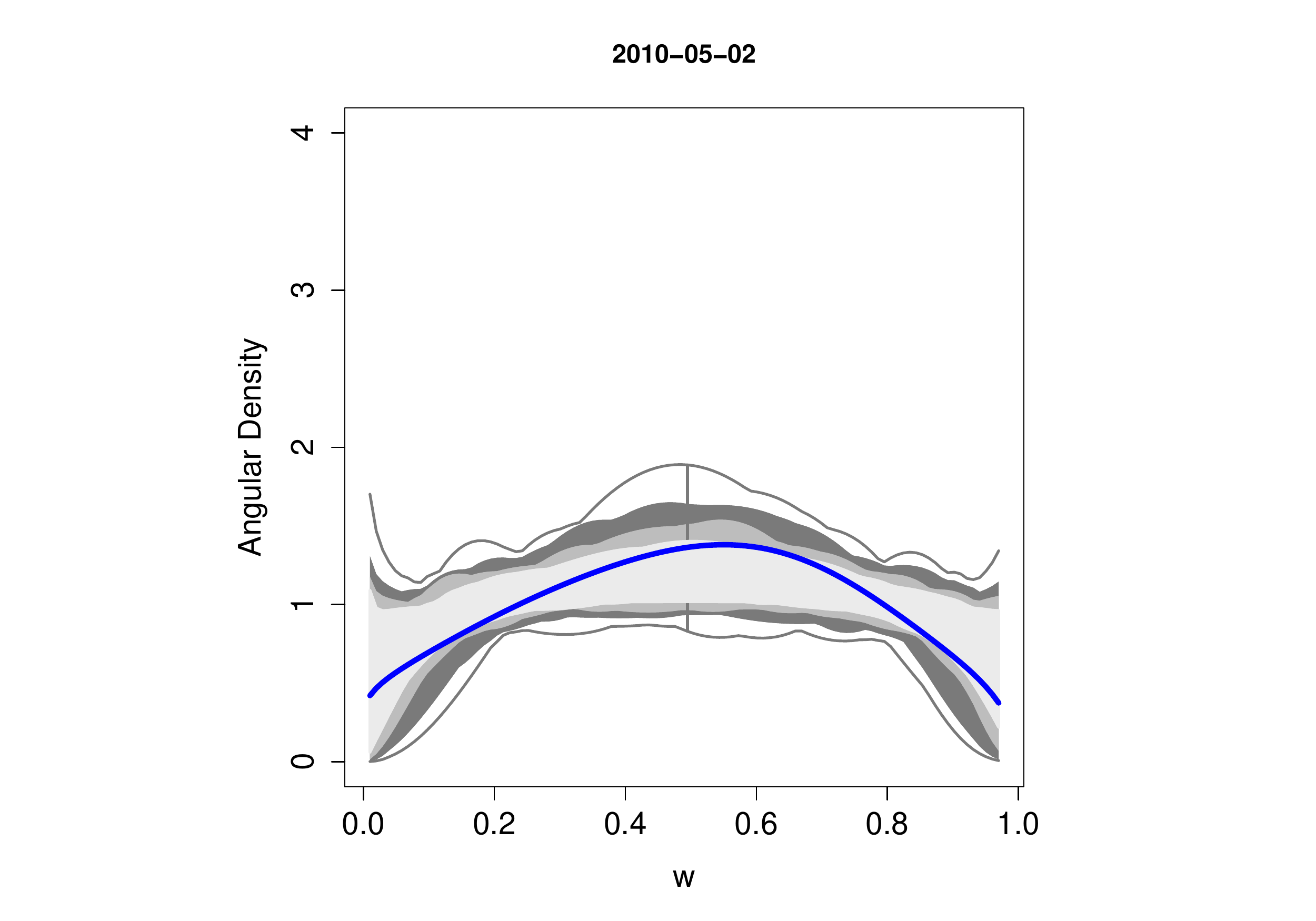}
  \end{minipage}%
  \\
  \vspace{-0.5 cm}
      \footnotesize \rotatebox{90}{\textbf{\hspace{-1cm}L--L weights}}
  \begin{minipage}[c]{0.4\linewidth}
    \hspace{-.1cm}
    \includegraphics[scale=\sizeA]{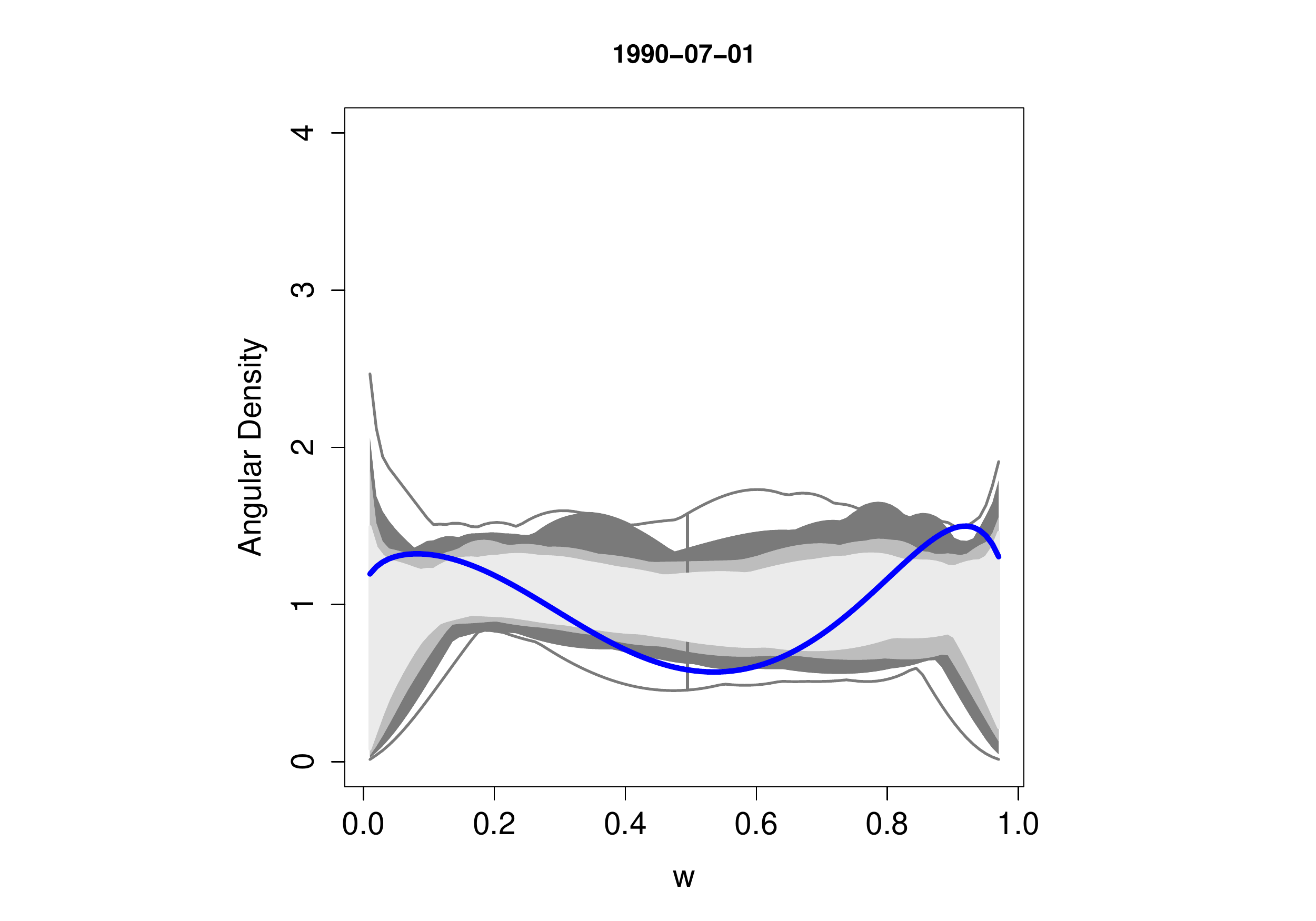}
  \end{minipage}%
  \hspace{-2cm}
  \begin{minipage}[c]{0.4\linewidth}
    \hspace{.2cm}
    \includegraphics[scale=\sizeA]{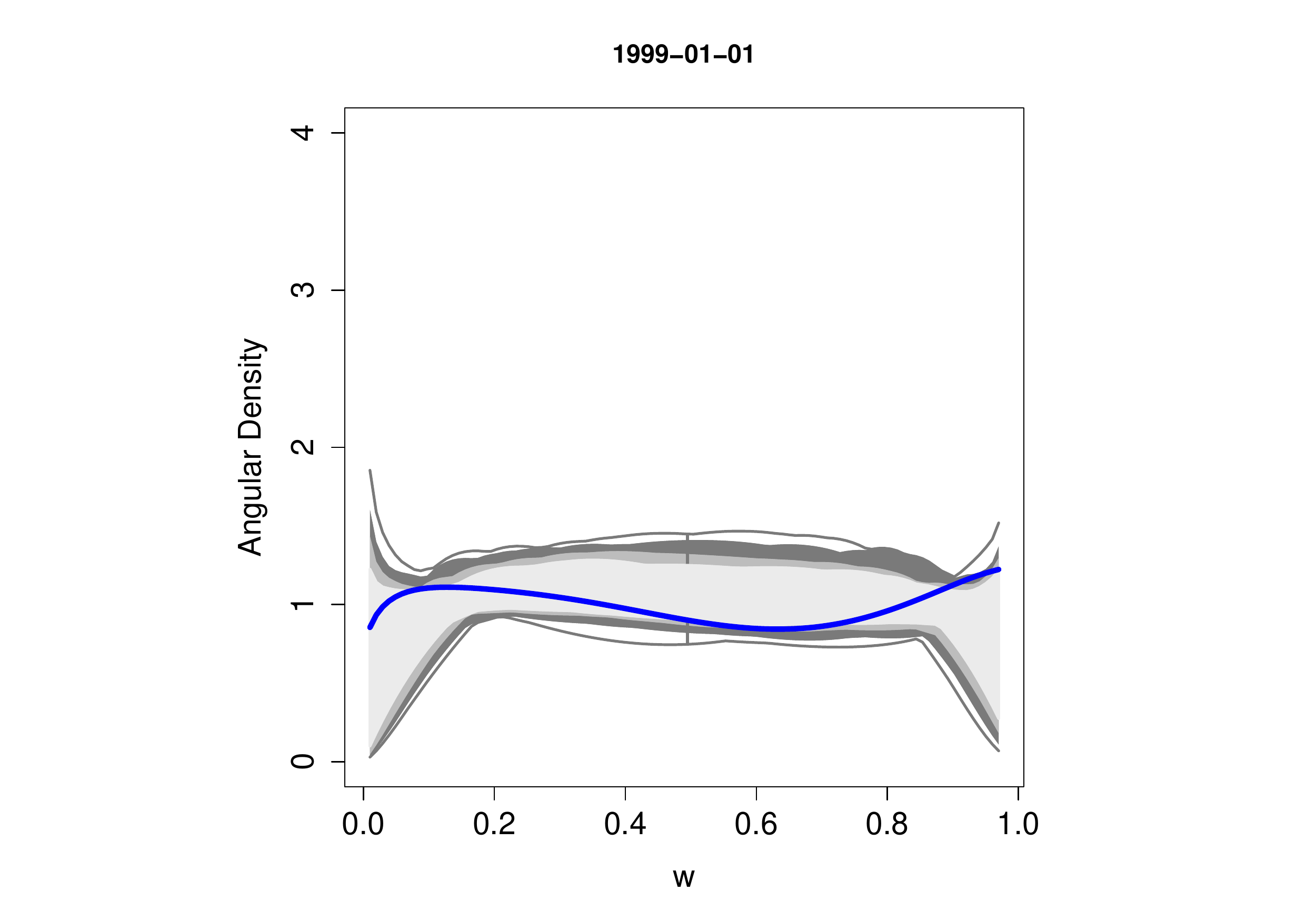}
  \end{minipage}%
  \hspace{-1.8cm}
  \begin{minipage}[c]{0.4\linewidth}
    \hspace{.2cm}
    \includegraphics[scale=\sizeA]{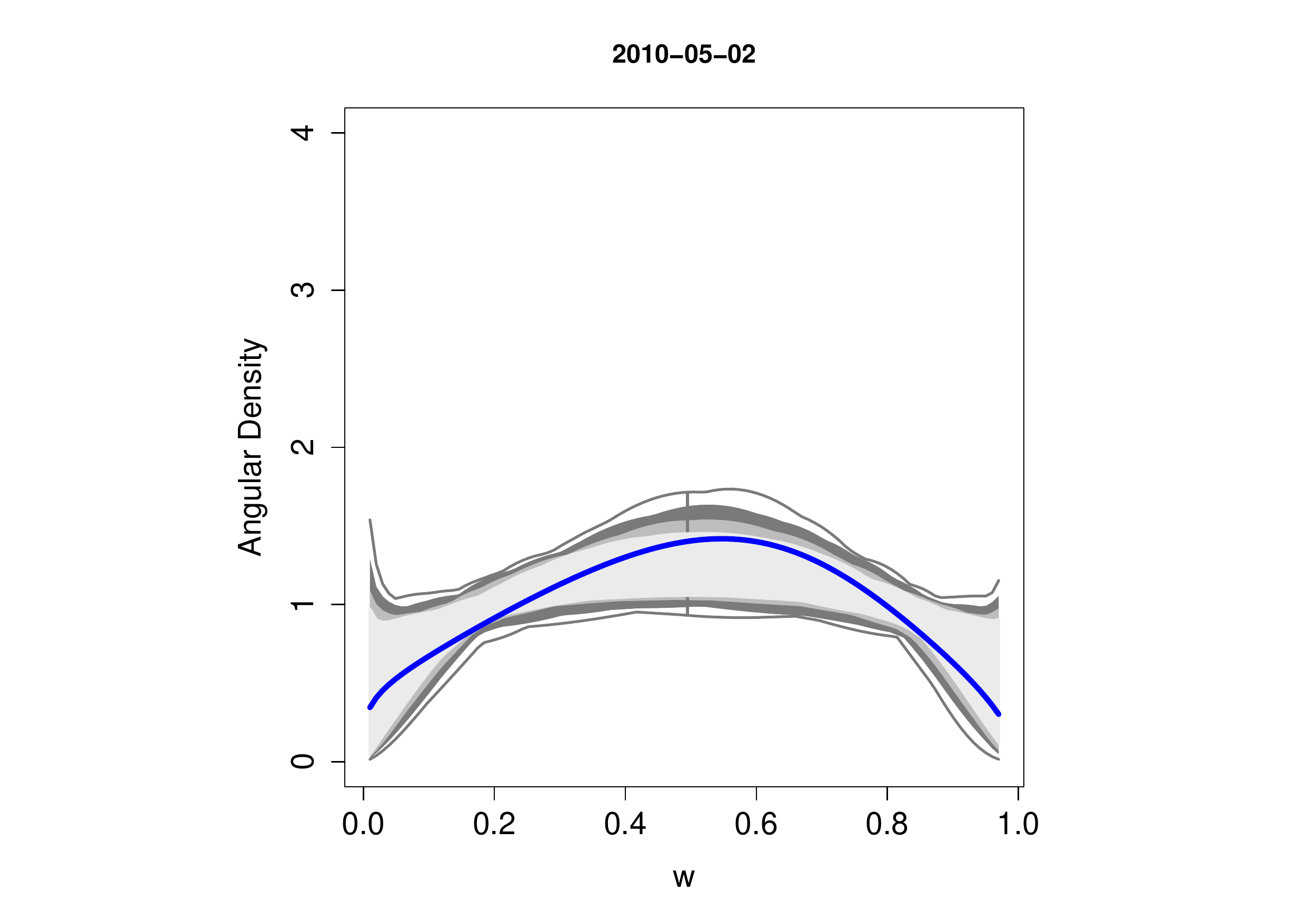}
  \end{minipage}%
 \caption{\footnotesize {Cross sections of angular surface estimates for CAC--DAX (top), FTSE--CAC (center), and FTSE--DAX (bottom) for Nadaraya--Watson and local linear weights (solid blue lines). The first column corresponds to the beginning of stage one of EMU (1 July, 1990), the second column corresponds to the beginning of stage three of EMU (1 January, 1999), and the third column corresponds to the time of activation of the assistance package for Greece (2 May, 2010). Functional boxplots (gray shadows) show the 50\%, 75\%, and 95\% central regions (as defined by~\cite{sun2012functional}) based on 1000 bootstrap samples.}}
  \label{obs-dens.pdf}
\end{figure}
In Figure~\ref{obs-dens.pdf} we plot cross sections of the angular surface estimate, using both Nadaraya--Watson and local linear weights as described in Section~\ref{Estimation and inference}, at three important periods on the EU agenda: \textsc{i}) Beginning of stage one of EMU (1 July, 1990); \textsc{ii}) beginning of stage three of EMU (1 January, 1999); \textsc{iii}) activation of the assistance package for Greece (2 May, 2010), the first country to be shut out of the bond market, which fostered the European sovereign debt crisis \citep{L12}. The choice of landmarks \textsc{i}--\textsc{iii}) is arbitrary, but recall that our main interest is in describing how extremal dependence may change, by comparing periods sufficiently apart in time. 
As can be observed from the first column in Figure~\ref{obs-dens.pdf}, at around 1990 the dependence between extreme losses for the three pairs were similar, exhibiting some evidence of extremal independence, that is also reflected in Figure~\ref{chi.pdf}. The second column in Figure~\ref{obs-dens.pdf} reveals that about a decade later this dynamic changed, and that extreme losses started to show some mild signs of extremal dependence. These signs become stronger, and 11 years later (third column in Figure~\ref{obs-dens.pdf}) we can clearly see evidence of extremal dependence of joint losses. Our findings may seem to contradict \cite{HAL06}---who claimed that the UK showed no increase in stock market integration---however we note that \cite{HAL06} did not assess extremal dependence. {The functional boxplots in Figure~\ref{obs-dens.pdf} were obtained following the bootstrap procedure detailed in Section~\ref{tuning}, with $B = 1000$ samples. We can clearly see some differences between the two types of estimators among the three pairs, but overall they report similar information in terms of the extremal dependence.}

Figure~\ref{obs-dens.pdf} provides only a few snapshots corresponding to landmarks \textsc{i}--\textsc{iii}). A more complete portrait of the temporal changes in extremal dependence is provided by the angular surface estimate in Figure~\ref{CacDaxFtse.pdf}, from which the cross-sections in Figure~\ref{obs-dens.pdf} are derived.
 
All in all, we can clearly see the change from weaker dependence around 1990 to strong dependence starting from 2005, thus suggesting that in recent decades there has been an increase in the extremal dependence in the losses for these leading European stock markets. The pair CAC--DAX is the one where extremal dependence peaks the most, thus suggesting a high level of synchronization and comovement of extreme losses in those markets over recent years.

\begin{figure}[h]
\begin{tabular}{ccc}
   \footnotesize \rotatebox{90}{\textbf{\hspace{-1cm}N--W weights}}
\begin{minipage}[c]{0.4\linewidth}
 \hspace{-1.3cm}
  \includegraphics[scale=0.275]{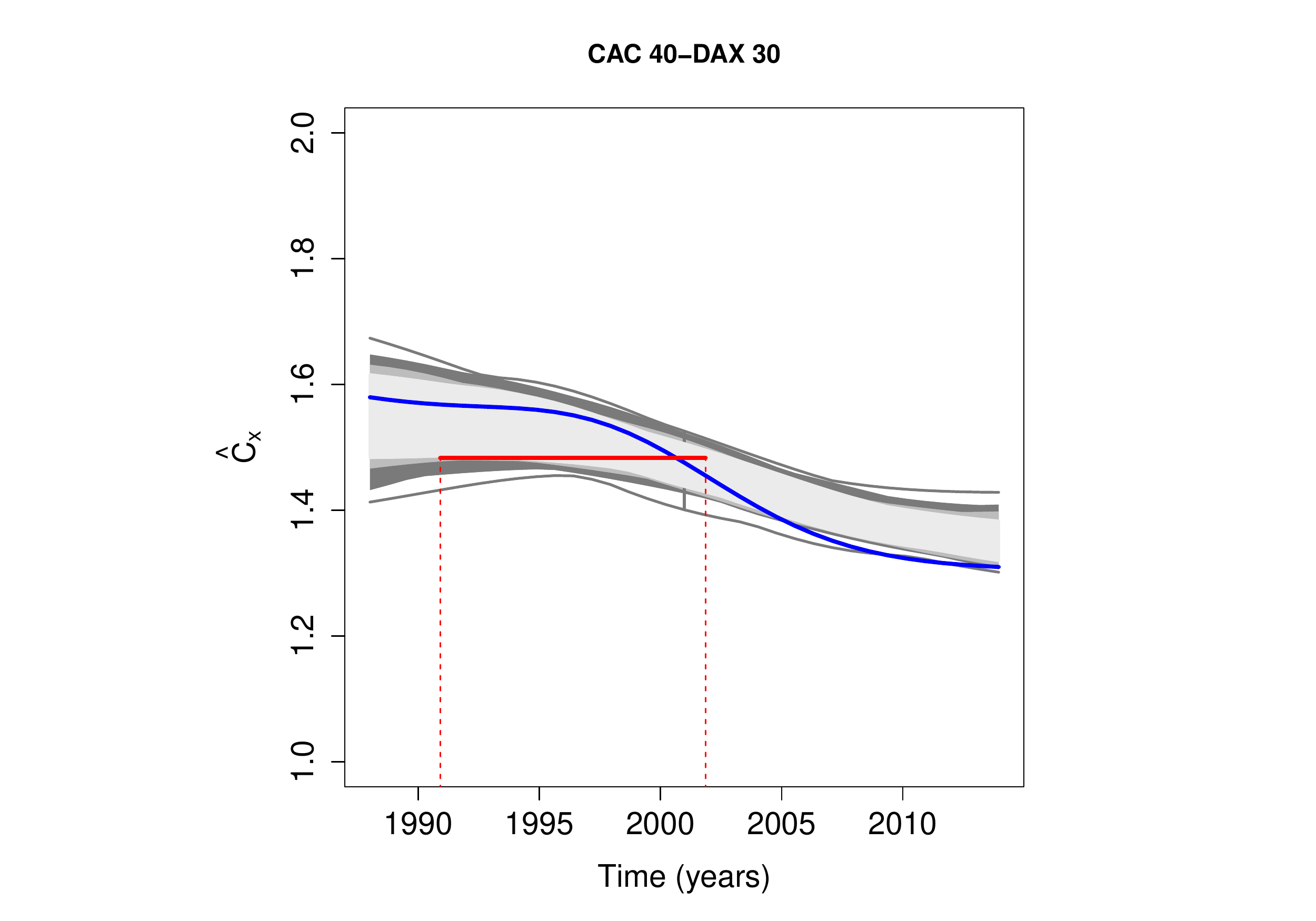}
 \end{minipage}&
\begin{minipage}[c]{0.4\linewidth}
\hspace{-2.8cm}
 \includegraphics[scale=0.275]{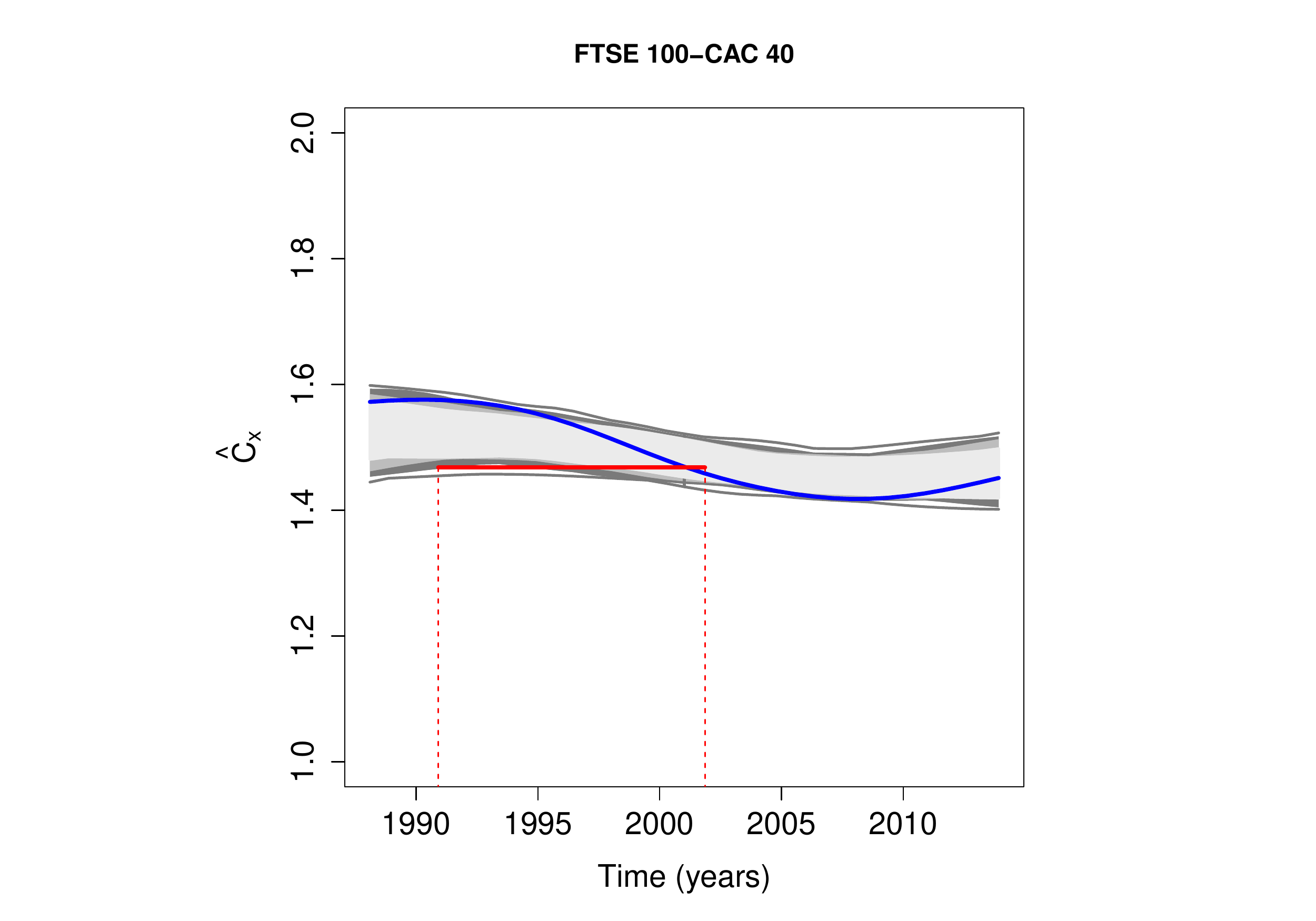}
\end{minipage}&
\begin{minipage}[c]{0.4\linewidth}
\hspace{-4.2cm}
 \includegraphics[scale=0.275]{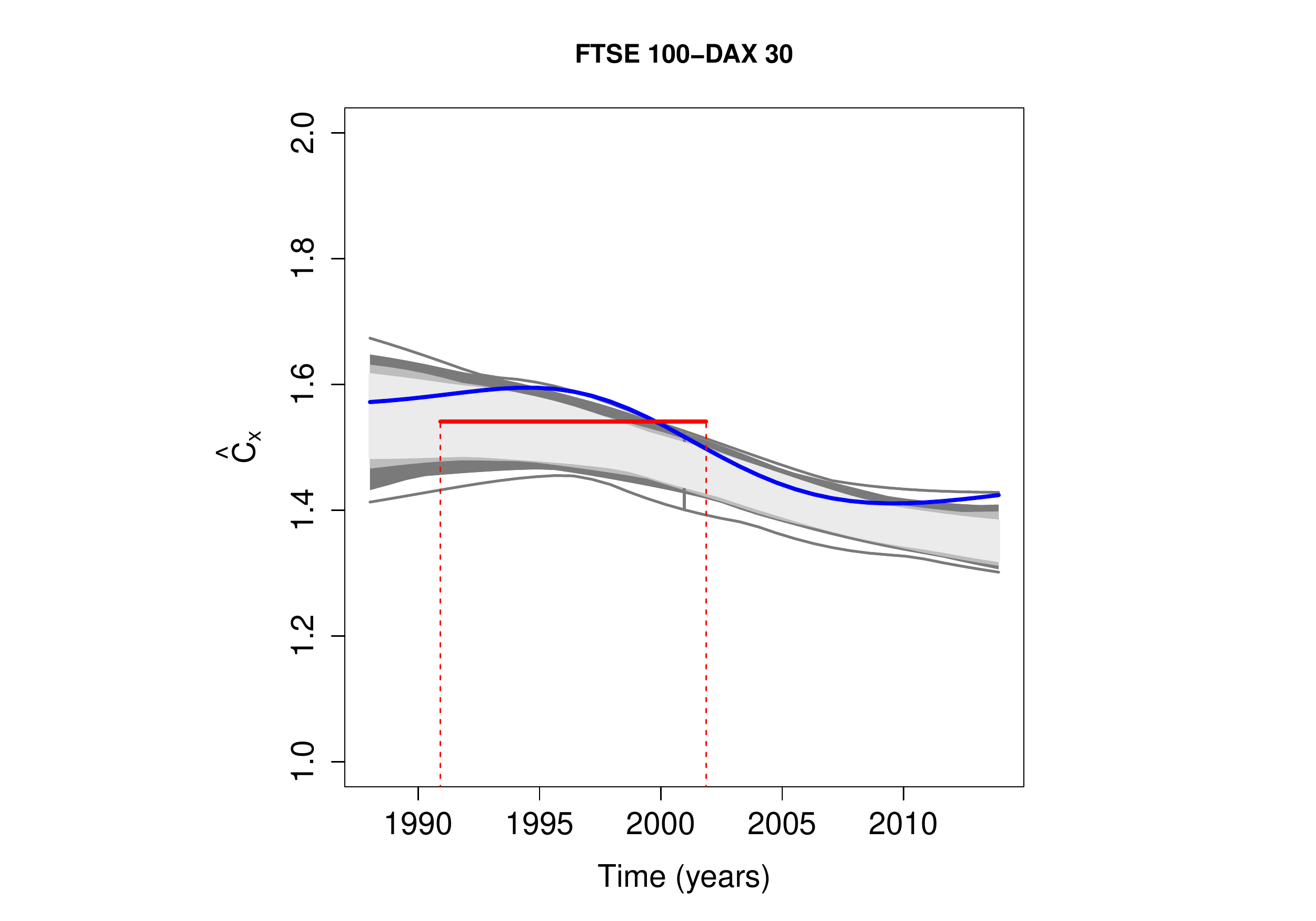}
\end{minipage}\\
      \footnotesize \rotatebox{90}{\textbf{\hspace{-1cm}L--L weights}}
\begin{minipage}[c]{0.4\linewidth}
 \hspace{-1.3cm}
  \includegraphics[scale=0.275]{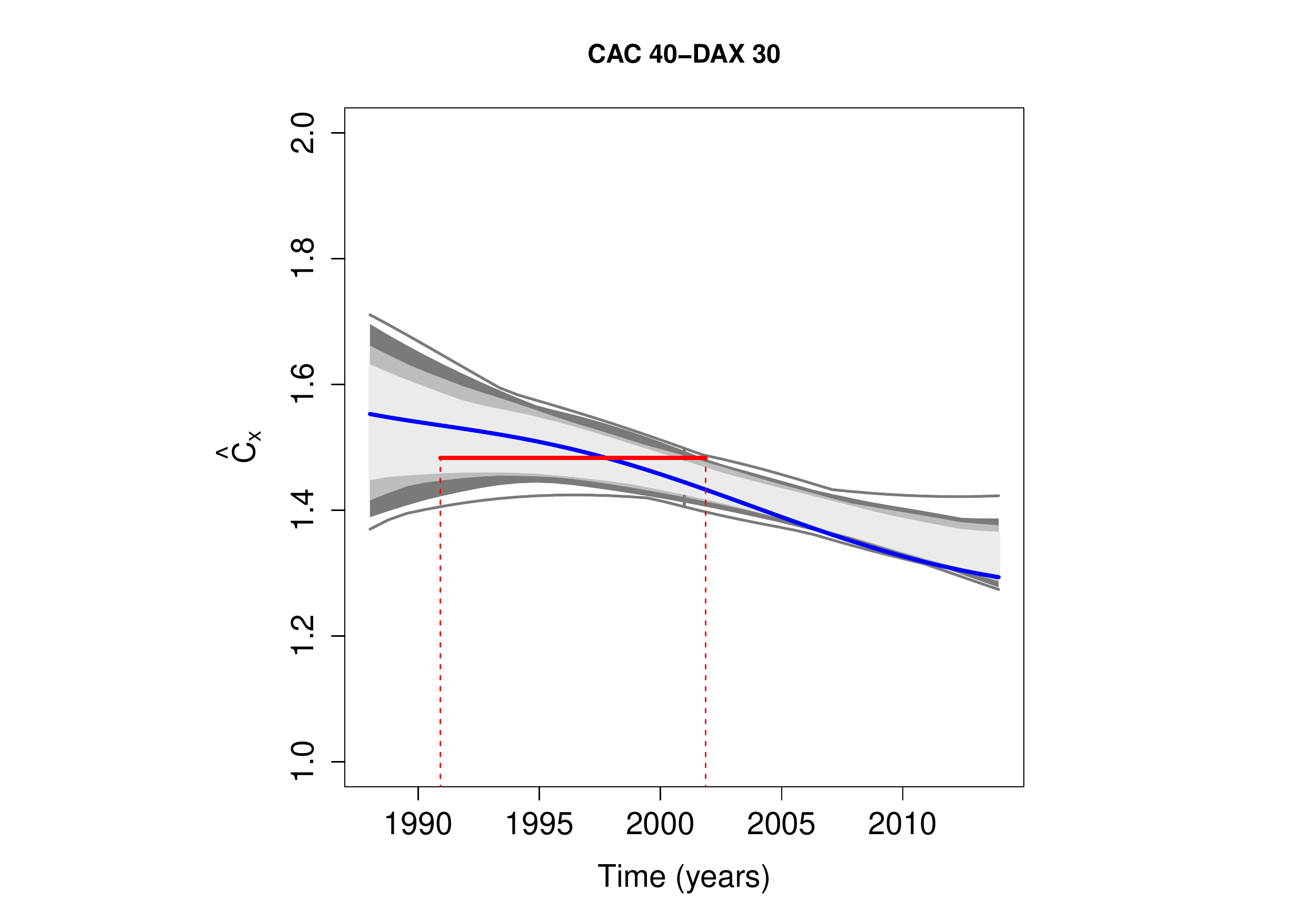}
 \end{minipage}&
\begin{minipage}[c]{0.4\linewidth}
\hspace{-2.8cm}
 \includegraphics[scale=0.275]{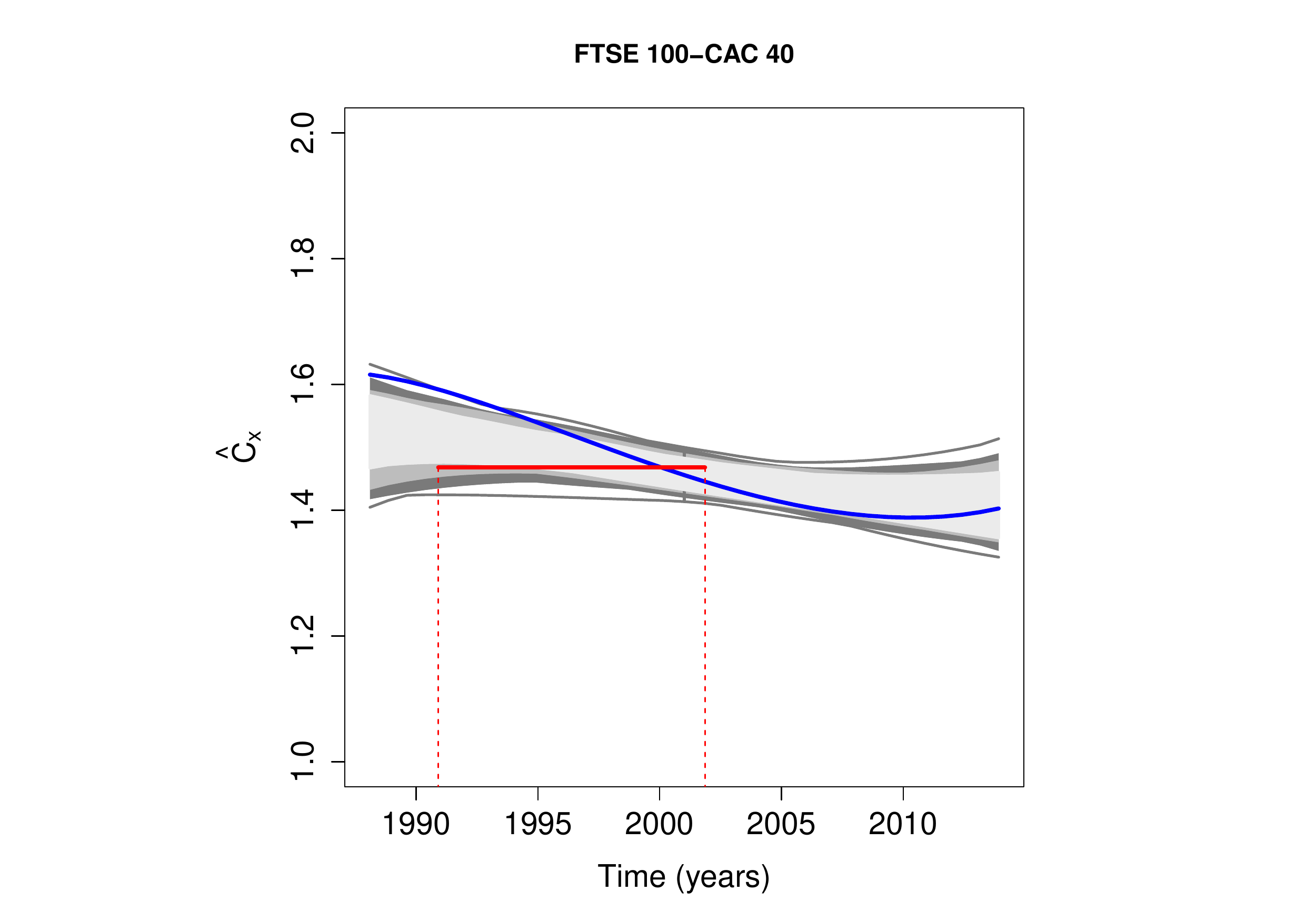}
\end{minipage}&
\begin{minipage}[c]{0.4\linewidth}
\hspace{-4.2cm}
 \includegraphics[scale=0.275]{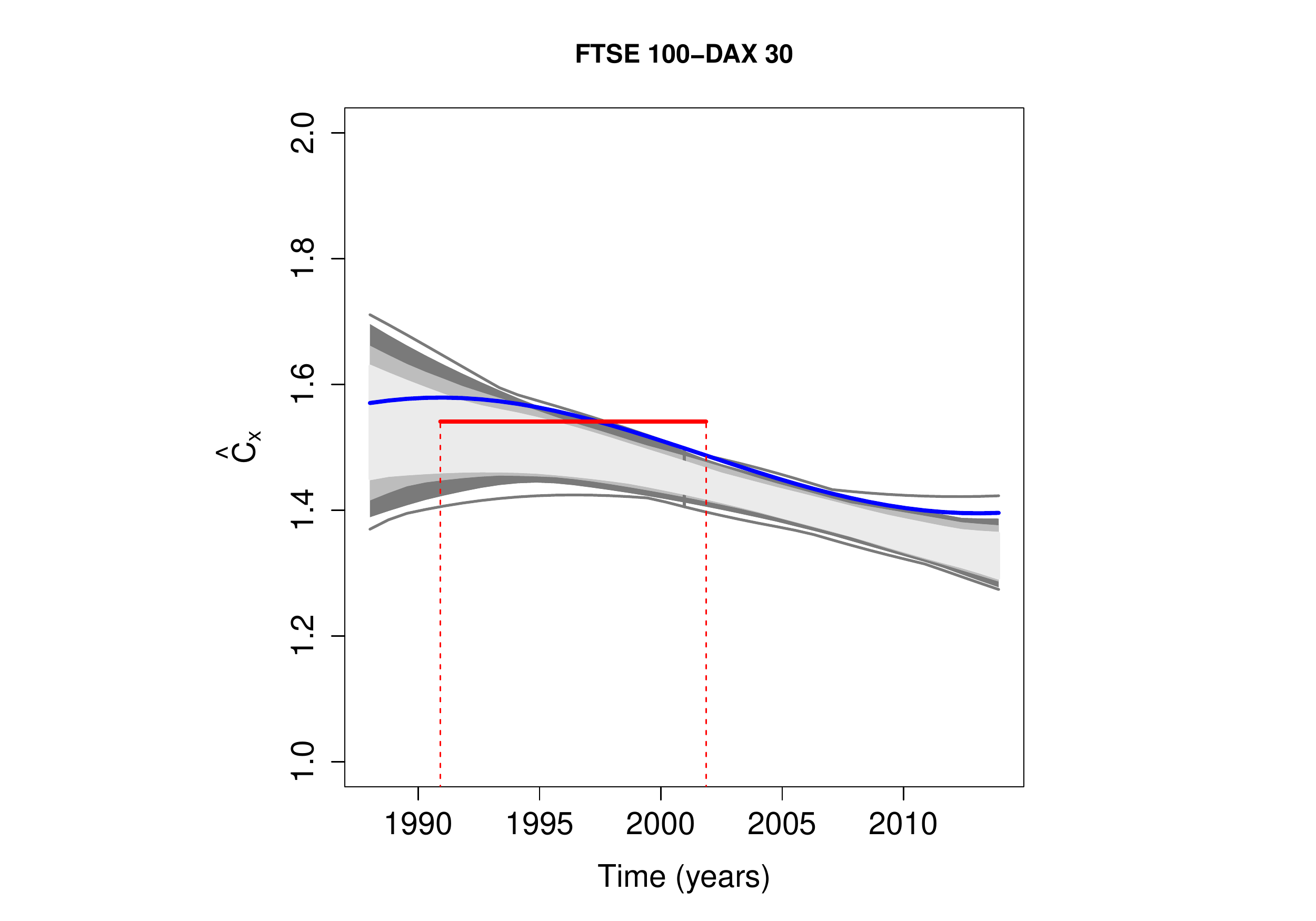}
\end{minipage}\\
 \end{tabular}
\caption{\footnotesize {Conditional extremal coefficients (solid blue lines) and functional boxplots (gray shadows) showing the 50\%, 75\%, and 95\% central regions (as defined by~\cite{sun2012functional}) based on 1000 bootstrap samples. The red lines represent the values from the analysis of Poon et al.~for the subperiod November 1990--November 2001 \citep[cf][Table 3]{PAL04}.}}
  \label{extcoe}
\end{figure}

Similar conclusions can be drawn from Figure~\ref{extcoe}, where we plot the conditional extremal coefficient, as defined in Section~\ref{Related conditional objects of interest}. The extremal coefficient is equal to $2-\chi$, and as such is equal to 2 under asymptotic independence, and takes values in $[1,2)$ under asymptotic dependence. Figure~\ref{extcoe} permits comparison with the results of \cite{PAL04}, who calculated $\chi$ over subperiods. The red lines in Figure~\ref{extcoe} represent the values from the analysis of Poon et al.~for the subperiod November 1990--November 2001 \citep[cf][Table 3]{PAL04}. Specifically, Poon et al.~report the following values of $\chi$ for: CAC--DAX, 0.517 (0.037); FTSE--CAC, 0.532 (0.035) and FTSE--DAX, 0.459 (0.039), with standard errors in parentheses. As can be seen from Figure~\ref{extcoe}, the magnitudes of the extremal coefficients estimated by Poon et al.~are in reasonable agreement with the ones computed with our methods when uncertainty is taken into account.

\section{Final comments}\label{Discussion}
This paper develops methods for modeling nonstationary extremal dependence structures, motivated by the need to assess the comovement of extreme losses in some leading European stock markets over recent years. Although there are many studies analyzing stock market integration over time \citep[see for example][]{KAL94, LS95, LS01, KS96, FR02, BN04, BN05, RN09}, few attempts have been made to assess the dynamics of extreme value dependence of stock market returns over time. An exception in this regard is the paper of \citet{PAL03}, which provides evidence suggesting increasing levels of extremal dependence for CAC, DAX and FTSE, although their analysis is essentially exploratory. The analysis performed in this paper reveals a more complete picture of this temporally-changing dependence. 

Two related approaches to the current work are the so-called spectral density ratio model of~\cite{CD14} and the spectral density regression model of~\cite{CC16}. While flexible, these approaches only apply to the setting where there are several pseudo-angles corresponding to the same value of the predictor---and thus they are inappropriate for our applied setting of interest. Our methods are more resilient in the sense that they do not require a sample of pseudo-angles for each value of the covariate, but apply more generally to a regression setting where each covariate value may only have a single corresponding pseudo-angle. Recent preprints of \citet{EAL16} and \citet{MAL17} suggest methods for estimating Pickands dependence function under covariate dependence, offering alternative approaches to those presented herein.

Computational experiments suggest that U-shaped angular surfaces are much more difficult to fit. Whilst absolute errors may become large at the boundaries when we have an unbounded density, when this is translated to other quantities ($H_x$ or $A_x$), the errors will be much less noticeable. Our methods have been developed with the setting of asymptotic dependence in mind, but certainly there is room for developing methodology for conditional modeling under asymptotic independence. Indeed, in common with any approach based on multivariate extreme value distributions, a limitation with our methods is that they will overestimate risk if data are asymptotically independent. Figure~\ref{chi.pdf} gave some indication of possible asymptotic independence near the beginning of the analysis period. As such, the need for developing conditional models able to cope with both asymptotic dependence and asymptotic independence is of utmost importance. 

\section*{Acknowledgements}
  We thank the Editor, AE, and two anonymous referees. We extend our thanks to Ant\'onio Rua, Vanda~In\'acio de Carvalho, and Claudia~Wehrhahn for helpful discussions. Part of this work was written while D.~C.~was visiting the University of Cambridge---Statistical Laboratory, and while M.~de.~C.~was visiting \textit{Banco de Portugal}.  The research was partially funded by by Funda\c c\~ao para a Ci\^encia e a Tecnologia, through UID/MAT/00006/2013 and by the Chilean National Science Foundation through Fondecyt 11121186, ``Constrained Inference Problems in Extreme Value Modeling.''

\invisiblesection{Supplementary material}
\sppart{supplementary material}
\textbf{Supplementary Monte Carlo evidence and empirical reports}.~The supplement includes additional simulation results, descriptive statistics for daily stock index negative returns, and further empirical analysis using the NGARCH-filtered residuals and LSCV bandwidths.
\vspace{-.5cm}

\invisiblesection{References}
\sppart{references} 
\begin{footnotesize}
\begingroup
\renewcommand{\section}[2]{}%

\endgroup
\end{footnotesize}

\begin{thebibliography}{9}
  \bibitem[Acar et~al.(2011)]{AAL11} 
    \textsc{Acar, E.~F., Craiu, R.~V.} and \textsc{Fang, Y.}~(2011).  
    \newblock Dependence calibration in conditional copulas: A nonparametric approach.
    \newblock \textit{Biometrics} \textbf{67} 445--453.

  \bibitem[Beirlant et~al.(2004)]{BAL04} 
    \textsc{Beirlant, J., Goegebeur, Y., Segers, J.} and \textsc{Teugels, J.}~(2004).  
    \newblock \textit{Statistics of Extremes: Theory and Applications}. 
    \newblock  Wiley, New York.


\bibitem[Brooks and Del Negro(2004)]{BN04}
  \textsc{Brooks, R.}~and~\textsc{Del Negro, M.} (2004). 
  \newblock The rise in comovement across national stock markets: market integration or IT bubble? 
  \textit{J. Empir. Finance} \textbf{11} 659--680.

\bibitem[Brooks and Del Negro(2005)]{BN05}
  \textsc {Brooks, R.}~and~\textsc{Del Negro, M.} (2005). 
  \newblock Country versus region effects in international stock returns. 
  \newblock \textit{J. Portfolio Management} \textbf{31} 67--72. 

  \bibitem[B{\"u}ttner and Hayo(2011)]{BH11}
    \textsc{B{\"u}ttner, D.} and \textsc{Hayo, B.} (2011).
    \newblock Determinants of European stock market integration. 
    \newblock \textit{Econ. Sys.} \textbf{35} 574--585.

  \bibitem[Castro Camilo and de Carvalho(2016)]{CC16}
    \textsc{Castro Camilo, D.} and \textsc{de Carvalho, M.}~(2016).
    \newblock Spectral density regression for bivariate extremes.
    \newblock \textit{Stoch. Env. Res. Risk Ass.} DOI: 10.1007/s00477-016-1257-z.


  \bibitem[Chavez-Demoulin and Davison(2005)]{CD05}
    \textsc{Chavez-Demoulin, V.} and \textsc{Davison, A.~C.} (2005).
    \newblock Generalized additive modelling of sample extremes. 
    \textit{J. R. Statist. Soc.}~C \textbf{54} 207--222.

\bibitem[Chen(1999)]{C99}
  \textsc{Chen, S.~X.} (1999).
  \newblock Beta kernel estimators for density functions.
  \newblock \textit{Comput. Statist. Data Anal.}~\textbf{31} 131--145.

\bibitem[Cline(1988)]{C88}
  \textsc{Cline, D.~B.~H.} (1988).
  \newblock Admissible kernel estimators of a multivariate density.
  \newblock \textit{Ann. Statist.}~\textbf{16} 1421--1427.

  \bibitem[Coles(2001)]{C01}
    \textsc{Coles, S.~G.} (2001). 
    \newblock \textit{An Introduction to the Statistical Modeling of Extreme Values}. 
    \newblock Springer, London.

  \bibitem[Coles and Tawn(1991)]{CT91}
    \textsc{Coles, S.~G.} and \textsc{Tawn, J.~A.} (1991).
    \newblock Modelling extreme multivariate events.
    \textit{J. R. Statist. Soc.}~B \textbf{53} 377--392.

  \bibitem[DasGupta(2008)]{dasgupta2008asymptotic}
	\textsc{DasGupta, A.} (2008). 
  	\newblock \textit{Asymptotic Theory of Statistics and Probability}.
	\newblock New York, Springer.


  \bibitem[Davison and Smith(1990)]{DS90}
    \textsc{Davison, A.~C.} and \textsc{Smith, R.~L.} (1990).
    \newblock Models for exceedances over high thresholds. 
    \newblock \textit{J. R. Statist. Soc.}~B \textbf{52} 393--442.

  \bibitem[de~Carvalho(2016)]{C16}
    \textsc{de~Carvalho, M.} (2016).
    \newblock Statistics of extremes: Challenges and opportunities. 
    \newblock In: \textit{Handbook of Extreme Value Theory and its Applications to Finance and Insurance}, Eds F.~Longin. 
    \newblock Wiley, Hoboken.

  \bibitem[de~Carvalho and Davison(2014)]{CD14}
    \textsc{de~Carvalho, M.} and \textsc{Davison, A.~C.} (2014).
    \newblock Spectral density ratio models for multivariate extremes. 
    \newblock \textit{J. Amer. Statist. Assoc.} \textbf{109} 764--776.

  \bibitem[de~Carvalho et~al.(2013)]{CAL13}
    \textsc{de~Carvalho, M., Oumow, B., Segers, J.} and \textsc{Warcho\l, M.} (2013).
    \newblock A Euclidean likelihood estimator for bivariate tail dependence.
    \newblock \textit{Comm. Statist. Theo. Meth.} \textbf{42} 1176--1192.

  \bibitem[de Haan and Resnick(1977)]{HR77}
    \textsc{de Haan, L.} and \textsc{Resnick, S.~I.} (1977).
    \newblock Limit theory for multivariate sample extremes. 
    \newblock \textit{Zeitschrift f{\"u}r Wahrscheinlichkeitstheorie und verwandte Gebiete} \textbf{40} 317--377.

  \bibitem[Eastoe and Tawn(2009)]{ET09}
    \textsc{Eastoe, E.} and \textsc{Tawn, J.} (2009).
    \newblock Modelling non-stationary extremes with application to surface level ozone. 
    \newblock \textit{J. R. Statist. Soc.}~C \textbf{58} 22--45.

  \bibitem[Eastoe(2009)]{E09}
    \textsc{Eastoe, E.~F.} (2009).
    \newblock A hierarchical model for non-stationary multivariate extremes: 
    A case study of surface-level ozone and NO$_\text{x}$ data in the UK. 
    \newblock \textit{Environmetrics} \textbf{20} 428--444.

    
  \bibitem[Engle(1982)]{engle1982autoregressive}
    \textsc{Engle, R.~F.} (1982).
    \newblock Autoregressive conditional heteroscedasticity with estimates of the variance of United Kingdom inflation.
    \newblock \textit{Econometrica} \textbf{50}, 987--1007.

\bibitem[{Escobar-Bach et~al.(2016)Escobar-Bach, Goegebeur and
  Guillou}]{EAL16}
  \textsc{Escobar-Bach, M., Goegebeur, J.}~and~\textsc{Guillou, A.} (2016). 
  Local robust estimation of the Pickands dependence function.
  \newblock \textit{Submitted}.

  \bibitem[Fermanian and Marten(2012)]{FM12} 
    \textsc{Fermanian, J.-D.} and \textsc{Marten, H.}~(2012).  
    \newblock Time-dependent copulas.
    \newblock \textit{J. Mult. Anal.} \textbf{110} 19--29.

\bibitem[Forbes and Rigobon(2002)]{FR02}
  \textsc{Forbes, K.} and \textsc{Rigobon, R.} (2002).
  \newblock No contagion, only interdependence: Measuring stock market comovements. 
  \newblock \textit{J. Finance} \textbf{57} 2223--2261.

%
\bibitem[Fratzscher(2002)]{F02}
  \textsc{Fratzscher, M.} (2002). 
    \newblock Financial market integration in Europe: On the effects of EMU on stock markets.
    \newblock \textit{Int. J. Fin. Econ.} \textbf{7} 165--193.


\bibitem[Gudendorf and Segers(2010)]{GS10}
  \textsc{Gudendorf, G.} and \textsc{Segers, J.} (2010). 
    \newblock Extreme-value copulas.
    \newblock In: \textit{Copula Theory and Its Applications}, Eds P.~Jaworski et al.
    \newblock Springer, New York.

  \bibitem[Geenens(2014)]{G14}
    \textsc{Geenens, G.} (2014).
    \newblock Probit transformation for kernel density estimation on the unit interval.
    \newblock \textit{J. Amer. Statist. Assoc.} \textbf{109} 346--359.

  \bibitem[Hardle(1990)]{H90}
    \textsc{Hardle, W.} (1990).
    \newblock \textit{Applied Nonparametric Regression}. 
    \newblock Cambridge University Press, Cambridge UK.

  \bibitem[Hardouvelis et~al.(2006)]{HAL06}
    \textsc{Hardouvelis, G.~A., Malliaropulos, D.} and \textsc{Priestley, R.} (2006).
    \newblock EMU and European stock market integration. 
    \newblock \textit{J. Bus.} \textbf{79} 365--392.

  \bibitem[Hastie et~al.(2001)]{HAL01}
    \textsc{Hastie, T., Tibshirani, R.} and \textsc{Friedman, J.} (2001).
    \textit{The Elements of Statistical Learning: Data Mining, Inference and Prediction}. 
    \newblock Springer, New York.

  \bibitem[Heffernan(2000)]{H00}
    \textsc{Heffernan, J.~E.} (2000).
    \newblock Directory of coefficients of tail dependence.
    \newblock\textit{Extremes} \textbf{3} 279--290.
%
  \bibitem[Heffernan and Tawn(2004)]{HT04}
    \textsc{Heffernan, J.~E.} and \textsc{Tawn, J.~A.} (2004).
    \newblock A conditional approach for multivariate extreme values (with Discussion). 
    \newblock\textit{J. R. Statist. Soc.}~B \textbf{66} 497--546.

  \bibitem[Huser and Genton(2016)]{HG14}
    \textsc{Huser, R.} and \textsc{Genton, M.~G.} (2016).
    \newblock Non-stationary dependence structures for spatial extremes. 
    \newblock \textit{Journal of Agricultural, Biological, and Environmental Statistics} \textbf{21} 470--491.

  \bibitem[James(2012)]{J12}
    \textsc{James, H.} (2012).
    \newblock \textit{Making the {E}uropean {M}onetary {U}nion}. 
    \newblock Harvard University Press, Cambridge MA: 

  \bibitem[Jonathan et~al.(2014)]{JAL14}
    \textsc{Jonathan, P., Ewans, K.} and \textsc{Randell, D.} (2014).
    \newblock Non-stationary conditional extremes of northern north sea storm characteristics.
    \newblock \textit{Environmetrics} \textbf{25} 172--188.

  \bibitem[Jones and Henderson(2007)]{JH07}
    \textsc{Jones, M. C.} and \textsc{Henderson, D. A.} (2007).
    \newblock Kernel-type density estimation on the unit interval.
    \newblock \textit{Biometrika} \textbf{94} 977--984.

\bibitem[Longin and Solnik(1995)]{LS95} 
  \textsc{Longin, F.}~and~\textsc{Solnik, B.}~(1995).
  \newblock Is the correlation in international equity returns constant: 1960--1990? 
  \newblock \textit{J. Int. Money Finance} \textbf{14} 3--26.

\bibitem[Longin and Solnik(2001)]{LS01} 
  \textsc{Longin, F.}~and~\textsc{Solnik, B.}~(2001).
  \newblock Extreme correlation of international equity market
  \newblock \textit{J. Finance} \textbf{56} 649--676.

\bibitem[Karolyi and Stulz(1996)]{KS96}
  \textsc{Karolyi, G. A.} and \textsc{Stulz, R. M.} (1996). 
  \newblock Why do markets move together? An investigation of US--Japan stock return comovements. 
  \newblock \textit{J. Finance} \textbf{51} 951--986.

\bibitem[Kim et~al.(1998)]{kim1998stochastic}
    \textsc{Kim, S., Shephard, N.} and \textsc{Chib, S.} (1998).
    \newblock Stochastic volatility: Likelihood inference and comparison with ARCH models.
    \newblock \textit{Rev. Econ. Stud.} \textbf{65} 361--393.
    
  \bibitem[Kim et~al.(2005)]{KAL05}
    \textsc{Kim, S.~J., Moshirian, F.} and \textsc{Wu, E.} (2005).
    \newblock Dynamic stock market integration driven by the European monetary union: An empirical analysis.
    \newblock \textit{J. Bank. Fin.} \textbf{29} 2475--2502.

\bibitem[King et~al.(1994)]{KAL94}
    \textsc{King, M., Sentana, E.}~and~\textsc{Sushil, W.} (1994).
    \newblock Volatility and links between national stock markets.
    \newblock \textit{Econometrica} \textbf{62}, 901--933.

  \bibitem[Koenker(2005)]{K05}
    \textsc{Koenker, R.} (2005).
    \newblock \textit{Quantile Regression}.
    \newblock Cambridge University Press, Cambridge UK.

  \bibitem[Lane(2012)]{L12}
    \textsc{Lane, P.~R.} (2012).
    \newblock The European sovereign debt crisis. 
    \newblock \textit{J. Econ. Persp.} \textbf{26} 49--67.


\bibitem[{Mhalla et~al.(2017)Mhalla, {Chavez-Demoulin} and Naveau}]{MAL17}
  \textsc{Mhalla, L., Chavez-Demoulin, V.}~and~\textsc{Naveau, P.} (2017). 
  \newblock Non linear models for extremal dependence.
  \newblock \textit{Available at SSRN:} \url{https://ssrn.com/abstract=2836587}.

    
  \bibitem[Nadaraya(1964)]{N64}
    \textsc{Nadaraya, E.~A.} (1964).
    \newblock On estimating regression. 
    \newblock \textit{Theo. Prob. Appl.} \textbf{9} 141--142.

  \bibitem[Natural Environment Research Council(1975)]{NERC75}
    \textsc{Natural Environment Research Council} (1975).
    \newblock\textit{Flood Studies Report}.
    \newblock Natural Environment Research Council, London.


  \bibitem[Patton(2006)]{P06}
    \textsc{Patton, A. J.}~(2006). 
    \newblock Modelling asymmetric exchange rate dependence. 
    \newblock \textit{Int. Econ. Rev.} \textbf{47} 527--556.

  \bibitem[Pickands(1981)]{P81}
    \textsc{Pickands, J.} (1981).
    \newblock Multivariate extreme value distributions. In:
    \newblock \textit{Proceedings 43rd Session International Statistical Institute}, pp.~859--878.

{\bibitem[Polanski(2001)]{P01}
    \textsc{Polanski, A. M.}~(2001). 
    \newblock Bandwidth selection for the smoothed bootstrap percentile method. 
    \newblock \textit{Comput. Statist. Data Anal.} \textbf{36} 333--349.}

  \bibitem[Poon et~al.(2003)]{PAL03}
    \textsc{Poon, S.-H., Rockinger, M.} and \textsc{Tawn, J.~A.} (2003).
    \newblock Modelling extreme-value dependence in international stock markets. 
    \newblock \textit{Statist. Sinica} \textbf{13} 929--953.

  \bibitem[Poon et~al.(2004)]{PAL04}
     \textsc{Poon, S.-H., Rockinger, M.} and \textsc{Tawn, J.~A.} (2004).
    \newblock Extreme-value dependence in financial markets: Diagnostics, models and financial implications. 
    \newblock \textit{Rev. Fin. Stud.} \textbf{17} 581--610.

 \bibitem[Resnick(2007)]{R07}
   \textsc{Resnick, S.}~(2007).
   \newblock \textit{Heavy-Tail Phenomena: Probabilistic and Statistical Modeling}.
   \newblock Springer, New York.

  \bibitem[Rua and Nunes(2009)]{RN09} 
    \textsc{Rua, A.} and \textsc{Nunes, L.~C.}~(2009).  
    \newblock International comovement of stock market returns: A wavelet analysis.
    \newblock \textit{J. Emp. Finance} \textbf{16} 632--639.


  \bibitem[Silverman and Young(1987)]{SY87}
    \textsc{Silverman, B.~W.} and \textsc{Young, G.~A.}~(1987).
    \newblock The bootstrap: To smooth or not to smooth?
    \newblock \textit{Biometrika} \textbf{74} 469--479.

  \bibitem[Sun and Genton(2011)]{sun2012functional}
    \textsc{Sun, Y.} and \textsc{Genton, M.~G.}~(2011).
    \newblock Functional boxplots.
    \newblock \textit{J. Comp. Graph. Statist.} \textbf{20} 316--334.
    
  \bibitem[Veraverbeke et~al.(2007)]{VAL07}
    \textsc{Veraverbeke, N., Omelka, M.} and \textsc{Gijbels, I.}~(2007).
    \newblock Estimation of a conditional copula and association measures.
    \newblock \textit{Scand. J. Statist.} \textbf{38} 766--780.

  \bibitem[Tawn(1988)]{T88}
    \textsc{Tawn, J.~A.}~(1988).
    \newblock Bivariate extreme value theory: Models and estimation. 
    \newblock \textit{Biometrika} \textbf{75} 397--415.

  \bibitem[Tawn(1990)]{T90}
    \textsc{Tawn, J.~A.}~(1990).
    \newblock Modelling multivariate extreme value distributions.
    \newblock \textit{Biometrika} \textbf{77} 245--253.

  \bibitem[{Wand and Jones(1995)}]{WJ95}
    \textsc{Wand, M.~P.} and \textsc{Jones, M.~C.} (1995).
    \newblock \textit{Kernel Smoothing}.
    \newblock Chapman \& Hall/CRC, Boca Raton FL.

  \bibitem[Watson(1964)]{W64}
    \textsc{Watson, G.~S.} (1964).
    \newblock Smooth regression analysis.
    \newblock \textit{Sankhy{\=a}}~A \textbf{26} 359--372.
\end{thebibliography}
\end{document}